\documentclass[english,twocolumn,prb]{revtex4-2}

\usepackage[T1]{fontenc}
\usepackage[utf8]{inputenc}
\usepackage{color}
\usepackage{babel}
\usepackage{verbatim}
\usepackage{refstyle}
\usepackage{amsmath}
\usepackage{amssymb}
\usepackage{graphicx}
\usepackage{wasysym}
\usepackage{esint}
\usepackage{microtype}
\usepackage[bookmarks=true,bookmarksnumbered=false,bookmarksopen=false,
 breaklinks=false,pdfborder={0 0 0},pdfborderstyle={},backref=false,colorlinks=true]
 {hyperref}
\hypersetup{pdftitle={Temperature dependence of the superfluid density},
 pdfauthor={Anton Khvalyuk, Thibault Charpentier, Nicolas Roch, Benjamin Sac\'ep\'e, Mikhail Feigel'man},
 pdfsubject={Superconductivity}}

\makeatletter

\AtBeginDocument{\providecommand\secref[1]{\ref{sec:#1}}}
\AtBeginDocument{\providecommand\appref[1]{\ref{app:#1}}}
\AtBeginDocument{\providecommand\subsecref[1]{\ref{subsec:#1}}}
\AtBeginDocument{\providecommand\figref[1]{\ref{fig:#1}}}
\providecommand{\tabularnewline}{\\}
\RS@ifundefined{subsecref}
  {\newref{subsec}{name = \RSsectxt}}
  {}
\RS@ifundefined{thmref}
  {\def\RSthmtxt{theorem~}\newref{thm}{name = \RSthmtxt}}
  {}
\RS@ifundefined{lemref}
  {\def\RSlemtxt{lemma~}\newref{lem}{name = \RSlemtxt}}
  {}

\usepackage{chngcntr} 

\newref{fig}{ refcmd={Figure~\ref{#1}} }
\newref{tab}{ refcmd={Table~\ref{#1}} }
\newref{eq}{ refcmd={(\ref{#1})} }
\newref{subsec}{ refcmd={Subsection~\ref{#1}} }
\newref{sec}{ refcmd={Section~\ref{#1}} }
\newref{app}{ refcmd={Appendix~\ref{#1}} }
\newref{part}{ refcmd={Part~\ref{#1}} }

\makeatother

\begin{document}
\title{Near-power-law temperature dependence of the superfluid stiffness
in strongly disordered superconductors }
\author{Anton V. Khvalyuk}
\email{anton.khvalyuk@lpmmc.cnrs.fr}

\affiliation{LPMMC, Université Grenoble Alpes, 38000 Grenoble, France}
\affiliation{L.D. Landau Institute for Theoretical Physics, 142432 Chernogolovka,
Moscow region, Russia}
\author{Thibault Charpentier}
\affiliation{Université Grenoble Alpes, CNRS, Grenoble INP, Institut Néel, 38000
Grenoble, France}
\author{Nicolas Roch }
\affiliation{Université Grenoble Alpes, CNRS, Grenoble INP, Institut Néel, 38000
Grenoble, France}
\author{Benjamin Sacépé}
\email{benjamin.sacepe@neel.cnrs.fr}

\affiliation{Université Grenoble Alpes, CNRS, Grenoble INP, Institut Néel, 38000
Grenoble, France}
\author{Mikhail V. Feigel'man}
\email{mvfeigel@gmail.com}

\affiliation{Nanocenter CENN, 1000 Ljubljana, Slovenia}
\affiliation{Floralis \& LPMMC, Université Grenoble Alpes, 38000 Grenoble, France}
\affiliation{L.D. Landau Institute for Theoretical Physics, 142432 Chernogolovka,
Moscow region, Russia}
\date{\today{}}
\begin{abstract}
In BCS superconductors, the superfluid stiffness is virtually constant
at low temperature and only slightly affected by the exponentially
low density of thermal quasiparticles. Here, we present an experimental
and theoretical study on the temperature dependence of superfluid
stiffness $\Theta\left(T\right)$ in a strongly disordered pseudo-gaped
superconductor, amorphous $\text{InO}_{x}$, which exhibits non-BCS
behavior. Experimentally, we report an unusual power-law suppression
of the superfluid stiffness $\delta\Theta\left(T\right)\propto T^{b}$
at $T\ll T_{c}$, with $b\sim1.6$, which we measured via the frequency
shift of microwave resonators. Theoretically, by combining analytical
and numerical methods to a model of a disordered superconductor with
pseudogap and spatial inhomogeneities of the superconducting order
parameter, we found a qualitatively similar low-temperature power-law
behavior with exponent $b\sim1.6-3$ being disorder-dependent. This
power-law suppression of the superfluid density occurs mainly due
to the broad distribution of the superconducting order parameter that
is known to exist in such superconductors~\citep{Sacepe_2011_for-pair-preformation},
even moderately far from the superconductor-insulator transition.
The presence of the power-law dependence $\delta\Theta\left(T\right)\propto T^{b}$
at low $T\ll T_{c}$ demonstrates the existence of low-energy collective
excitations; in turn, it implies the presence of a new channel of
dissipation in inhomogeneous superconductors caused by sub-gap excitations
that are not quasiparticles. Our findings have implications for the
use of strongly disordered superconductors as superinductance in quantum
circuits.
\end{abstract}
\maketitle

\section{Introduction\protect\label{sec:Introduction}}

Superconducting superinductors, proposed about decade ago~\citep{Mooij05,Mooij06,Doucot12,Kitaev13,Groszowski18,grunhaupt2019granular}
as important elements of quantum circuits, now constitute an intensively
developing sub-field in the physics of strongly disordered superconductors,
as some selected examples~\citep{Astafiev12,Mooij16,DeGraaf18,Bylander19,Gershenson19,Astafiev22}
demonstrate. Superconducting films used for the construction of superinductors
must combine high kinetic inductance per square $L_{K}$ with low
dissipation in the microwave frequency range. Large $L_{K}$ corresponds
to small 2D superfluid stiffness $\Theta=(\hbar c/2e)^{2}/L_{K}$,
and the latter can be achieved close to the Superconducting-Insulator
Transition (SIT). The condition of low losses naturally points to
a family of superconducting materials in which the SIT occurs without
closing the single-particle spectral gap~\citep{SFK-review-2020}:
Indium Oxide~\citep{Sacepe_2011_for-pair-preformation}, Titanium
Nitride~\citep{Sacepe10}, Niobium Nitride~\citep{NbN}, and, possibly,
granular Aluminum~\citep{GrAl}. Indeed, the absence of low-energy
quasiparticles naturally decreases the absorption of microwave electromagnetic
field. However, it does not guarantee the absence of other channels
of dissipation.

Far away from the SIT, moderately dirty superconductors described
by the semiclassical BCS-like theory have a sharp gap $\Delta$ in
the excitation spectrum, leading to exponentially low density of quasiparticles
at low temperatures, i.e. $n\left(T\right)\propto\exp(-\Delta/T)$.
One then expects similar temperature dependence in all other physical
quantities, e.g., $\Theta\propto\int dE\,\nu_{\text{qp}}\left(E\right)\,E\,e^{-E/T}\sim T\,n\left(T\right)$,
where $\nu_{\text{qp}}\left(E\right)$ stands for the single-particle
density of states (DoS). %
However, such a connection between the single-particle DoS and the
temperature dependence of $\Theta$ is not observed experimentally
at strong disorder. In particular, it was found in Ref.~\citep{Klapwijk2013}
that the temperature dependence of the kinetic inductance per square
$L_{K}\propto1/\Theta$ in strongly disordered TiN films is much stronger
than the one predicted within usual Mattis-Bardeen model~\citep{MattisBardeen}
for the single-particle DoS extracted by Scanning Tunneling Spectroscopy
in the same experiment.

In the present paper we report even more striking behavior of the
low-temperature suppression of superfluid stiffness, $\delta\Theta(T)=\Theta(0)-\Theta(T)$,
in strongly disordered amorphous $\text{InO}_{x}$, as deduced from
the dispersion law of one-dimensional plasmon waves in a long superconducting
stripe. Namely, we observe the power-law-like temperature dependence
$\delta\Theta/\Theta\sim\left(T/T_{0}\right)^{b}$ with $b\sim1.6-1.7$
and $T_{0}\sim\left(1-7\right)\times10^{1}\,\text{K}$. This unusual
dependence as well as the large magnitude of the effect confidently
defy any semiclassical explanation based on mean thermal quasiparticle
density. 

Addressing $\delta\Theta$ by a semiclassical approach is additionally
hindered by the fact that strongly disordered $\text{InO}_{x}$ is
known~\citep{Sacepe_2011_for-pair-preformation,Dubouchet_2018_Preformation-of-Pairs}
to posses a hard gap $\Delta_{P}\sim5\text{ K}$ in the single-particle
DoS \emph{even above} the transition temperature $T_{c}\sim2\text{ K}$.
As a result, the electron-hole quasiparticle excitations can be safely
neglected at $T\ll T_{c}$ and thus cannot account for the experimentally
observed suppression of the superfluid stiffness~$\Theta$ with temperature
reported in this paper.

To understand the latter, one instead should turn to the properties
of the collective modes. While certain contribution to $\delta\Theta$
comes from the aforementioned 1D long-wavelength plasmonic excitations,
a simple calculation presented in~\secref{Experiment} below quickly
demonstrates that the associated effect is too weak at low temperatures.
One should thus address the behavior of short-range collective excitations.
This latter issue was initially considered within the approximate
analytical theory of Ref~\citep{Feigelman_Microwave_2018}, where
it was found that low-energy collective excitations are expected to
exist in a broad range of disorder, not necessarily close to the SIT.
Unfortunately, the approximation of the space-independent order parameter~$\Delta$
employed in Ref.~\citep{Feigelman_Microwave_2018} was later found
to be to rather crude~\citep{op-distribution-paper}, prompting the
issue of short-range low-energy collective modes to be reconsidered.

In the present paper, we show that a proper account of strong inhomogeneity
of the order parameter~$\Delta(\mathbf{r})$ in a specific model
of a pseudo-gaped superconductor allows one to describe near-power-law
temperature dependence of the superfluid stiffness~$\Theta$, with
the order of magnitude of the effect comparable to that in the experimental
data. We find that the exponent $b$ of the power law decreases with
the increase of disorder, with $b\in(1.5,3)$ in a wide range of disorder
parameters. We predict such behavior in a broad range of low temperatures,
$\lambda\Delta_{0}\leq T\leq\Delta_{0}/2$, where $\Delta_{0}$ is
the typical energy scale of the order parameter, and $\lambda\ll1$
is the dimensionless Cooper coupling constant. The broad distribution
of the order parameter $P(\Delta)$, similar to the one found in Ref.~\citep{op-distribution-paper},
plays a crucial role in our theoretical description.

The paper is organized as follows: \secref{Experiment} presents the
experimental results; \secref{Model-and-Theory} formulates the theoretical
model; the results of the calculations (both numerical and analytical)
are presented in \secref{Results}. Qualitative comparison between
experimental and theoretical results, discussions of our findings,
and conclusions are present in \secref{Discussion}. The paper is
supplemented by a number of Appendices containing various technical
details of the analytical approach.

\section{Suppression of superfluid stiffness in strongly disordered amorphous
indium oxide resonators\protect\label{sec:Experiment}}

\subsection{Experimental results}

The low-temperature evolution of superfluid stiffness in strongly
disordered superconductors can be probed experimentally by studying
the shift in resonance frequency of a microwave resonator due to the
increase of kinetic inductance $L_{K}\propto1/\Theta$ with temperature~\citep{Klapwijk2013,Grunhaupt2018},
a method originally developed for the field of microwave kinetic inductance
detectors (MKIDs)~\citep{Day2003}.

In this work we fabricated open-ended microstrip resonators made from
five strongly disordered amorphous indium oxide thin films of different
disorder and constant thickness $d=40~\mathrm{nm}$. The film disorder
is characterized \emph{in~situ} at cryogenic temperatures using a
co-deposited Hall bar and is shown to be increasingly strong, as evidenced
by the significant reduction of critical temperature $T_{c}$ and
enhancement of normal-state resistance above the superconducting transition.
Details of sample disorder are summarized in Table~\ref{tab:exp_data}.
Importantly the disorder range shown here is known to be characterized
by the presence of a pseudogap and spatial inhomogeneities of the
order parameter~\citep{Sacepe_2011_for-pair-preformation}.

The microwave resonators are long ($L=3.5~\mathrm{mm}$) and narrow
($w=1~\mathrm{\mu m}$) indium oxide strips deposited on a silicon
dielectric substrate under which a gold metallic layer acts as a ground
plane. Through capacitive coupling to a microwave feedline, a collective
motion of Cooper pairs in the resonator can be excited by an AC drive,
giving rise to the propagation of plasmon waves with velocity $v=1/\sqrt{lc}$,
(where $l$ and $c$ are kinetic inductance and capacitance per unit
length respectively)~\citep{Kulik1973,Mooij1985,Camarota2001}.

The open boundary conditions at each ends of the strip lead to Fabry-Pérot-like
standing wave resonances with nearly linear dispersion relation $f(k_{n})\sim(1/2\pi)vk_{n}$,
where $k_{n}=n\pi/L$ is the wave~vector for mode $n$.

The superfluid stiffness of a given $\text{InO}_{x}$ resonator can
be extracted from the velocity of plasmons since $v\propto1/\sqrt{l}\propto\sqrt{\Theta}$.
Upon increase of temperature the superfluid density decreases, leading
to a decrease of resonance frequency. The evolution of $\Theta(T)$
with temperature can therefore be extracted from the relative frequency
shift $\delta f(T)/f=\left[f(20~\mathrm{mK})-f(T)\right]/f(20~\mathrm{mK})\approx(1/2)\,\delta\Theta(T)/\Theta$
defined with respect to the lowest achievable temperature. Further
details on the experimental setup and samples can be found in~\citep{Charpentier2023_thesis}.

\begin{figure*}
\begin{centering}
\includegraphics[scale=0.6]{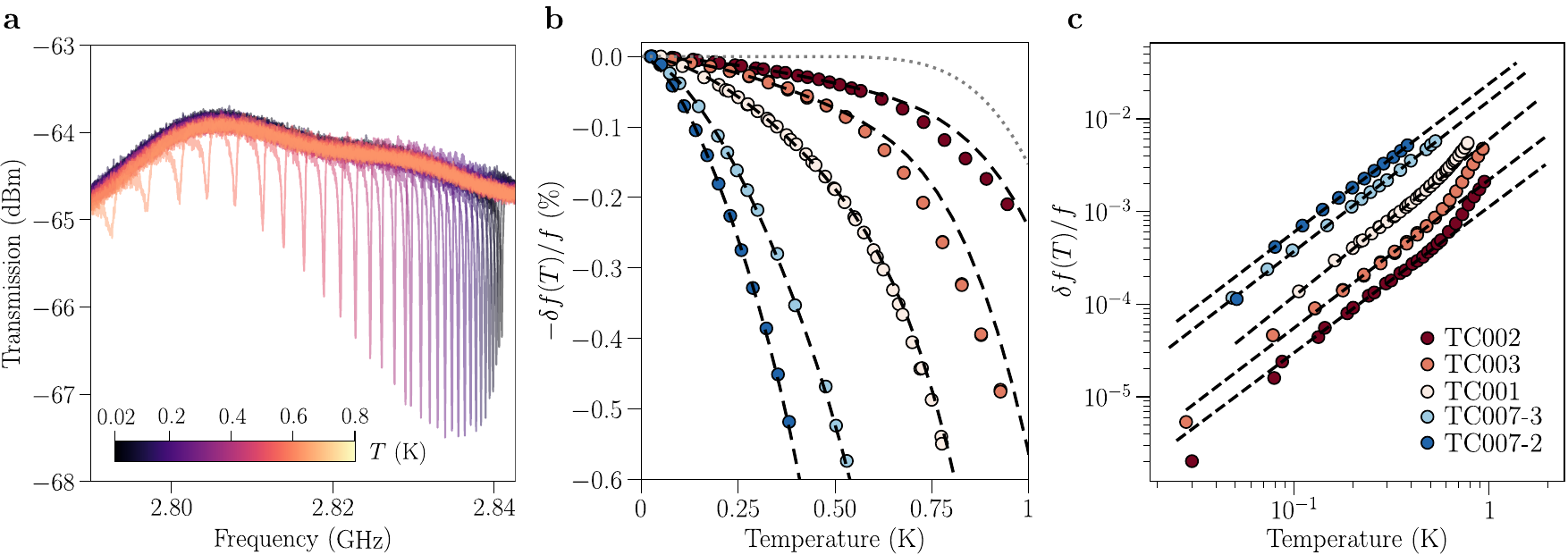}
\par\end{centering}
\caption{\protect\label{fig:data} Temperature frequency shift of amorphous
indium oxide microwave resonators. \textbf{a}.~Transmission resonance
as a function of frequency of sample TC007-3 for different temperatures.
Upon increasing temperature, the resonance shifts towards lower frequencies.
\textbf{b}.~Relative frequency shift $\left[f(20~\mathrm{mK})-f(T)\right]/f(20~\mathrm{mK})$
for the five samples of Table~\ref{tab:exp_data}. The low-temperature
part of the data departs from the standard Mattis-Bardeen (MB) behavior
(shown in grey dotted line for $\Delta_{1}=6.8\,\text{K}$) but can
be fitted by adding a power-law contribution as in Eq.~(\ref{eq:power_law}),
as shown by the black dashed lines. The power law exponent is $1.57\protect\leq b\protect\leq1.69$.
\textbf{c.~}Frequency shift in log-log scale. At the lowest temperatures
all curves show a power law with exponent $b\sim1.6$. Above $0.5\,\text{K}\sim0.2~T_{c}$
the MB mechanism kicks in and the frequency shift retrieves an exponential
behavior for the moderately disordered samples (TC002, TC003 and TC001). }
\end{figure*}

The main experimental data are shown in Fig.~\ref{fig:data}\textbf{a}
which displays the transmission as a function of temperature for sample
TC007-3. The resonance exhibits a frequency shift upon increasing
the temperature from $20\,\text{mK}$ up to $0.8\,\text{K}$. The
resulting relative frequency shift $\delta f(T)/f$ is plotted in
Fig.~\ref{fig:data}\textbf{b} together with that of the other four
samples listed in Table~\ref{tab:exp_data}. We readily see that
$\delta f(T)/f$ is not exponentially suppressed at low temperatures,
in stark contrast with usual BCS dirty superconductors in which the
superfluid density is suppressed as $\delta\Theta(T)\propto\exp\left\{ -\Delta/T\right\} $
due to thermally activated quasiparticles. In log-log scale shown
in Fig.~\ref{fig:data}\textbf{c}, these data exhibit a nearly power-law
dependence, $\delta f(T)/f\propto\left(T/T_{0}\right)^{b}$, at the
lowest temperature with an exponent $b\sim1.7$ independent of disorder.
This non-BCS power-law dependence of the frequency shift is the central
result of this work.

\begin{table}
\caption{\protect\label{tab:exp_data} Critical temperature, normal-state resistance
and zero-temperature superfluid stiffness of disordered indium oxide
samples measured in this work. Power law exponent $b$ and temperature
scale $T_{0}$ extracted from the temperature-induced frequency shift
are also displayed. Where possible, the energy gap $\Delta_{1}$ corresponding
to the best fit to Mattis-Bardeen theory at higher temperatures is
presented.}
\begin{ruledtabular} %
\begin{tabular}{lllllcl}
Sample & $T_{c}\,(\text{K})$ & $R_{n}~\mathrm{(k\Omega/\square)}$ & $\Theta(0)\,(\text{K})$ & $T_{0}\,(\text{K})$ & $b$ & $\Delta_{1}\,(\text{K})$\tabularnewline
TC002 & 3.2 & 1.45 & 13.3 & 74.5 & 1.57 & 6.8\tabularnewline
TC003 & 2.8 & 2.0 & 8.6 & 45.8 & 1.60 & 6.0\tabularnewline
TC001 & 2.2 & 3.4 & 4.4 & 20.5 & 1.69 & 4.9\tabularnewline
TC007-3 & 1.6 & 5.95 & 1.9 & 12.9 & 1.62 & --\tabularnewline
TC007-2 & 1.4 & 7.47 & 1.4 & 10.3 & 1.60 & --\tabularnewline
\end{tabular}\end{ruledtabular}
\end{table}

Inspecting Fig.~\ref{fig:data}\textbf{c} in the high temperature
range, we see that deviations from this power law with a stronger
frequency shift occur for temperatures above about $0.5\,\text{K}$,
as shown by the upward curvature of the data in Fig.~\ref{fig:data}\textbf{c}.
We conjecture that these high-temperature deviations most likely relate
to the thermally activated quasiparticles and can be phenomenologically
described by standard Mattis-Bardeen (MB) theory~\citep{MattisBardeen}
to account for thermal activation with energy scale $\Delta_{1}$.
We thus describe the entire temperature dependence of the frequency
shift with:
\begin{equation}
\frac{\delta f(T)}{f}=\left(\frac{T}{T_{0}}\right)^{b}+\left(\frac{\delta f(T)}{f}\right)_{\mathrm{MB}},
\label{eq:power_law}
\end{equation}
where $(\delta f(T)/f)_{\mathrm{MB}}\propto\exp\left\{ -\Delta_{1}/T\right\} $
is the MB contribution accounting for the high-$T$ deviations. The
resulting dashed lines in Fig.~\ref{fig:data}\textbf{b} fit well
data with $b=1.6-1.7$ and $\Delta_{1}\sim2~T_{c}$, which is a reasonable
approximation of the single particle gap for the moderately disordered
samples~\citep{Sacepe2015}. For the samples TC007-3 and TC007-2
(in blue and light blue on Fig.~\ref{fig:data}\textbf{b}~and~\textbf{c}),
there is no MB contribution in the measured temperature range.

\begin{figure}
\begin{centering}
\includegraphics[scale=0.5]{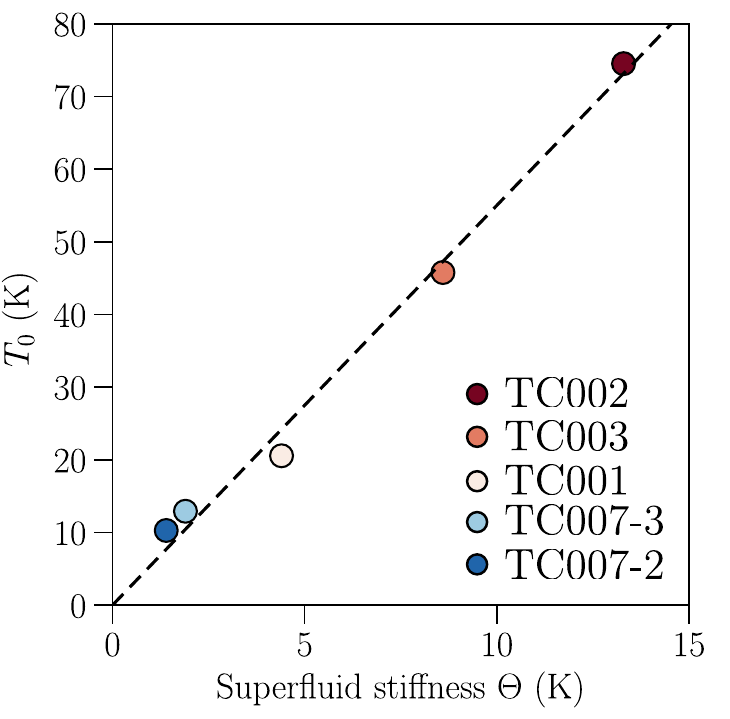} 
\par\end{centering}
\caption{\protect\label{fig:T0_vs_Theta} Dependence of the energy scale $T_{0}$
obtained from the fit of the frequency shift with Eq.~(\ref{eq:power_law})
as a function of the zero-temperature superfluid stiffness $\Theta(0)$.
Dashed line is a linear fit indicating that $T_{0}=5.5~\Theta(0)$.}
\end{figure}

Interestingly, we found that the $T_{0}$ values resulting from the
fits scale linearly with the low temperature superfluid stiffness
$\Theta(0)$ that we extracted from the plasmon dispersion of the
resonators. Fig.~\ref{fig:T0_vs_Theta} shows $T_{0}$ as a function
of $\Theta(0)$ together with a linear fit of slope~$5.5$. This
particular dependence points to a phase-fluctuation origin of this
power law suppression of the frequency shift and thus of the superfluid
stiffness. 

These experimental observations suggest the existence of low-energy
excitations with energies $E\ll T_{c}$ which cannot be captured by
the semiclassical BCS-like theory. Indeed, for a given density of
bosonic excitation states $\nu\left(E\right)$, the frequency shift
is given by $\hbar\delta f\left(T\right)\sim\int dE\,\nu\left(E\right)\,E/\left(e^{\beta E}-1\right)$,
so for a power-law dependence of $\delta f(T)\sim T^{b}$ with $b\sim2$
one needs a finite density of states at $E\sim T$, viz., $\nu\left(E\sim T\right)\sim E^{b-2}$.
Otherwise, all thermal effects are exponentially suppressed.

\subsection{Thermal excitation of one-dimensional plasma waves}

One possible source of low-energy excitations is provided, in principle,
by one-dimensional plasmon waves as we discuss below. We first describe
the behavior of plasmon excitations at $T=0$, and then calculate
the effect of the latter on the superfluid stiffness at low $T$.

It is well known that the application of an electromagnetic drive
on a superconductor in the dirty limit leads to a non-linear current-phase
relation~\citep{Maki1964} (see for instance, Eqs.~(12-14) in Ref.~\citep{Maki1964},
derived in the framework of the Gor'kov's equations). For a one-dimensional
superconducting wire in the dirty limit and at low frequencies $\omega\ll\Delta$
such non-linearity translates into the appearance of higher-order
terms in the expansion of the 2D current density $\boldsymbol{j}_{\text{2D}}$
with respect to vector potential $\boldsymbol{A}$: 
\begin{equation}
\boldsymbol{j}_{\text{2D}}=-\frac{\rho_{S}d}{c}\boldsymbol{A}\left(1-\alpha\xi^{2}\left(\frac{2e}{\hbar c}\boldsymbol{A}\right)^{2}\right),
\label{eq:current_phase_relation}
\end{equation}
where $\alpha=\frac{\pi}{4}+\frac{3}{4\pi}\approx1.02$, $\rho_{S}$
is the superfluid density, $d$ is the film thickness, and $\xi$
is the dirty-limit superconducting coherence length. Eq.~(\ref{eq:current_phase_relation})
is analogous to the current-phase relation for a Josephson junction
$I(\varphi)=I_{c}\sin\varphi\approx I_{c}\left(\varphi-\varphi^{3}/6\right)$,
where $\varphi$ is the superconducting phase-difference across the
junction and $I_{c}$ is the critical current.

The Hamiltonian describing long-wavelength plasmons along the wire
can be split into two parts: $H=H_{0}+\delta H$. $H_{0}$ is the
Hamiltonian related to the linear part of the current-phase relation,
and can be diagonalized in a basis of normal modes as $H_{0}=\sum_{n}\hbar\omega_{n}a_{n}^{\dagger}a_{n}$
where $a_{n}^{\dagger}$ and $a_{n}$ are bosonic creation and annihilation
operators. The effect of nonlinearity in Eq.~(\ref{eq:current_phase_relation})
translates into the perturbation $\delta H=-\frac{\alpha}{4}\Theta\xi^{2}\int d^{2}\boldsymbol{r}\,|\nabla\varphi|^{4}$,
where $\nabla\varphi$ is the superconducting phase gradient.

The relevant part of the Hamiltonian then takes the form (the derivation
is presented in \appref{Kerr-effect}):

\begin{equation}
H_{0}+\delta H=\sum_{n}\hbar\omega_{n}^{\prime}\,a_{n}^{\dagger}a_{n}-\frac{\hbar}{2}\sum_{n,m}K_{nm}\,a_{n}^{\dagger}a_{n}\,a_{m}^{\dagger}a_{m},
\label{eq:Kerr_Hamiltonian}
\end{equation}
with $\omega_{n}^{\prime}=\omega_{n}-\left(K_{nn}+\sum_{m}K_{nm}\right)/4$
being the renormalized frequency, and the Kerr coefficients $K_{nm}$
for the strip geometry defined as 
\begin{equation}
K_{nm}=3\alpha\left(1-\frac{1}{4}\delta_{nm}\right)\frac{\xi^{2}}{Lw}\,\frac{\hbar\omega_{n}}{\Theta(0)}\,\omega_{m},
\label{eq:Kerr_coeff}
\end{equation}
where $L$ and $w$ are the strip length and width, respectively.
Eq.~(\ref{eq:Kerr_Hamiltonian}) is well-known in quantum optics
and is used to describe the interaction of a given mode $n$ with
itself (through the self-Kerr coefficient $K_{nn}$), and with another
mode $m$ (via the cross-Kerr coefficient $K_{nm}$). The corresponding
\emph{Kerr effect} is seen as the reduction of a normal mode's frequency
due to occupation of other modes:

\begin{equation}
\omega_{n}\to\omega_{n}^{\prime}-\frac{1}{2}\sum_{m}K_{nm}N_{m},
\label{eq:cross_Kerr_shift}
\end{equation}
where $N_{m}=\langle a_{m}^{\dagger}a_{m}\rangle$ is the bosonic
occupation number of a given normal mode.

A result similar to Eq.~(\ref{eq:Kerr_Hamiltonian}) for a chain
of Josephson junctions can be found in Refs.~\citep{weissl2015kerr,Krupko2018},
where authors find a good agreement between experimental and theoretical
Kerr coefficients. Both Eq.~(\ref{eq:Kerr_coeff}) and the model
of~\citep{Krupko2018} with short-range capacitive coupling give
the exact same result if one identifies $\Theta(0)$ with the Josephson
junction energy $E_{J}$, and $Lw/(2\xi^{2})$ with the number of
junctions in the chain $N$, and sets $\alpha=1/6$, corresponding
to the coefficient of the cubic term in the expansion of $\sin\varphi$.

We emphasize that the results above are applicable to a homogeneous
superconducting strip made of a moderately dirty superconductor in
the diffusive limit. The presence of a pseudogap and other non-trivial
consequences of strong disorder are, therefore, completely ignored.

We now discuss how one-dimensional plasmons induce a frequency shift
as a function of temperature. Upon increase of temperature, the thermal
population of plasmonic modes is increased, following the Bose-Einstein
distribution $N_{m}=\left[e^{\hbar\beta\omega_{m}}-1\right]^{-1}$,
where $\beta=1/k_{B}T$. As a result, the frequency of a given mode
$n$ is reduced due to the interaction with other thermally populated
modes. Using Eqs.~(\ref{eq:cross_Kerr_shift})~and~(\ref{eq:Kerr_coeff}),
the total frequency shift with temperature can be expressed as
\begin{equation}
\frac{\delta f(T)}{f}=\frac{3}{2}\alpha\,\frac{\xi^{2}}{Lw}\sum_{m\geq1}\frac{\hbar\omega_{m}/\Theta(0)}{e^{\hbar\beta\omega_{m}}-1}.
\label{eq:Kerr-effect_frequency-shift}
\end{equation}
To calculate the sum over modes, the dispersion law $\omega\left(k\right)$
of the plasmonic modes with the wave number $k$ is required, and
the latter depends on the electrostatic properties of the system.
While generally one expects logarithmic corrections to the linear
dispersion law due to long-range Coulomb interaction~\citep{Mooij1985},
in the present experimental setup the plasmonic modes with $kh\ll1$
are not affected due to the screening from the ground plane at distance
$h\sim300\text{ \ensuremath{\mu}m}$, see \appref{Kerr-effect} for
details. For the lowest temperatures, one can thus approximate the
plasmonic spectrum by a purely linear dispersion relation, corresponding
to $\omega_{m}=m\omega_{1}$, where $\omega_{1}$ is the fundamental
mode's angular frequency. For temperatures higher than $\hbar\omega_{1}/k_{B}$,
the sum in Eq.~(\ref{eq:Kerr-effect_frequency-shift}) then evaluates
to
\begin{equation}
\frac{\delta f(T)}{f}\approx\frac{3}{2}\,\alpha\,\frac{\xi^{2}}{Lw}\,\frac{\hbar\omega_{1}}{\Theta(0)}\,\frac{\pi^{2}}{6}\,\left[\frac{\hbar\omega_{1}}{k_{B}T}\right]^{-2}=\left(\frac{T}{T_{\mathrm{K}}}\right)^{2},
\label{eq:frequency-shift_plasmons}
\end{equation}
leading to a power-law frequency shift with the temperature scale
$T_{\mathrm{K}}$ given by
\begin{equation}
k_{B}T_{\mathrm{K}}=\frac{2}{\pi\sqrt{\alpha}}\,\frac{\sqrt{Lw}}{\xi}\,\sqrt{\Theta(0)\,\hbar\omega_{1}}.
\end{equation}
Since $\omega_{1}\propto\sqrt{\Theta(0)}$, $T_{K}$ should scale
with the superfluid stiffness as $T_{K}\propto\left[\Theta(0)\right]^{3/4}$.
Eq.~(\ref{eq:frequency-shift_plasmons}) is applicable while $\hbar\omega_{1}\ll k_{B}T\ll\hbar\omega_{1}\,L/h\sim10\,\hbar\omega_{1}$,
which translates to $0.02\,\text{K}\apprle T\apprle0.2\,\text{K}$
for the parameters of the present experimental setup. At higher temperatures,
the logarithmic corrections to the plasmonic spectrum should be taken
into account, resulting in a weak (logarithmic in $T$) enhancement
of the effect. Nevertheless, Eq.~(\ref{eq:frequency-shift_plasmons})
allows one to correctly estimate the magnitude of the frequency shift
due to the plasmonic resonances.

In particular, the estimate Eq.~(\ref{eq:frequency-shift_plasmons})
predicts a power-law frequency shift at low temperatures with an exponent
$b=2$ close to the experimental value of $\sim1.6$. However, the
magnitude of the effect is much smaller than observed experimentally:
using a reasonable estimate for the coherence length in disordered
Indium Oxide $\xi\approx5\,\text{nm}$~\citep{Sacepe2015} and the
experimental values of $\Theta(0)$ and $\omega_{1}/(2\pi)\sim0.4\,\text{GHz}$,
one obtains $T_{\mathrm{K}}\sim3900\,\text{K}$ for the lowest disorder
and $T_{K}\sim1300\,\text{K}$ for the highest disorder, both of which
are two orders of magnitude larger than the values of $T_{0}$ observed
experimentally (see table~\ref{tab:exp_data} and Fig.~\ref{fig:T0_vs_Theta}).

We now discuss theoretically how the features of strong disorder can
lead to collective modes that suppress the superfluid stiffness at
low temperatures.

\section{Model and Theoretical Approach\protect\label{sec:Model-and-Theory}}

In the present section, we outline the theoretical approach that consistently
describes the superfluid stiffness $\Theta$ in a strongly disordered
superconductor. Subsections~\ref{subsec:Model-Hamiltonian}~and~\ref{subsec:Current-operator}
present the Hamiltonian of a pseudogapped superconductor and the corresponding
current operator; \subsecref{Macroscopic-superfluid-stiffness} discusses
the problem of macroscopic electromagnetic response of a disordered
superconductor and presents a semi-phenomenological connection between
$\Theta$ and the statistics of the \emph{microscopic} current response;
\subsecref{Local-current-response} then provides a way to calculate
the latter in a particular disorder realization by means of a certain
generalization of Belief Propagation, and \subsecref{Statistical-properties-of-the-current-response}
describes the numerical procedure for calculating the statistics of
the microscopic response. As a result, one obtains a controllable
approach to calculate the temperature dependence $\Theta\left(T\right)/\Theta\left(0\right)$
for various disorder strengths.

\subsection{Model Hamiltonian\protect\label{subsec:Model-Hamiltonian}}

As demonstrated both experimentally~\citep{Sacepe_2011_for-pair-preformation,Dubouchet_2018_Preformation-of-Pairs}
and theoretically~\citep{Feigelman_Fractal-SC_2010}, the materials
in question feature localized Cooper pairs even above the transition
temperature, whereas the single-particle spectrum exhibits a spectral
gap $\Delta_{P}$ several times larger than the bulk transition temperature.
Quasiparticle excitation are thus practically absent at low temperatures,
and the low-energy physics is governed by hopping of Cooper pairs
between \emph{localized }single-particle states. This can be captured
by the following pseudo-spin Hamiltonian:

\begin{equation}
H=-\sum_{i}2\xi_{i}S_{i}^{z}-\sum_{\left\langle ij\right\rangle }4J_{ij}\left(S_{i}^{x}S_{j}^{x}+S_{i}^{y}S_{j}^{y}\right).
\label{eq:pseudo-spin_Hamiltonian}
\end{equation}
Here, $i$ is the index of the single-particle state, $S_{i}^{\alpha}$
are the pseudo-spin operators derived from the fermionic operators
as $2S_{i}^{z}=a_{i\downarrow}^{\dagger}a_{i\downarrow}+a_{i\uparrow}^{\dagger}a_{i\uparrow}-1$,
$S_{i}^{+}=a_{i\uparrow}^{\dagger}a_{i\downarrow}^{\dagger}$, $S_{i}^{-}=a_{i\downarrow}a_{i\uparrow}$.
Then, $\xi_{i}$ are random energies of single-particles states, with
probability density at the Fermi level $\xi=0$ expressed in terms
of the true single-particle density of states~$\nu_{0}$ \emph{per
spin projection }as $P\left(\xi=0\right)=\nu_{0}/n$, with $n$ being
the electron concentration, and the summation $\left\langle ij\right\rangle $
goes over all pairs of states that interact due to Cooper attraction
with amplitude $J_{ij}$. The latter is given by the corresponding
matrix element of the form $J_{ij}=g\intop dr\,\left|\psi_{i}\left(\boldsymbol{r}\right)\right|^{2}\left|\psi_{j}\left(\boldsymbol{r}\right)\right|^{2}$,
with $g$ being an interaction constant, and $\psi_{i}\left(\boldsymbol{r}\right)$
corresponding to the single-particle wave function of state~$i$.
While the value of $J_{ij}$ generally vanishes for sites that are
localized sufficiently far apart from each other (further than several
localizations lengths $\xi_{\text{loc}}$), the randomness of the
$\psi_{i}\left(\boldsymbol{r}\right)$ renders the magnitudes of the
matrix elements between spatially close states random. To simplify
this situation, one adopts three approximations: \emph{i)}~all pairs
of states are classified as either strongly interacting or not interacting
at all, \emph{ii)}~each state interacts with a large constant number
$K+1$ of other states, such that $1\ll K+1\ll N_{\text{loc}}$, with
$N_{\text{loc}}=nV_{\text{loc}}$ being the total number of other
states available within the localization volume and \emph{iii)}~the
interaction amplitude can be replaced by a constant value $J_{ij}=\lambda n/2\nu_{0}K$,
where $\lambda\ll1$ is the dimensionless Cooper coupling constant.
While these approximations might seem too crude, a detailed analysis
shows~\citep{op-distribution-paper} that the ignored effects mostly
amount to renormalization of physical quantities and inessential corrections,
with one notable exception of approximation \emph{iii)} discussed
later. An extended discussion and derivation of this model can be
found in \citep{Feigelman_Fractal-SC_2010,op-distribution-paper},
while Ref.~\citep{Feigelman_SIT_2010} directly addresses the Superconductor-Insulator
Transition in this model.

The proposed structure of interaction matrix elements~$J_{ij}$ suggests
the notion of interaction graph, in which each single-particle state
represents a vertex, while each pair of interacting states $\left\langle ij\right\rangle $
corresponds to an edge. In essence, Hamiltonian~(\ref{eq:pseudo-spin_Hamiltonian})
reduces the problem to hopping of Cooper pairs (equivalent to hard-core
bosons) along this interaction graph, with each site representing
a point in real 3D space with some approximate coordinate $\boldsymbol{r}_{i}$
(e.g. the center of mass of the corresponding localized wave function
$\psi_{i}$), and each edge $\left\langle ij\right\rangle $ describing
the tunneling amplitude between the two ``points'' $\boldsymbol{r}_{i}$
and $\boldsymbol{r}_{j}$. Crucially, this graph features locally
tree-like topology, i.e. for a given site $i$ the neighborhood of
size $m_{\text{tree}}\sim\ln N_{\text{loc}}/\ln K\gg1$ hops along
the graph is unlikely to contain any loops, which is the direct consequence
of the sparsity of the graph controlled by the parameter $K/N_{\text{loc}}\ll1$~\citep{bollobas2001random}.
As a result, the graph is indistinguishable from a portion of the
Bethe lattice as long as any local quantity is concerned, allowing
one to apply the rich palette of methods available for analysis of
tree-like systems (as done e.g. in Ref.~\citep{Feigelman_SIT_2010}).
On the other hand, at distance $r\sim\xi_{\text{loc}}\sqrt{m_{\text{tree}}}\gg\xi_{\text{loc}}$
from a given point $\boldsymbol{r}_{i}$ the actual 3D structure of
the model reveals itself via the presence of many loops of length
$\ge2m_{\text{tree}}$ hops containing the chosen site $i$ in the
interaction graph. This connection between the structure in the real
3D space and the interaction graph allows one to calculate spatially
resolved quantities, as demonstrated e.g. in Ref.~\citep{Mirlin_1991},
where a similar graph model was used in the framework of single-particle
Anderson localization problem.

The model~(\ref{eq:pseudo-spin_Hamiltonian}) possesses a natural
superconducting energy scale $\Delta_{0}\approx e^{-1/\lambda}\,n/\nu_{0}$,
as explained in~\citep{op-distribution-paper}. In what follows,
all energy quantities in the problem are expressed in units of $\Delta_{0}$.
Along with this scale goes the dimensionless disorder strength $\kappa$,
also introduced in \citep{op-distribution-paper}:
\begin{equation}
\kappa=\frac{\lambda n}{2\nu_{0}\Delta_{0}K}=\frac{\lambda e^{1/\lambda}}{2K}.
\end{equation}
We expect that all qualitative properties of the model are defined
by physical parameters, such as temperature~$T$ and the dimensionless
strength of disorder~$\kappa$. In particular, the value of other
microscopic parameters, such as $\lambda$ and $K$, are not essential
as long as the values of $\kappa$ and $T$ (in units of $\Delta_{0}$)
are set.

\subsection{Current operator\protect\label{subsec:Current-operator}}

The only possible way to transfer charge in the system described by
the Hamiltonian of Eq.~(\ref{eq:pseudo-spin_Hamiltonian}) is hopping
of the Cooper pairs between different states. This implies that the
current density operator is described as 
\begin{equation}
\boldsymbol{j}\left(r,t\right)=\frac{1}{2}\sum_{e}I_{e}\boldsymbol{D}_{e}\left(r\right),
\label{eq:real-space-current_via_edges-currents}
\end{equation}
with the sum going over all \emph{directed} edges of the interaction
graph, hence the factor 1/2. Here, $I_{e}$ is the operator of the
current through a given directed edge $e=i\rightarrow j$:
\begin{equation}
I_{e}=2eV_{e}-\frac{4e^{2}}{c}N_{e}A_{e},
\label{eq:current-operator_definition}
\end{equation}
where $V_{e}$ plays the role of the velocity operator
\begin{equation}
V_{i\rightarrow j}=J_{ij}\left(S_{i}^{x}S_{j}^{y}-S_{j}^{x}S_{i}^{y}\right),
\label{eq:discrete-velocity_definition}
\end{equation}
and the second term represents the diamagnetic contribution to the
current:
\begin{align}
 & N_{ij}=J_{ij}\left(S_{i}^{x}S_{j}^{x}+S_{i}^{y}S_{j}^{y}\right),
\label{eq:diamagnetic-term_definition}\\
 & A_{e}=\intop d^{3}r\;\left(\boldsymbol{D}_{e}\left(r\right),\boldsymbol{A}\left(r\right)\right)
\label{eq:edge-vector-potential_definition}
\end{align}
with $\boldsymbol{A}\left(\boldsymbol{r}\right)$ being the vector
potential. Both in $\boldsymbol{j}\left(r\right)$ and in the diamagnetic
term, $\boldsymbol{D}_{e}\left(r\right)$ is a short-range vector
field that translates the graph topology to the real space, i.e. describes
the distribution of the current density $\boldsymbol{j}\left(\boldsymbol{r}\right)$
induced by the process of hopping of a Cooper pair from one site to
the other. The exact value of $\boldsymbol{D}_{e}\left(\boldsymbol{r}\right)$
is expressed via the response of the interaction matrix elements $J_{ij}$
to external vector potential and it thus also inherits the randomness
of the matrix elements themselves (see \subsecref{Model-Hamiltonian}).
Importantly for us, $\boldsymbol{D}_{e}$ is antisymmetric w.r.t the
edge direction, viz. $\boldsymbol{D}_{i\rightarrow j}=-\boldsymbol{D}_{j\rightarrow i}$,
and it also obeys the following exact identity due to charge conservation
in real space:
\begin{equation}
2\text{div}\boldsymbol{D}_{i\rightarrow j}\left(\boldsymbol{r}\right)=\left|\psi_{i}\left(\boldsymbol{r}\right)\right|^{2}-\left|\psi_{j}\left(\boldsymbol{r}\right)\right|^{2},
\label{eq:div-D_condition}
\end{equation}
where $\psi_{i}\left(\boldsymbol{r}\right)$ are single-particle wave
functions.

\subsection{Macroscopic superfluid stiffness\protect\label{subsec:Macroscopic-superfluid-stiffness}}

We determine the macroscopic superfluid stiffness at low frequencies
via the relation $\Theta=\left(\hbar/2e\right)^{2}\rho_{S}d$, where
$d$ is the film thickness, and $\rho_{S}$ is the superfluid density
entering the London's equation:
\begin{equation}
\overline{\boldsymbol{j}}=-\frac{1}{c}\rho_{S}\overline{\boldsymbol{A}_{\text{ext}}},\,\,\,\text{div}\boldsymbol{A}_{\text{ext}}=0,
\end{equation}
where $\overline{\boldsymbol{A}_{\text{ext}}}$ and $\overline{\boldsymbol{j}}$
are, respectively, the external vector potential and the current density
averaged over a macroscopically large region. The current density
is, in turn, determined from the response equation at vanishing frequency
$\omega\rightarrow0$:
\begin{equation}
j^{\alpha}=-\intop d^{3}\boldsymbol{r}'\,Q_{0}^{\alpha\beta}\left(\boldsymbol{r},\boldsymbol{r}'\right)\,A^{\beta}\left(\boldsymbol{r}'\right),
\label{eq:current-response_real-space}
\end{equation}
with $Q_{\omega}$ being the Fourier transform of the microscopic
current response to external vector potential $Q^{\alpha\beta}\left(t-t';\boldsymbol{r},\boldsymbol{r}'\right)=-\delta\left\langle j^{\alpha}\left(t,\boldsymbol{r}\right)\right\rangle /\delta A^{\beta}\left(t',\boldsymbol{r}'\right)$.
Due to the microscopic disorder, the $Q$~kernel depends on both
coordinates $\boldsymbol{r},\boldsymbol{r}'$ rather than on their
difference~$\boldsymbol{r}-\boldsymbol{r}'$ and has a nontrivial
tensor structure. As a result, the current density induced by the
external field $\boldsymbol{A}_{\text{ext}}$ does not automatically
satisfy the charge conservation,
\begin{equation}
\text{div}\boldsymbol{j}=0,
\label{eq:charge-conservation_continuos}
\end{equation}
so the current $\boldsymbol{j}$ induces additional electromagnetic
field $\boldsymbol{A}_{\text{int}}$ such that the current response~(\ref{eq:current-response_real-space})
to the total field $\boldsymbol{A}_{\text{ext}}+\boldsymbol{A}_{\text{int}}$
satisfies Eq.~(\ref{eq:charge-conservation_continuos}).

Consider first the case of weak disorder with diffusive transport
in the normal state characterized by $k_{F}l\gg1,$ where $l$ is
the mean free path, and $k_{F}$ is the Fermi wave number. The average
current response $\overline{\boldsymbol{j}}$ to a smooth external
field $\boldsymbol{A}_{\text{ext}}$ obeying $\text{div}\boldsymbol{A}_{\text{ext}}=0$
already satisfies the charge conservation~(\ref{eq:charge-conservation_continuos}),
while the disorder-induced deviation $\delta\boldsymbol{j}=\boldsymbol{j}-\overline{\boldsymbol{j}}$
turns out to be small~\citep{spivak_1988_mesoscopic_rho-s_fluctuations}:
$\delta\boldsymbol{j}^{2}/\overline{\boldsymbol{j}}^{2}\sim\frac{\xi_{0}/l}{(k_{F}l)^{2}}\ll1$,
where $\xi_{0}$ is the zero-temperature coherence length. This allows
one to neglect the contribution of $\boldsymbol{A}_{\text{int}}$
to current response and calculate $\rho_{S}$ simply as $\rho_{S}\delta^{\alpha\beta}=c\,\intop d^{3}\boldsymbol{r}'\,\overline{Q_{0}^{\alpha\beta}\left(\boldsymbol{r},\boldsymbol{r}'\right)}$,
where $\overline{\bullet}$ denotes average over disorder, which in
this case is carried out by means of the impurity technique~\citep{AGD}. 

In our model, on the other hand, statistical and spatial fluctuations
$\boldsymbol{j}-\overline{\boldsymbol{j}}$ of the response to a smooth
external field are much larger than $\overline{\boldsymbol{j}}$ itself,
necessitating a consistent account of the induced field $\boldsymbol{A}_{\text{int}}$.
However, we can still assume that the spatial scale at which the total
response $\boldsymbol{j}$ convergence to its average value is much
smaller than the London's penetration length $\lambda_{L}=\sqrt{c^{2}/4\pi\rho_{S}}$.
In this case, one can neglect the induced magnetic field $B=\text{rot}\boldsymbol{A}_{\text{int}}$
at the relevant length scales, so $\boldsymbol{A}_{\text{int}}$ is
dominated by its potential component, viz. $\boldsymbol{A}_{\text{int}}\approx-\frac{c}{i\omega}\nabla\phi$.
The current distribution is then described by Eq.~(\ref{eq:current-response_real-space}),
where the induced component $\boldsymbol{A}_{\text{int}}$ of total
electromagnetic field $\boldsymbol{A}_{\text{ext}}+\boldsymbol{A}_{\text{int}}$
is found self-consistently from the following system of equations:

\begin{equation}
\text{div}\boldsymbol{j}=0,\,\,\,\text{rot}\boldsymbol{A}_{\text{int}}=0.
\label{eq:macro-rho-S_problem}
\end{equation}
In particular, due to the assumed division of scales, one can set
$\boldsymbol{A}_{\text{ext}}=\overline{\boldsymbol{A}_{\text{ext}}}=\text{const}$.
A more detailed derivation of (\ref{eq:macro-rho-S_problem}) is presented
in \appref{disordered-sc_electrodynamics}.

Due to Eqs.~(\ref{eq:real-space-current_via_edges-currents})~and~(\ref{eq:div-D_condition}),
the problem~(\hphantom{}\ref{eq:current-response_real-space}\nobreakdash-\ref{eq:macro-rho-S_problem}\hphantom{})
is equivalent to the following \emph{discrete} system of equations
on the values of the onsite electric potential $\phi_{i}$:
\begin{align}
 & \sum_{j\in\partial i}I_{i\rightarrow j}=0,\,\,\,E_{i\rightarrow j}=-\left(\phi_{j}-\phi_{i}\right),\nonumber \\
 & I_{i\rightarrow j}=\frac{c}{i\omega}Q_{ij}\left(\omega=0\right)E_{i\rightarrow j},
\label{eq:macro-response_discrete-problem}
\end{align}
with the sum in the first equation going over all neighbors of site
$i$, and $Q_{ij}\left(\omega\right)=\intop dt\,e^{i\omega t}Q_{ij}\left(t\right)$
being the Fourier transform of the local current response $Q_{ij}\left(t\right)=-\delta I_{i\rightarrow j}\left(t\right)/\delta A_{i\rightarrow j}\left(0\right)$
along the directed graph edge $i\rightarrow j$ to the \emph{discrete}
vector potential, Eq. (\ref{eq:edge-vector-potential_definition}),
on the same edge:
\begin{equation}
Q_{ij}\left(t\right)=-\frac{4e^{2}}{c}\left[\left\langle \left[V_{i\rightarrow j}\left(t\right),V_{i\rightarrow j}\left(0\right)\right]\right\rangle -\left\langle N_{ij}\right\rangle \delta\left(t\right)\right].
\label{eq:discrete-current-response}
\end{equation}
While one would expect non-local current response $\delta I_{e}/\delta A_{e'}$
with $e\neq\pm e'$ to be also present in Eq.~(\ref{eq:macro-response_discrete-problem}),
all such contributions vanish due to locally tree-like structure of
the underlying graph, as explained in detail in \appref{Current-response_to_potential-field}.
Eq.~(\ref{eq:macro-response_discrete-problem}) should be solved
for the values of $\phi_{i}$ for all sites $i$ of the system, which
then also yields the values of the edge currents $I_{i\rightarrow j}$.
The distribution of the electric potential in the real space $\phi\left(r\right)$
and the associated electric field $\boldsymbol{E}\left(\boldsymbol{r}\right)=-\nabla\phi\left(\boldsymbol{r}\right)$
are then restored by inverting the following relation:
\begin{equation}
\phi_{i}=\intop d^{3}r\,\phi\left(r\right)\,\left|\psi_{i}\left(r\right)\right|^{2},
\label{eq:discrete-potential-def}
\end{equation}
where $\psi_{i}\left(\boldsymbol{r}\right)$ are the single-electron
wave functions, see \subsecref{Model-Hamiltonian}. Upon also computing
the current density $\boldsymbol{j}\left(\boldsymbol{r}\right)$ with
the help of Eq.~(\ref{eq:real-space-current_via_edges-currents}),
one is able to calculate the true superfluid density by means of the
last relation in Eq.~(\ref{eq:macro-rho-S_problem}).

One can, however, avoid the procedure of calculating the fields $\phi\left(\boldsymbol{r}\right)$
and $\boldsymbol{j}\left(\boldsymbol{r}\right)$ in real space altogether
by noting that Eq.~(\ref{eq:macro-response_discrete-problem}) is
structurally identical to the Kirchhoff's law and the Ohm's law for
the interaction graph, with $Q_{ij}c/i\omega$ playing the role of
graph edge conductance and $\rho_{S}/i\omega$ corresponding to macroscopic
conductance. Macroscopic Ohm's law then suggests that for a system
with the geometry of a brick with macroscopical sizes $L\times w\times d$
the total ``conductance'' along the $L$ direction is equal to $\Delta\phi/I_{\text{total}}=\left(\rho_{S}/i\omega\right)\,\left(L/wd\right)$,
where $I_{\text{total}}$ is the total current through all boundary
sites and $\Delta\phi$ is the potential difference between the two
boundaries. This constitutes a way to calculate the superfluid stiffness
numerically for a given realization of disorder, as done in \appref{Verification-of-Dykhne-law}.

The resemblance of Eq.~(\ref{eq:macro-response_discrete-problem})
to the local Ohm's law also bears certain physical meaning: according
to the second Josephson's relation, the quantity $\varphi_{i}=-\frac{2e}{\hbar}\phi_{i}/i\omega$
is precisely the superconducting phase of a given site, whereas the
last two relations in Eq.~(\ref{eq:macro-response_discrete-problem})
express the first Josephson's relation linearized w.r.t the phase
difference. The local structure of Eq.~(\ref{eq:macro-response_discrete-problem})
thus becomes a direct consequence of the fact that the system conducts
via coherent tunneling of the Cooper pairs. It is important, however,
that the linear local current response in Eq.~(\ref{eq:macro-response_discrete-problem})
\emph{cannot} be extended to the Josephson's sinusoidal current-phase
relation of the form $I=I_{0}\,\sin\left(\varphi_{2}-\varphi_{1}\right)$.
Indeed, as it is explained in \appref{Current-response_to_potential-field},
this would be an unjustified simplification of a rather complex problem:
neither is the local response sinusoidal, nor is the full response
guaranteed to be local, as it conveniently happened in the linear
response. We thus leave this question for future study.

We are interested in estimating the macroscopic superfluid stiffness
analytically, and there are currently no means to do this for a general
setting, especially given that the values of the current response~$Q_{ij}$
are randomly distributed across many orders of magnitude. However,
our numerical experiments on the solution of~(\ref{eq:macro-response_discrete-problem})
for 2D systems have shown (see \appref{Verification-of-Dykhne-law}
for details) that the following approximate expression is applicable
to our problem for $\kappa\ge1$:
\begin{equation}
\Theta\left(T\right)\approx C\,\exp\left\{ \left\langle \ln Q_{ij}\left(\omega=0\right)\right\rangle \right\} =:C\,Q_{\text{typ}}\left(T\right),
\label{eq:macro-rho-S_via_typical-current-response}
\end{equation}
where $C$ is a prefactor that depends only on the details of the
graph structure (such as $K$, $\xi_{\text{loc}}$ and concentration
$n$), but not on temperature~$T$. As it is explained in \appref{superfluid-stiffness_small-disorder},
one can estimate $C\sim nL_{0}^{2}dK$, where $L_{0}\sim\xi_{\text{loc}}$
is the interaction length scale discussed in \subsecref{Model-Hamiltonian},
$n$ is site concentration, and $d$ is the thickness of the film.
The exact value of $C$ is additionally modified by the presence of
short-range correlations in $Q_{ij}$, but this does not change the
order of magnitude of the answer, as explained in \appref{Verification-of-Dykhne-law}.
We therefore \emph{approximate} the temperature change of the superfluid
stiffness by that of the typical current response $Q_{\text{typ}}$
defined in Eq.~(\ref{eq:macro-rho-S_via_typical-current-response}).

The qualitative origin of the result~(\ref{eq:macro-rho-S_via_typical-current-response})
could be understood by noting that both the original problem~(\hphantom{}\ref{eq:macro-rho-S_problem}\nobreakdash-\ref{eq:current-response_real-space}\hphantom{})
and its discrete reformulation~(\hphantom{}\ref{eq:macro-response_discrete-problem}\nobreakdash-\ref{eq:discrete-current-response}\hphantom{})
are similar to the problem of macroscopic conductivity of disordered
media studied in the seminal paper~\citep{dykhne1971} by A.M. Dykhne.
In~Ref.~\citep{dykhne1971} it was shown that the macroscopic conductance
$\sigma_{\text{eff}}$ of a 2D random medium is equal to the typical
value of the microscopic conductance $\sigma_{\text{typ}}=\exp\left\{ \left\langle \ln\sigma\left(\boldsymbol{r}\right)\right\rangle \right\} $
provided that $\ln\sigma\left(\boldsymbol{r}\right)$ is distributed
symmetrically around its mean value. At first glance, our problem
is different: it is three-dimensional, and there are no \emph{a~priori}
reasons to assume that either the actual current response $Q_{\omega}^{\alpha\beta}\left(r,r'\right)$
or its discrete counterpart $Q_{ij}$ satisfy the requirement on the
symmetry of the distribution (although for large enough $\kappa$
the distribution turns out to be sufficiently close to the symmetric
log-normal one, as shown in \appref{Verification-of-Dykhne-law}).
However, there are no qualitative reasons for $\Theta$ to differ
substantially from the typical current response (at least, in temperature
dependence). Indeed, while the calculation of Ref.~\citep{dykhne1971}
formally relies on certain duality properties of the 2D problem, it
still illustrates the general intuition for the conductance of random
media: regions with large conductance can be short-circuited, while
regions with small conductance do not conduct at all, rendering the
typical conductance to be the only relevant scale of the problem.
In the future, we plan to address the validity of the approximate
relation~(\ref{eq:macro-rho-S_via_typical-current-response}) in
more detail.

\subsection{Current response in a given disorder realization\protect\label{subsec:Local-current-response}}

The next step is to calculate the local current response $Q_{ij}$.
This is done by means of the extension of the Method of Belief Propagation
(BP) to quantum problems, which we birefly explain below. The standard
BP scheme is applicable to classical Hamiltonians with discrete local
degrees of freedom (e.g. Ising spins). These include several prominent
examples in the theory of spin glasses, where the BP method is also
known as the cavity method~\citep{MezardParisi_cavity0,MezardParisi_cavity1}.
In essence, the BP method tries to accurately capture the structure
of local two-point correlations in the Gibbs ensemble by tracking
the conditional probabilities $P_{i|j}$ of a local degree of freedom
at site $i$ for a given value of the local degree of freedom at a
neighboring site $j$. This results in a system of algebraic self-consistency
equations that describe $P_{i|j}$~\citep{yedidia_belief-propagation}.
For a classical model with Ising-like local degrees of freedom, $P_{i|j}$
can be parametrized by a single number $h_{j\rightarrow i}$ called
the cavity field, as it has the physical sense of the additional local
field acting on site~$i$ from site~$j$. 

Notably, the BP approach is known to be exact for the systems with
tree-like topology, i.e. the ones with no loops in the interaction
graph~\citep{yedidia_belief-propagation}. This latter property is
not sensitive to the particular structure of the local degrees of
freedom, which enables a generalization of the BP method to thermal
averages of quantum Hamiltonians, such as Eq.~(\ref{eq:pseudo-spin_Hamiltonian}).
However, the quantum nature of the problem implies replacing the algebraic
equations on $P_{i|j}$ or $h_{j\rightarrow i}$ with functional ones
in order to capture the quantum noise, which makes the problem intractable
in general. In some cases of disordered ``all to all'' interactions~\citep{BiroliCulia},
this quantum noise can be fully characterized by a common pairwise
correlator in imaginary time, thus allowing for efficient analysis
of the problem. 

In our problem, quantum corrections to the static approximation, by
which we mean describing the additional action of site $i$ from site
$j$ by a single static field $h_{i\rightarrow j},$ can be shown
to be proportional to the interaction constant $\lambda\ll1$ and
definitely negligible at $\kappa\leq1$. Although it is not evident
that these quantum corrections are irrelevant at large $\kappa$ as
well, the numerical analysis provided in Sec.~IIIG of Ref.~\citep{Feigelman_SIT_2010}
indicates they are weak even very close to SIT. In the present paper
we will use static approximation, leaving the account of quantum noise
for future work. One can interpret the proposed Approximate Quantum
Belief Propagation (AQBP) scheme as the mean field theory of the same
type as that of Ref.~\citep{op-distribution-paper} but with the
Onsager reaction terms taken into account, essentially representing
the analogue of the classical Thouless-Anderson-Palmer equations~\citep{TAP_1977}
for our system.

The central object of the AQBP scheme is the set of local fields $h_{k\rightarrow i}^{\alpha},\,\,\alpha=x,y$
that describe the contribution of spin $k$ to the action of spin~$i$,
so the latter is described by the effective Hamiltonian:
\begin{equation}
H_{\text{eff}}=-2\xi_{i}S_{i}^{z}+\frac{1}{Z_{i}-1}\sum_{j\in\partial i}\left[2h_{j\rightarrow i}^{x}S_{i}^{x}+2h_{j\rightarrow i}^{y}S_{i}^{y}\right],
\label{eq:effective_single-site_Hamiltonian}
\end{equation}
where $Z_{i}$ is the number of neighbors of the given site $i$ (equal
to $K+1=\text{const}$ within our model). In full analogy with the
classical BP method, the AQBP provides the self-consistency equations
on the values of those fields:
\begin{equation}
h_{k\rightarrow i}^{\alpha}=\sum_{j\in\partial i\backslash\left\{ k\right\} }J_{ij}\left\langle 2S_{j}^{\alpha}\right\rangle _{j\left(i\right)},\,\,\,\alpha=x,y,
\label{eq:h-equations}
\end{equation}
where the summation now goes over all neighbors of $i$ except $k$
(i.e. $Z_{i}-1$ summation terms), thus excluding the ``self-action''
of spin $i$, and $\left\langle 2S_{j}^{\alpha}\right\rangle _{j\left(i\right)}$
is the thermal expectation of yet another single-spin Hamiltonian:
\begin{equation}
H_{j\left(i\right)}=-2\xi_{i}S_{i}^{z}-2h_{j\rightarrow i}^{x}S_{j}^{x}-2h_{j\rightarrow i}^{y}S_{j}^{y},
\label{eq:single-spin_Ham}
\end{equation}
\begin{equation}
\left\langle 2S_{j}^{\alpha}\right\rangle _{j\left(i\right)}:=\frac{h_{i\rightarrow j}^{\alpha}}{B_{i\rightarrow j}}\tanh\beta B_{i\rightarrow j},
\label{eq:spin-average}
\end{equation}
with $B_{i\rightarrow j}^{2}=\xi_{j}^{2}+\left(h_{i\rightarrow j}^{x}\right)^{2}+\left(h_{i\rightarrow j}^{y}\right)^{2}$.

Given a solution to this system of algebraic equations for the $h$
fields for a given disorder realization, one can then express physical
observables via solutions to effective one- or two-site quantum problems.
In particular, the values of the local current response $Q_{ij}$
at frequency $\omega$ are computed as quantum averages over two-spin
Hamiltonians that depend on the $h$ fields:
\begin{widetext}
\begin{equation}
\frac{Q_{ij}\left(\omega\right)}{\left(2e\right)^{2}/c}=\sum_{m,n=0}^{3}\left|\left\langle m\left|V_{i\rightarrow j}\right|n\right\rangle \right|^{2}\frac{1}{Z}\frac{\left(\lambda_{n}-\lambda_{m}\right)\left(e^{-\beta\lambda_{n}}-e^{-\beta\lambda_{m}}\right)}{\left(\lambda_{n}-\lambda_{m}\right)^{2}-\left(\omega+i0\right)^{2}}+\sum_{n=0}^{3}\frac{e^{-\beta\lambda_{n}}}{Z}\left\langle n\left|N_{ij}\right|n\right\rangle ,\,\,\,Z=\sum_{n=0}^{3}e^{-\beta\lambda_{n}}.
\label{eq:discrete-current-response_QBP}
\end{equation}
\end{widetext}

\noindent Here $V_{i\rightarrow j}$ and $N_{ij}$ are given by~(\ref{eq:discrete-velocity_definition})
and (\ref{eq:diamagnetic-term_definition}), and $\left\{ \lambda_{n},\left|n\right>\right\} ,\,n=0,1,2,3$
is the eigensystem of the following two-spin Hamiltonian:
\begin{align}
H_{\left\langle ij\right\rangle } & =-\sum_{n=i,j}2\xi_{n}S_{n}^{z}-4J_{ij}\left(S_{i}^{x}S_{j}^{x}+S_{i}^{y}S_{j}^{y}\right)\nonumber \\
 & -\sum_{\alpha=x,y}\left(2h_{i\rightarrow j}^{\alpha}S_{j}^{\alpha}+2h_{j\rightarrow i}^{\alpha}S_{i}^{\alpha}\right).
\label{eq:two-sping_effective_Hamiltonian}
\end{align}

At this point we can identify the main reason behind the temperature
dependence of $Q_{\text{typ}}$ in Eq.~(\ref{eq:macro-rho-S_via_typical-current-response}).
Almost by definition, this quantity is contributed by the most probable
disorder configurations. Now, the majority of the disorder configurations
have $\left|\xi_{i}\right|,\left|\xi_{j}\right|\gg J,h,T$, hence
the spectral problem for Hamiltonian~(\ref{eq:two-sping_effective_Hamiltonian})
can be addressed perturbatively, rendering $\lambda\approx\pm\xi_{1}\pm\xi_{2}$,
with corrections being small as $J/\xi,h/\xi$. As a result, the value
of $Q_{ij}$ for such disorder configurations at low temperatures
can be estimated as
\begin{equation}
\frac{Q_{ij}}{\left(2e\right)^{2}/c}\approx-2\sum_{n=1}^{3}\frac{\left|\left\langle n\left|V_{i\rightarrow j}\right|0\right\rangle \right|^{2}}{E_{n0}-\left(\omega+i0\right)^{2}/E_{n0}}+\left\langle 0\left|N_{ij}\right|0\right\rangle ,
\end{equation}
with $n=0$ denoting the ground state, $n>0$ enumerating the excited
states, and $E_{n0}=\lambda_{n}-\lambda_{0}\sim E_{F}\gg\Delta_{0},J$.
Generally, temperature influences the values of $h$, thus modifying
both the matrix elements and the spectral gaps $E_{n0}.$ The latter,
however, are only shifted by a quantity of the order $\Delta_{0},J\ll E_{F}$,
hence this effect can be discarded along with the frequency dependence,
as we are interested in $\omega\sim\Delta_{0}$. The main temperature
dependence is thus given by the matrix elements, and by means of the
perturbation theory it can be estimated as
\begin{equation}
\frac{Q_{ij}}{\left(2e\right)^{2}/c}\approx\frac{4Jh_{1}h_{2}}{E_{10}E_{20}},
\label{eq:local-response-estimation}
\end{equation}
from which it immediately follows that
\begin{equation}
\frac{Q_{\text{typ}}\left(T\right)}{Q_{\text{typ}}\left(T=0\right)}\approx\frac{h_{\text{typ}}^{2}\left(T\right)}{h_{\text{typ}}^{2}\left(T=0\right)}.
\label{eq:typical-response-estimation}
\end{equation}
To obtain this latter Eq.~(\ref{eq:typical-response-estimation}),
we used the fact that $h_{i\rightarrow j},h_{j\rightarrow i}$ are
all uncorrelated (see below), so the r.h.s of Eq.~(\ref{eq:typical-response-estimation})
is expressed via the typical value of the $h$ field. Essentially,
this implies that the relative change of the superfluid stiffness
mirrors that of the typical order parameter at low temperatures.%

\subsection{Statistical properties of the current response\protect\label{subsec:Statistical-properties-of-the-current-response}}

According to the Eq.~(\ref{eq:macro-rho-S_via_typical-current-response}),
the information about the temperature dependence of the superfluid
stiffness is encoded in the statistical distribution of the local
current response $Q_{ij}$ in the form of the typical value $Q_{\text{typ}}$.
It is thus our aim to compute the average of this quantity over various
realizations of disorder. As Eq.~(\ref{eq:discrete-current-response_QBP})
suggests, the value of $Q_{ij}$ is expressed via the values of $\xi_{i},\xi_{j}$
on the neighboring sites and via the pair of values $h_{i\rightarrow j},h_{j\rightarrow i}$.
Crucially, all four quantities are statistically independent. Indeed,
for region of parameters in question, the solution to Eq.~(\ref{eq:h-equations})
is short-correlated due to the large number of summation terms in
the right-hand side, as discussed in~\citep{op-distribution-paper}.
As a result, the statistics of $h$ on a given edge does not depend
on whether this edge is a part of a tree or a locally tree-like interaction
graph that is realized in our case. Iteratively expanding the r.h.s
of Eq.~(\ref{eq:h-equations}) then reveals, that these equations
possess a directed structure, in contrast to similar equations of
Ref.~\citep{op-distribution-paper}. In other words, the value of
$h$ on edge $i\rightarrow j$ gathers statistical information only
from the finite branch rooted at $i$ and not containing edge $i\rightarrow j$
or sites $i$ or $j$. From this, it follows straightforwardly that
all four quantities $h_{i\rightarrow j},h_{j\rightarrow i},\xi_{i},\xi_{j}$
are uncorrelated.

As a result, we can substantially simplify the procedure of calculating
the value of $Q_{\text{typ}}$: instead of solving the system~(\ref{eq:h-equations})
on a finite size instance of the interaction graph, we can only track
the distribution of the $h$ fields with subsequent averaging of the
$Q_{ij}$ value~(\ref{eq:discrete-current-response_QBP}) over the
the distribution of $h$ and $\xi$. The distribution of $h$ can,
in turn, be found using the Method of Population Dynamics (MPD), which
boils down to claiming that the two sides of Eq.~(\ref{eq:h-equations})
are equal \emph{in distribution}, which follows from the aforementioned
directed structure of Eq.~(\ref{eq:h-equations}) and the short range
of correlations in the solution. MPD then allows one to efficiently
prepare large number of samples from this distribution: given a large
initial pool of $h$ values, one updates each value by replacing it
with the value of the r.h.s of Eq.~(\ref{eq:h-equations}), where
the values of $\xi$ are sampled randomly for each term, and the values
of $h$ are randomly selected from the current pool. After a number
of such updates, the pool of $h$ values converges to a large sample
from the target distribution. The required number of iterations is
not large because the corresponding distribution does not have fat
tails in the given range of parameters~\citep[Sec. III C]{op-distribution-paper}.
In our implementation, we terminated the process once the mean value
$\left\langle h\right\rangle $ has converged. The pool sizes should
simply be sufficient to capture the rare events responsible for the
formation of the relevant parts of the distribution. In particular,
for all simulations presented below, pools of size $10^{7}$ or larger
were used to guarantee the convergence of low-value tails of the distribution.

It is worth noting that the assumption of short correlation distance
is crucial for this procedure and is not guaranteed in general. In
particular, the work~\citep{Feigelman_SIT_2010} analyzes Eq.~(\ref{eq:h-equations})
in the context of the disorder-driven SIT and finds that for $K\le K_{\text{RSB}}=\lambda\exp\left\{ 1/2\lambda\right\} $
the distribution of the $h$ fields becomes fat-tailed, and the associated
spatial configuration is not at all short-correlated. We are thus
interested in the interval $K>K_{\text{RSB}}$, which still contains
rich physics, as shown in~\citep{op-distribution-paper}.

\section{Theoretical results\protect\label{sec:Results}}

Upon having designed the numerical AQBP scheme to address the superfluid
density of the model described in \secref{Model-and-Theory}, we now
present the key results in \subsecref{Numerical-results}. In particular,
we discuss the observed connection between the low-temperature dependence
of the superfluid stiffness~$\Theta$ with that of the superconducting
order parameter~$\Delta$, with the latter being up until this point
a purely theoretical quantity without a corresponding physical observable.
We further employ this observation in \subsecref{Analytical-approach}
to describe $\Theta\left(T\right)$ analytically at low temperatures
via a finite-temperature extension of the technique developed earlier~\citep{op-distribution-paper}
for the order parameter itself. These results allow us to infer some
information about the nature of the low-energy excitations that cause
the suppression of the superfluid density.

\subsection{Numerical results\protect\label{subsec:Numerical-results}}

\begin{figure}
\begin{centering}
\includegraphics[scale=0.3]{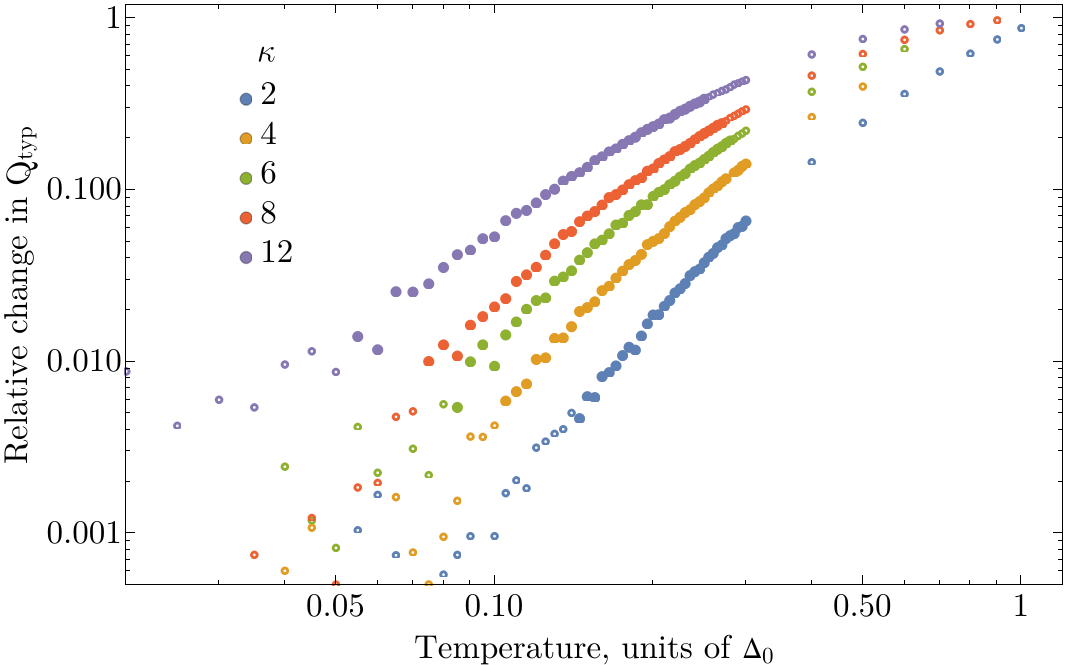}
\par\end{centering}
\caption{The temperature dependence of the relative change of typical local
current response $1-Q_{\text{typ}}\left(T\right)/Q_{\text{typ}}\left(T=0\right)$
in logarithmic scale along both axes for various dimensionless disorder
strength~$\kappa$ and $K=15$. The temperature is measured in units
of $\Delta_{0}$. The points with solid filling are selected according
to the empirically chosen criteria $T\in\left[\lambda\left\langle \Delta\right\rangle ,\lambda\left\langle \Delta\right\rangle +0.2\Delta_{0}\right]$
with $\left\langle \Delta\right\rangle $ being the mean order parameter
at zero temperature. These highlighted points are used for power-law
fitting on \figref{typical-Q-exponent_disorder-dependence}. \protect\label{fig:typical-Q_numerical-data}}
\end{figure}

\begin{figure}
\begin{centering}
\includegraphics[scale=0.3]{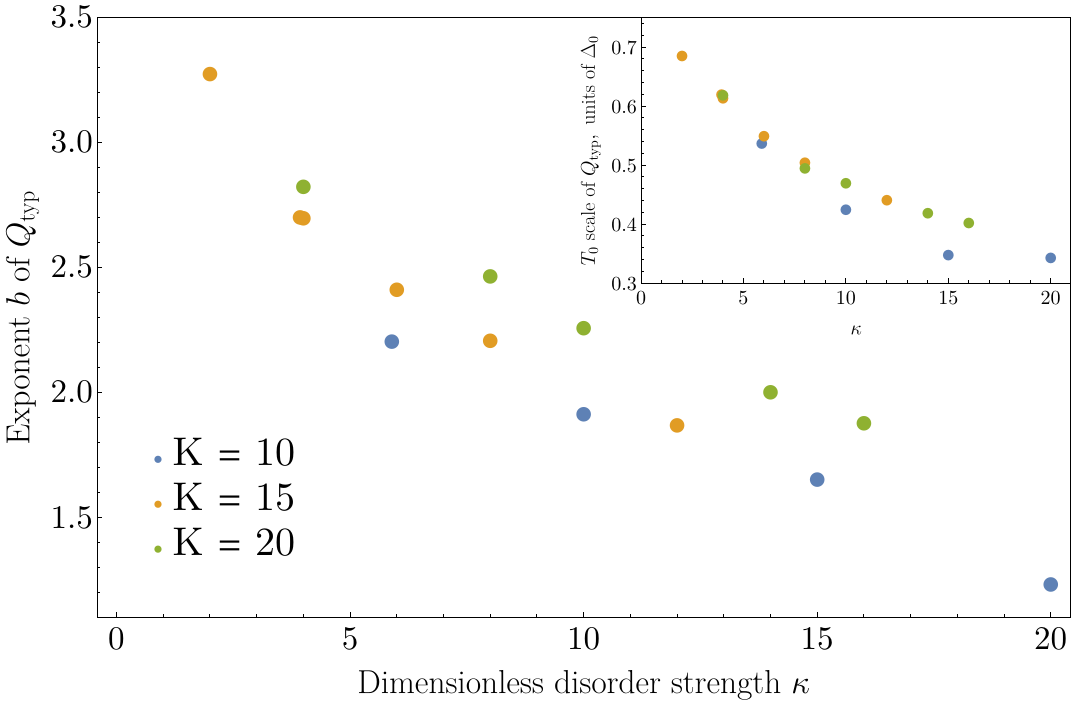}
\par\end{centering}
\caption{Dependence of the parameters $b$ (main panel) and $T_{0}$ (inset)
of the power-law fit $\left(T/T_{0}\right)^{b}$ of the portion of
numerical data for $\delta Q_{\text{typ}}\left(T\right)/Q_{\text{typ}}\left(0\right)$.
For each set of parameters, the points for fitting are selected according
to the empirically chosen criteria $T\in\left[\lambda\left\langle \Delta\right\rangle ,\lambda\left\langle \Delta\right\rangle +0.2\Delta_{0}\right]$,
with $\left\langle \Delta\right\rangle $ being the mean order parameter
at zero temperature, so the dataset for $K=15$ corresponds to highlighted
points on \figref{typical-Q_numerical-data}. The value of $\kappa$
was varied by changing the value of $\lambda$ while keeping $K$
constant, and various colors correspond to various values of $K$.
\protect\label{fig:typical-Q-exponent_disorder-dependence}}
\end{figure}

We start by discussing the qualitative shape of the temperature dependence
of the local current response presented on~\figref{typical-Q_numerical-data}.
There are three temperature regimes: \emph{i)~}at very low temperatures,
one observes exponentially small change in the superfluid stiffness,
with the numerical method being unable to properly resolve these values.
\emph{ii)~}At moderately low temperatures $T\sim0.05\,\Delta_{0}-0.3\,\Delta_{0}$
the dependence resembles a power law, although the exponent decreases
with temperature. Crucially, the apparent exponent of the power law
also decreases with disorder, as shown on~\figref{typical-Q-exponent_disorder-dependence}.
\emph{iii)}~At higher temperatures $0.3\,\Delta_{0}\le T<\Delta_{0}$
the local current response continues to decrease until it vanishes
at the transition point. We do not analyze the region of higher temperatures
$T\sim\Delta_{0}$, which might also be influenced by quaisparticles
due to finite value of the single-particle pseudogap $\Delta_{P}$.

Another important observation from the numerical results is that the
relative change in the typical current response $\delta Q_{\text{typ}}/Q_{\text{typ}}$
goes in line with that of the mean order parameter $\delta\left\langle \Delta\right\rangle /\left\langle \Delta\right\rangle $.
In particular, it is true that
\begin{equation}
\frac{\delta\left\langle \Delta\right\rangle /\left\langle \Delta\right\rangle }{\delta Q_{\text{typ}}/Q_{\text{typ}}}=0.5\pm0.05,
\label{eq:delta-q_to_delta-op}
\end{equation}
for all points highlighted on \figref{typical-Q_numerical-data},
regardless of the model parameters. This result is in agreement with
the qualitative prediction~(\ref{eq:typical-response-estimation}),
as our numerical experiments show that $\Delta_{\text{typ}}$ and
$\left\langle \Delta\right\rangle $ differ by a temperature-independent
factor. In addition, this numerical relation is roughly consistent
with the result of Ref.~\citep{feigelman_superfluid}, where the
relation $\rho_{S}\propto\Delta^{2}\sim h^{2}$ is claimed. While
our numerical experiments clearly imply that such a relation does
not hold literally, it is apparently correct as far as the relative
temperature variations are concerned. Eq.~(\ref{eq:delta-q_to_delta-op})
then allows us to calculate the temperature dependence of $Q_{\text{typ}}$
from that of $\left\langle \Delta\right\rangle $. The latter is amenable
for analytical description, as presented below.

\subsection{Analytical analysis of low-$T$ behavior\protect\label{subsec:Analytical-approach}}

While calculating $Q_{ij}$ requires diagonalization of the two-spin
Hamiltonian~(\ref{eq:two-sping_effective_Hamiltonian}), the order
parameter $\Delta$ is almost directly accessible from the statistics
of the $h$~fields. Indeed, from~(\ref{eq:effective_single-site_Hamiltonian})
and~(\ref{eq:h-equations}) one obtains:
\begin{equation}
\Delta_{i}=\sum_{j\in\partial i}J_{ij}\left\langle 2S_{j}^{\alpha}\right\rangle _{j\left(i\right)},
\label{eq:delta-via-h}
\end{equation}
which coincides with the Eq.~(\ref{eq:h-equations}) for $h$ itself
for up to the difference in the summation number of terms in the r.h.s
($K+1$ instead of $K$).

In what follows, we will compute the expectation $\left\langle h\right\rangle $
of the $h$ field, which coincides with the average order parameter
$\left\langle \Delta\right\rangle $ up to an insignificant factor
$\left(K+1\right)/K$, as can be seen from Eq.~(\ref{eq:delta-via-h}).
At the cost of additional technical effort, our theory also allows
us to calculate the typical value $h_{\text{typ}}=\exp\left\{ \left\langle \ln h\right\rangle \right\} $,
which is more relevant for the value of the superfluid stiffness according
to Eq.~(\ref{eq:typical-response-estimation}). However, this appears
to be unnecessary: the statistical data for the $h$ field obtained
by the MPD (see \subsecref{Statistical-properties-of-the-current-response})
demonstrates that the relation $\left\langle h\right\rangle /h_{\text{typ}}$
is nearly temperature-independent up until the transition region $T\sim\Delta_{0}$.
This also explains why we chose to use the mean value of $\left\langle \Delta\right\rangle $
rather than the typical value $\Delta_{\text{typ}}$ in the numerical
relation~(\ref{eq:delta-q_to_delta-op}): the former is easier to
calculate analytically while being proportional to the latter with
a nearly temperature-independent coefficient in the relevant range
of temperatures.

The distribution of $h$~fields can be computed straightforwardly
by means presented earlier in~\citep{op-distribution-paper}, although
the calculations are considerably simpler due to the absence of the
self-action. Indeed, the work~\citep{op-distribution-paper} dealt
with the necessity to disentangle the mutual correlation between the
values of $\Delta$ on neighboring sites due to their equivalence
in the self-consistency equation. On the other hand, the value of
$h_{j\rightarrow i}$ only collects information about the values of
$\xi_{k}$ and $h_{l\rightarrow m}$ in one direction (safe for the
presence of loops, which, however, do not influence the value of $h$
due to rapid decay of correlations with distance on the graph). This
feature has made this problem amenable to the MPD in the first place,
and the latter is equivalent to claiming that the two sides of the
Eq.~(\ref{eq:h-equations}) are equal in distributional sense, with
all random variables in the r.h.s being statistically independent:
\begin{widetext}
\begin{equation}
P_{h}\left(h\right)=\prod_{k=1}^{K}\intop_{0}^{\infty}P_{h}\left(h_{k}\right)dh_{k}\int P_{\xi}\left(\xi_{k}\right)d\xi_{k}\,\delta\left(h-\sum_{k=1}^{K}f\left(h_{k},\xi_{k}\right)\right),
\end{equation}
\end{widetext}

\noindent where $f\left(h_{k},\xi_{k}\right)$ is the term in the
r.h.s of Eq.~(\ref{eq:h-equations}) under the summation sign. As
soon as the equation on distribution of $h$ is obtained in such form,
the approach described below is similar to that of~\citep{op-distribution-paper}.
In particular, at zero temperature one finds for the cumulant generating
function $m\left(s\right)$:
\begin{equation}
m\left(s\right):=\ln\left\langle e^{ish}\right\rangle =is\left\langle h\right\rangle +\frac{\lambda\left\langle h\right\rangle }{\kappa}F\left(\kappa s\right),\,\,\,\text{Im}s\ll e^{1/\lambda},
\label{eq:m-function_zero-temp}
\end{equation}
\begin{equation}
F\left(\sigma\right)=\intop_{0}^{1}dw\frac{e^{i\sigma w}-1-i\sigma w}{w^{2}\sqrt{1-w^{2}}},
\end{equation}
with $\left\langle h\right\rangle $ determined self-consistently
from
\begin{equation}
\left\langle h\ln h\right\rangle =0.
\label{eq:mean-value-equation_zero-temp}
\end{equation}
Here, the l.h.s is calculated by using the expression~(\ref{eq:m-function_zero-temp})
for $m\left(s\right)$, thus representing a function of $\left\langle h\right\rangle $,
and Eq.~(\ref{eq:mean-value-equation_zero-temp}) has to be solved
for $\left\langle h\right\rangle $. In the same way, one can show
that the statistics of the order parameter is described by the following
generating function:
\begin{equation}
\ln\left\langle \exp\left\{ is\Delta\right\} \right\rangle =\frac{K+1}{K}m\left(s\right),
\end{equation}
and we once again remind that both $h$ and $\Delta$ are now measured
in units of $\Delta_{0}$. We are interested in $K\gg1$, hence the
prefactor in front of $m$, arising from the different number of summation
terms in Eqs.~(\ref{eq:h-equations})~and~(\ref{eq:delta-via-h}),
can be discarded. 

It is expected that any effect that is larger than the BCS-like exponential
dependence of the form $e^{-2\left\langle \Delta\right\rangle /T}$
will be produced by anomalously low values of the order parameter
(the unusual factor 2 in the exponent is due to the absence of single-electron
quasiparticles in our model). In other words, we are interested in
the probability of the order parameter $\Delta$ and the $h$ field
to attain values of the order of temperature, with latter being much
smaller than the mean order parameter~$\left\langle \Delta\right\rangle $.
Quantitatively, the extreme value statistics of $h$ is encoded in
the asymptotic of the $m$ function at $s=it,t\gg1$, as shown in~\citep{op-distribution-paper}.
This asymptotic is given by
\begin{equation}
m\left(it\right)\approx-At+Bt\ln t,\,\,\,1\ll t\ll t^{*}\sim e^{A/B-1},
\label{eq:m-function_asymptotic-behavior}
\end{equation}
\begin{equation}
A=\left\langle h\right\rangle \left\{ 1-\lambda\left[\gamma-1+\ln2\kappa\right]\right\} ,\,\,\,B=\lambda\left\langle h\right\rangle ,
\label{eq:A-B-values_zero-temp}
\end{equation}
where $\gamma=0.577...$ is the Euler-Mascheroni constant, and the
estimation for $t^{*}$ follows from the fact that expression~(\ref{eq:m-function_asymptotic-behavior})
is only valid until it no longer represents a decreasing function.
The distribution function $P\left(h\right)$ for $h<\left\langle h\right\rangle $
behaves as follows:%
\begin{equation}
P\left(h\right)\approx\sqrt{\frac{\zeta\left(h\right)}{2\pi B^{2}}}\exp\left\{ -\zeta\left(h\right)\right\} ,
\label{eq:h-probability_small-values}
\end{equation}
{} 
\begin{equation}
\zeta\left(h\right)=B\exp\left\{ \frac{A-h}{B}-1\right\} .
\end{equation}
For instance, at $T=0$, one uses Eq.~(\ref{eq:A-B-values_zero-temp})
to obtain $\zeta\left(h\right)=\frac{\lambda\left\langle h\right\rangle e^{-\gamma}}{2\kappa}\exp\left\{ \frac{1}{\lambda}\left(1-h/\left\langle h\right\rangle \right)\right\} $.
Note that while the exponent in $\zeta\left(h\right)$ features a
large factor $\sim1/\lambda\gg1$, the prefactor $\sim\lambda\left\langle h\right\rangle /2\kappa$
in front of this exponent is actually small in the region of interest,
rendering the probability density~(\ref{eq:h-probability_small-values})
a complicated function of $h$. 

Eq.~(\ref{eq:h-probability_small-values}) ceases to work at sufficiently
small $h$, where the relevant value of $t$ exceeds $t^{*}$, rendering
Eq.~(\ref{eq:m-function_asymptotic-behavior}) inapplicable. This
corresponds to $\zeta\left(h\right)$ reaching the value of $Bt^{*}$,
which happens at
\begin{equation}
h\sim h_{\min}=B=\lambda\left\langle h\right\rangle .
\end{equation}
Physically, $h_{\min}$ plays the role of the minimum value of the
$h$ field in the sense that the average $\left\langle e^{-2h/T}\right\rangle $
for $T\ll2h_{\min}$ is essentially given by $e^{-2h_{\min}/T}$,
which can also be expressed by the following description of the $m$
function for $t\ge t^{*}$ :
\begin{equation}
m\left(it\right)\sim-h_{\min}t,\,\,t\gg t^{*}.
\label{eq:m-function_largest-t-asymptotic}
\end{equation}
The value of $h_{\min}$ is then consistent with the continuity of
$m$, i.e. the values of $m\left(it^{*}\right)$ given by the two
asymptotic expressions~(\ref{eq:m-function_asymptotic-behavior})~and~(\ref{eq:m-function_largest-t-asymptotic})
coincide. 

We also note that Eq.~(\ref{eq:m-function_asymptotic-behavior})
is only valid for sufficiently weak fluctuations of the interaction
matrix elements $J_{ij}$ of the original Hamiltonian~(\ref{eq:pseudo-spin_Hamiltonian}),
viz. $\delta J/J\ll\sqrt{\lambda/\kappa}$~\citep{op-distribution-paper},
whereas our model assumes $J_{ij}=\text{const}$ for all connected
sites (see approximation \emph{iii)} in \subsecref{Model-Hamiltonian}).

The same analysis that lead to Eq.~(\ref{eq:m-function_zero-temp})
shows that for finite temperatures the value of $m\left(it\right)$
at $t\gg1$ acquires a correction
\begin{equation}
\delta m\left(it\right)\approx+\lambda t\frac{\partial}{\partial\beta}\intop_{0}^{1}dw\frac{e^{m\left(2i\beta/w\right)}}{\sqrt{1-w^{2}}},
\label{eq:m-correction}
\end{equation}
which is produced by the leading term in the expansion of $\tanh\beta\sqrt{\xi^{2}+\Delta^{2}}$
in Eq.~(\ref{eq:spin-average}) in powers of $\exp\left\{ -2\beta\sqrt{\xi^{2}+\Delta^{2}}\right\} $.
As already anticipated, the integral in the r.h.s is only sensitive
to the asymptotic of the $m$ function at $s=it,t\gg1$. The correction
itself amounts to a renormalization of the $A$ coefficient in Eq.~(\ref{eq:m-function_asymptotic-behavior}):
\begin{equation}
A\left(T\right)-A\left(0\right)=-\lambda\frac{\partial}{\partial\beta}\intop_{0}^{1}dw\frac{e^{m\left(2i\beta/w\right)}}{\sqrt{1-w^{2}}}.
\label{eq:A-correction}
\end{equation}
Eq.~(\ref{eq:mean-value-equation_zero-temp}) for the value of $\left\langle h\right\rangle $
also changes to
\begin{equation}
\left\langle h\ln h\right\rangle \approx+\frac{\partial}{\partial\beta}\intop_{0}^{1}dw\frac{e^{m\left(2i\beta/w\right)}}{\sqrt{1-w^{2}}},
\label{eq:mean-h-equation_correction}
\end{equation}
and contains exactly the same $w$ integral as the one in~(\ref{eq:A-correction}).
One has to solve the new equation for $\left\langle h\right\rangle $,
similarly to the zero-temperature case.

It is then natural to replace $m\left(it\right)$ in Eqs.~(\hphantom{}\ref{eq:A-correction}\nobreakdash-\ref{eq:mean-h-equation_correction}\hphantom{})
with its intermediate asymptotic expression~(\ref{eq:m-function_asymptotic-behavior})
to obtain a closed system of equations on $A$ and $B=\lambda\left\langle h\right\rangle $.
The emerging divergence of the integrals at small values of $w$ should
be cut at $w\sim w^{*}=2\beta/t^{*}$, in accordance with the limit
of applicability of the intermediate asymptotic~(\ref{eq:m-function_asymptotic-behavior}).
The remaining region $w\le w^{*}$ produces an exponentially small
contribution because for $t\ge t^{*}$ the value of $e^{m\left(it\right)}$
is exponentially small: $e^{m\left(it\right)}\sim e^{-h_{\text{min}}t}\ll1$.
If the temperatures are also exponentially small, i.e. $T\le T^{*}=2/t^{*}=e^{\gamma}e^{-1/\lambda}/\kappa\ll1$
one obtains $w^{*}\ge1$, so a more accurate calculation of the $w$~integral
has to be performed in this case. However, the same Eq.~(\ref{eq:m-function_largest-t-asymptotic})
still implies that the value of the $w$~integrals is of the order
of $e^{-2h_{\min}/T}\le e^{-h_{\min}t^{*}}\ll1$, implying an exponentially
small change in all physical quantities, which is consistent with
the numerical data on \figref{typical-Q_numerical-data}.

We also note that Eqs.~(\hphantom{}\ref{eq:m-correction}\nobreakdash-\ref{eq:mean-h-equation_correction}\hphantom{})
are applicable while the approximation $1-\tanh x\approx2e^{-2x}$
is applicable for the typical values of $x=\beta\sqrt{\xi^{2}+h^{2}}$,
which is true for $T\le T_{\max}\sim2\left\langle h\right\rangle $.

\begin{figure}
\begin{centering}
\includegraphics[scale=0.3]{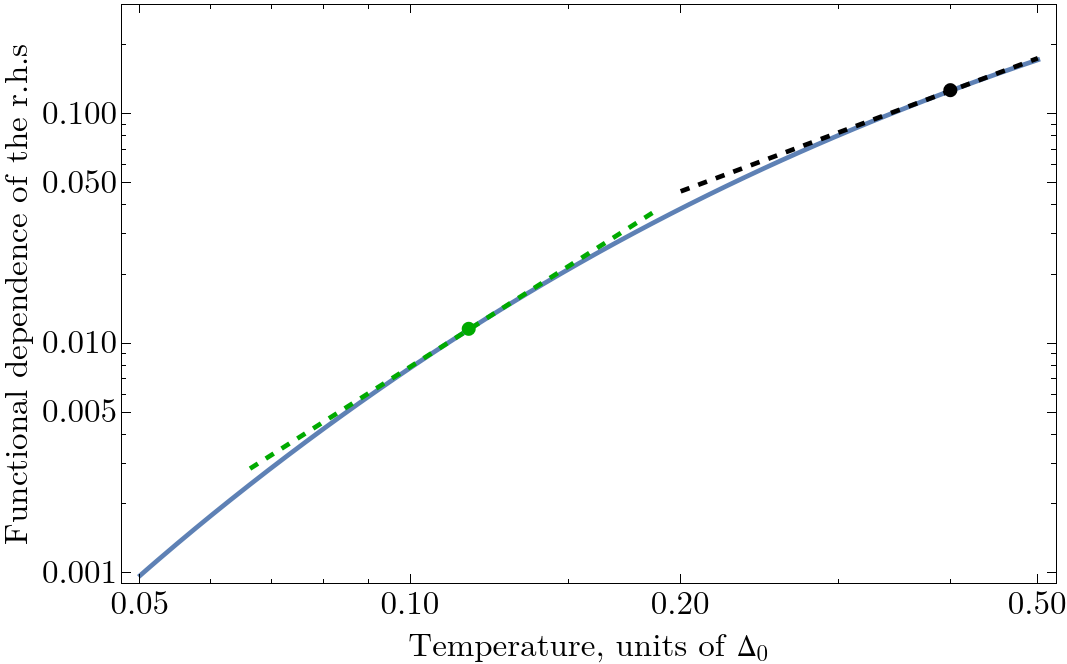}
\par\end{centering}
\caption{Temperature dependence of the r.h.s of~(\ref{eq:mean-value-correction_perturbation-theory})
for $\left\langle h\right\rangle =0.46\,\Delta_{0},\,\,\kappa=10,\,\,\lambda=0.123$.
The dashed straight lines visualize the log-derivative $b=d\ln I\left(T\right)/d\ln T$,
\emph{green}: $T=T_{\min}=2\lambda\left\langle h\right\rangle $,
$b=2.48$, \emph{black}: $T=0.4\,\Delta_{0}$, $b=1.46$. \protect\label{fig:perturbative-estimate-plot}}
\end{figure}

The qualitative behavior of the solution to Eqs.~(\hphantom{}\ref{eq:A-correction}\nobreakdash-\ref{eq:mean-h-equation_correction}\hphantom{})
can be understood by employing the direct perturbation theory in $T$.
Namely, one treats the change in $A$ and $\left\langle h\right\rangle $
as a perturbation, and the leading order of the latter is given by:
\begin{equation}
\delta h\propto I\left(T\right)=\frac{\partial}{\partial\beta}\intop_{w^{*}}^{1}dw\frac{e^{-A\frac{2\beta}{w}+B\frac{2\beta}{w}\ln\frac{2\beta}{w}}}{\sqrt{1-w^{2}}},\,\,w^{*}=\frac{2}{T}e^{1-A/B},
\label{eq:mean-value-correction_perturbation-theory}
\end{equation}
where the values $A,B$ are taken at $T=0$, i.e., from~(\ref{eq:A-B-values_zero-temp}),
and the omitted positive coefficient of proportionality is temperature-independent
(but does depend on $\kappa,\lambda$). The r.h.s of this equation
is negative, and the plot of its absolute value for a reasonable choice
of the parameters is shown on~\figref{perturbative-estimate-plot}.
It clearly indicates the same type of power-law-like behavior as the
one shown on~\figref{typical-Q_numerical-data}, although the power
varies with $T$, as can be see e.g. by the values of the log-derivative
$d\ln I/d\ln T$. 

\begin{figure}
\begin{centering}
\includegraphics[scale=0.5]{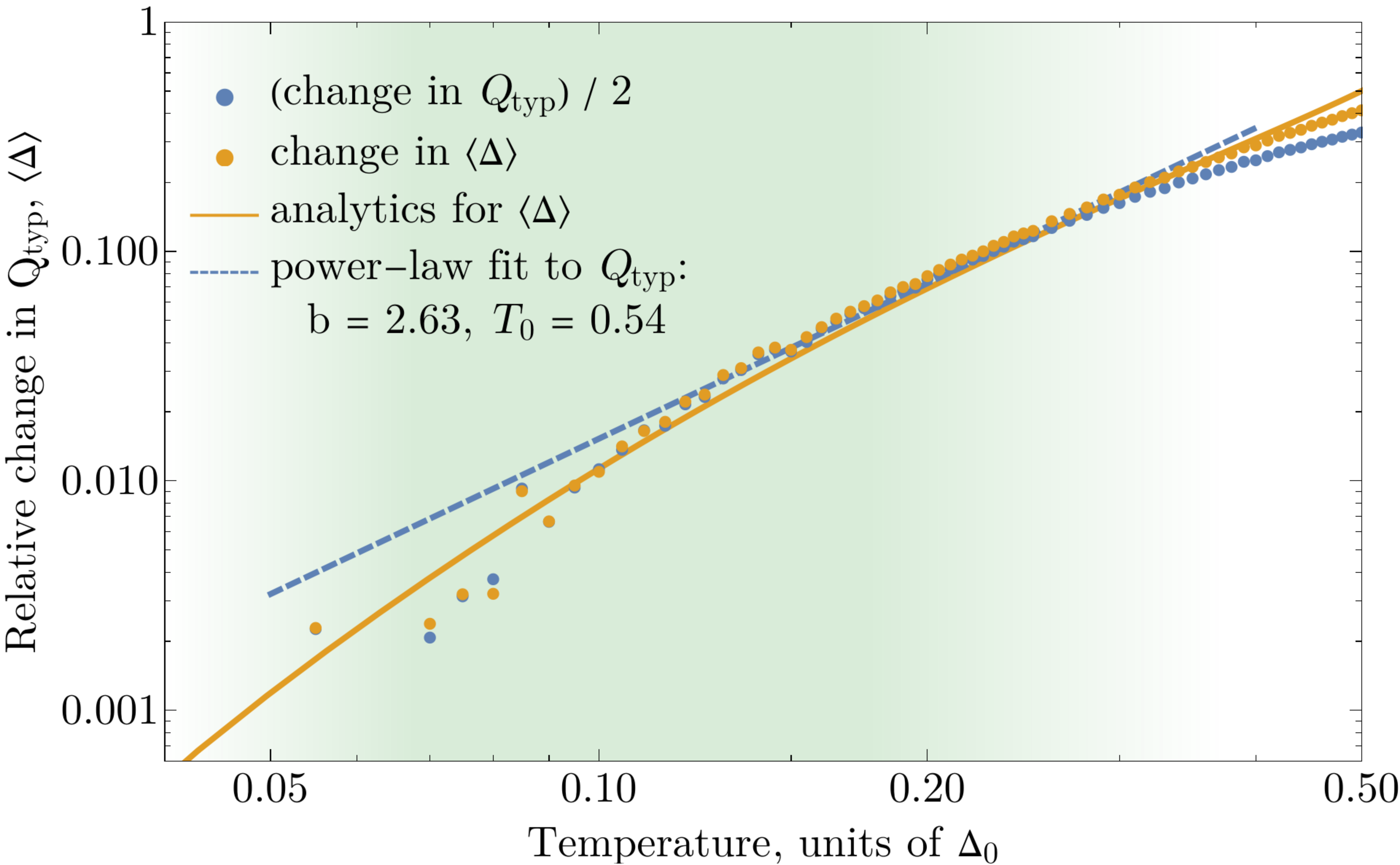}
\par\end{centering}
\caption{Low-temperature dependence of the relative changes in the typical
local current response $\delta_{Q}=1-Q_{\text{typ}}\left(T\right)/Q_{\text{typ}}\left(T=0\right)$
and the mean order parameter $\delta_{\Delta}=1-\left\langle \Delta\right\rangle \left(T\right)/\left\langle \Delta\right\rangle \left(T=0\right)$
according to the numerical MPD (solid dots) and theoretical description
given by Eqs.~(\protect\hphantom{}\ref{eq:A-correction}\protect\nobreakdash-\ref{eq:mean-h-equation_correction}\protect\hphantom{})
(solid line) in logarithmic scale along both axes. The values of $\delta_{Q}$
are divided by two, which makes relation~(\ref{eq:delta-q_to_delta-op})
apparent. The dashed line corresponds to the power-law fit $\left(T/T_{0}\right)^{b}$
of numerical data for $\delta_{Q}$ in the range $T\in\left[\lambda\left\langle \Delta\right\rangle ,\lambda\left\langle \Delta\right\rangle +0.2\right]$,
with $\left\langle \Delta\right\rangle =0.49\,\Delta_{0}$ (the values
of the fitting curve are also divided by two). The parameters of the
model are $K=20,\,\kappa=10$. The green region delineates the limits
of applicability of the theoretical approach corresponding to $\lambda\left\langle \Delta\right\rangle \le T\le\left\langle \Delta\right\rangle /2$.
\protect\label{fig:mean-order-parameter_numerics-vs-theory}}
\end{figure}

The comparison of the approximate theory with the numerical calculation
by means of the MPD is shown on~\figref{mean-order-parameter_numerics-vs-theory}
and is rather satisfying. Discrepancies are only visible at high temperature,
where the expressions~(\hphantom{}\ref{eq:m-correction}\nobreakdash-\ref{eq:mean-h-equation_correction}\hphantom{})
are no longer applicable due to higher powers of $e^{-\beta B}$ in
the low-temperature expansion of $\tanh\beta B$. This theory allows
us to give more quantitative description to the claims above: 
\begin{enumerate}
\item At low temperatures $T\le T_{\max}\sim\left\langle \Delta\right\rangle $
the temperature dependence of the superfluid stiffness $\delta\Theta\left(T\right)=\Theta\left(T=0\right)-\Theta\left(T\right)$
follows that of the mean order parameter:
\begin{equation}
\frac{\delta\Theta\left(T\right)}{\Theta\left(T=0\right)}\approx2\,\frac{\delta\left\langle \Delta\right\rangle \left(T\right)}{\left\langle \Delta\right\rangle \left(T=0\right)}.
\end{equation}
\item For $\lambda\left\langle \Delta\right\rangle \sim T_{\min}\le T\le T_{\max}\sim\left\langle \Delta\right\rangle $
a complicated profile of the temperature dependence of the superfluid
stiffness is observed. It can be roughly described by a power law
\begin{equation}
\frac{\delta\Theta}{\Theta}\approx\frac{\delta Q_{\text{typ}}}{Q_{\text{typ}}}\approx2\frac{\delta\left\langle \Delta\right\rangle }{\left\langle \Delta\right\rangle }\sim\left(T/T_{0}\right)^{b},
\label{eq:stiffness-defect_power-law}
\end{equation}
although the power $b$ gradually decreases with $T$, which also
allows other descriptions of the data (see \appref{Other-plots-for-the-data}).
The common trend is that $b$ decreases with disorder strength $\kappa$.
The qualitative shape of the dependence is described by the integral
of the form~(\ref{eq:mean-value-correction_perturbation-theory}).
We also note that our analysis suggests that the shape of this dependence
is sensitive to certain details of the microscopic model as the latter
determine the statistics of low values of the order parameter.
\item At $T\le T_{\min}$ the dependence of all physical quantities on temperature
roughly follows an activation profile:
\begin{equation}
\frac{\delta\Theta}{\Theta}\approx\frac{\delta Q_{\text{typ}}}{Q_{\text{typ}}}\approx2\frac{\delta\left\langle \Delta\right\rangle }{\left\langle \Delta\right\rangle }\sim\exp\left\{ -\frac{2T_{\min}}{T}\right\} .
\end{equation}
\end{enumerate}

\begin{figure}
\begin{centering}
\includegraphics[scale=0.5]{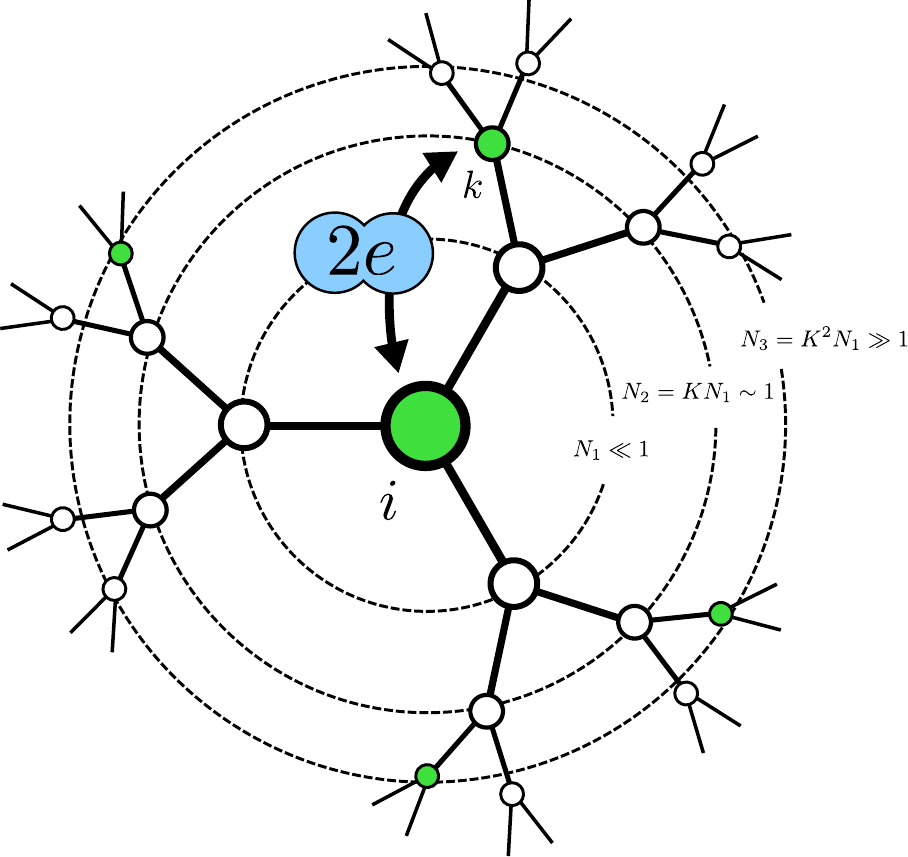}
\par\end{centering}
\caption{A sketch of the local two-level system due to hopping of a Cooper
pair between two states with low total onsite field $B_{i}=\sqrt{\xi_{i}^{2}+\Delta_{i}^{2}}$.
Each node represents a localized single-particle state, with green
ones satisfying $B_{i}\apprle T$, and links correspond to interaction
matrix elements, as explained in \secref{Model-and-Theory}. For any
green site $i$, the typical number of similar sites $N_{d}$ is small
in the nearest neighborhood, but it grows rapidly with the graph distance
$d$ (denoted by dashed circles), quickly reaching the order of unity.
Due to thermal excitation, a Cooper pair can hop back and forth between
two close green states $i$ and $k$ with $\xi_{i}>0$ (occupied at
$T=0$) and $\xi_{k}<0$ (empty at $T=0$), arranging a dipole. \protect\label{fig:sketch-local-TLS}}
\end{figure}

The results above allow us to address the nature of the excitations
that are responsible for the low-temperature suppression of the superfluid
stiffness. As our analysis suggests, these excitations are also responsible
for the suppression of the order parameter. The analytical description
of the latter reflects the the probability~$\mathbb{P}$ of extremely
low total onsite fields $B_{i}=\sqrt{\xi_{i}^{2}+\Delta_{i}^{2}}\apprle T\ll\left\langle \Delta\right\rangle $,
implying that the excitations in question are localized around the
points with diminished values of both the order parameter $\Delta_{i}$
and onsite energy $\left|\xi_{i}\right|$. Each such configuration
with $\xi_{i}<0$ is occupied by a Cooper pair that can easily hop
away due to thermal fluctuations. Our speculation is that this hopping
is local: the Cooper pair hops to a neighboring unoccupied site~$k$
(i.e., $\xi_{k}>0$) with sufficiently low total field~$B_{k}$.
At first glance, this might look implausible, given that the number
of such neighbors at graph distance $d=1$ away from the initial site~$i$
is $N_{1}\sim K\,\mathbb{P}\left(B<T\right)\sim K\,\nu_{0}nT\,\mathbb{P}\left(\Delta\apprle T\right)\ll1$
(where $\nu_{0}$ stands for the single-particle DoS at the Fermi
level, and $n$ is the concentration). However, the typical number
$N_{d}$ of suitable neighbors quickly grows with $d$ as $N_{d}=N_{1}K^{d-1}$,
reaching the order of unity at $d_{\text{typ}}\sim\ln\left\{ 1/N_{1}\right\} /\ln K$.
While this distance is large for small disorder, for sufficiently
disordered systems ($\kappa\ge1$) and in the range of temperatures
where the theory is applicable, the estimate above renders $d_{\text{typ}}\sim2-3$.
This translates to the typical hopping distance in real space of about
$(2-3)\,\xi_{\text{loc}}$. Since a Cooper pair carries charge~$2e$,
such pairs of nearby states with low total onsite fields arrange fluctuating
dipoles, and those are responsible for the thermal suppression of
the superfluid stiffness. The whole picture is summarized in~\figref{sketch-local-TLS}.

\begin{figure}
\begin{centering}
\includegraphics[scale=0.3]{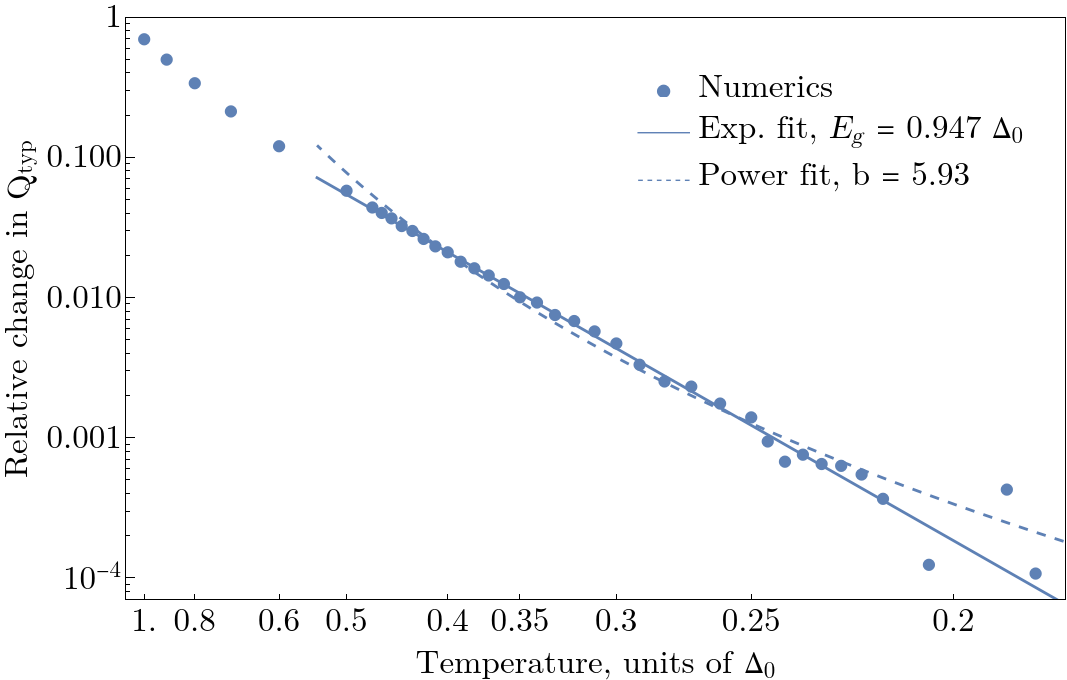}
\par\end{centering}
\caption{The temperature dependence of the relative change of typical local
current response $1-Q_{\text{typ}}\left(T\right)/Q_{\text{typ}}\left(T=0\right)$
for $\kappa=0.25,\,\,K=10$ (small disorder). The temperature is measured
in units of $\Delta_{0}$. The scale along the horizontal axis is
reciprocal ($1/T$), and the scale of the vertical axis is logarithmic.
The solid line corresponds to fit of the points with $T\le0.45\,\Delta_{0}$
with $A\exp\left\{ -2E_{g}/T\right\} $, resulting in $A=2.41,\,\,E_{g}=0.947\,\Delta_{0}$,
which is consistent with the standard BCS theory. Attempting to fit
the same points with a power law $\left(T/T_{0}\right)^{b}$ results
in $b\sim5.9$ (shown with the dashed line). \protect\label{fig:typical-Q_small-disorder}}
\end{figure}

We also note that the proposed description naturally incorporates
the small disorder limit $\kappa\ll1$. In this case, the distribution
of the order parameter is accurately described by a narrow Gaussian
distribution, leading to a BCS-like behavior of the form $e^{-2\left\langle \Delta\right\rangle /T}$
for both $\delta Q_{\text{typ}}$ and $\delta\left\langle \Delta\right\rangle $,
as demonstrated on~\figref{typical-Q_small-disorder}. In terms of
our analysis, this corresponds to replacing the $m$~function by
a Gaussian one:
\begin{equation}
m\left(it\right)\approx-\left\langle h\right\rangle t-\frac{\pi\kappa\lambda}{4}t^{2},
\end{equation}
which corresponds to small-$t$ expansion of the exact expression~(\ref{eq:m-function_zero-temp}).
This further reinforces the general conclusion: the nontrivial power-law-like
behavior of physical quantities is the direct consequence of the nontrivial
distribution of the order parameter~$\Delta$. We also emphasize
that the BCS-like result on \figref{typical-Q_small-disorder} is
obtained in the model with no quasi-particles (due to large pseudo-gap);
as a result, the activation energy is equal to \emph{twice} superconducting
gap, instead of the gap itself.

One may wonder if some modified activation-type behavior could describe
the data and our simulation results even at large $\kappa$. The corresponding
analysis is presented in \appref{Other-plots-for-the-data}. In particular,
\figref{typical-Q_numerical-data_reciprocal-scale} shows that the
activation-law fit $\delta\Theta(T)\sim\exp\left\{ -T_{1}/T\right\} $
works in a limited range of temperature, and the latter shrinks as
the dimensionless disorder strength~$\kappa$ increases. However,
for a larger value of the coordination number, $K=20$, we cannot
numerically resolve the distinction between power-law and exponential
fits, see \figref{mean-order-parameter_numerics-vs-theory_reciprocal-scale}.

\section{Discussion and Conclusions\protect\label{sec:Discussion}}

We presented both experimental data and analytical theory of the low-temperature
suppression of superfluid stiffness, $\delta\Theta(T)=\Theta(0)-\Theta(T)$,
in strongly disordered pseudogapped superconductors. 

The direct measurement of the superfluid stiffness at low temperatures
$T\ll T_{c}$ in amorphous $\text{InO}_{x}$ films revealed (see Fig.~\ref{fig:data}~\textbf{c})
a strong deviation from the activation-like behavior expected within
the semiclassical theory~\citep{MattisBardeen}. We observed instead
a power-law
\begin{equation}
\delta\Theta(T)/\Theta(0)\sim\left(T/T_{0}\right)^{b},
\label{eq:stiffness_approx-result}
\end{equation}
with an exponent~$b\sim1.6$ that is weakly sensitive to the disorder
strength and a characteristic scale $T_{0}$ that is roughly proportional
to the superfluid stiffness at low temperatures, viz., $T_{0}\sim5.5\,\Theta(0)$,
see Fig.~\ref{fig:T0_vs_Theta}. A systematic departure from~(\ref{eq:stiffness_approx-result})
towards higher values of $\delta\Theta$ is observed at higher temperatures
($T\apprge0.2T_{c}\approx0.5\,\text{K}$) for the less disordered
films, and might be attributed to the conventional Mattis-Bardeen
contribution due to quasiparticles.

To describe the experimental data, we developed an analytical approach
based on a microscopic model of strongly disordered superconductor
previously proposed in the literature~\citep{Ma_Lee_1985_Ref-to-pseudospins,Feigelman_Fractal-SC_2010,op-distribution-paper}.
The key ingredient of the model is the presence of a broad distribution
of the local pairing amplitude $\Delta(\mathbf{r})\neq\text{const}$.
The latter is known to exist~\citep{Sacepe_2011_for-pair-preformation}
in a relatively wide range of normal-state resistances and was described
theoretically in Ref.~\citep{op-distribution-paper}. The qualitative
behavior of the whole model is characterized primarily by the dimensionless
disorder strength $\kappa$. In particular, an estimation for $\kappa$
can be extracted from the observed shape of the pairing amplitude
distribution $P(\Delta)$ at large values~\citep{op-distribution-paper}.
The main challenge of the theoretical description is to connect theoretically
computable distribution $P\left(\Delta\right)$ with physically observable
quantities (e.g., the superfluid stiffness $\Theta$), as the latter
can no longer be expressed via the pairing amplitude in a simple fashion.

By combining analytical and numerical methods, we have shown that
this model also exhibits near-power-law suppression~(\ref{eq:stiffness_approx-result})
of the superfluid stiffness with temperature (see \figref{typical-Q_numerical-data}),
with disorder-dependent exponent $b\sim1.6-3$ in the region of large
disorder, $\kappa\gg1$. In particular, the exponent~$b$ decreases
with disorder and may become less than~2, as shown on~\figref{typical-Q-exponent_disorder-dependence}.
The characteristic temperature $T_{0}$ in Eq.~(\ref{eq:stiffness_approx-result})
is found to be of the order of the mean pairing amplitude $\left\langle \Delta\right\rangle $.
The dependence~(\ref{eq:stiffness_approx-result}) takes place in
the temperature range $\lambda\left\langle \Delta\right\rangle \leq T\leq\left\langle \Delta\right\rangle /2$,
and the temperature range itself reaches low temperatures $T\ll\left\langle \Delta\right\rangle $
due to the smallness of the dimensionless Cooper coupling constant
$\lambda$. However, even within this temperature interval the power-law
profile is only approximate, rendering the effective values of $b$
and $T_{0}$ temperature-dependent. In particular, \figref{perturbative-estimate-plot}
demonstrates that the \textquotedbl local\textquotedbl{} value of
$b$ decrease with the growth of temperature. At very low temperatures
$T\apprle\lambda\left\langle \Delta\right\rangle $, the value of
$\delta\Theta$ is better described by an activation law $\delta\Theta/\Theta\sim\exp\left\{ -2\lambda\left\langle \Delta\right\rangle /T\right\} $
with a diminished value of the gap in comparison to its semiclassical
value. This behavior results in lower values of $\delta\Theta$ than
those suggested by Eq.~(\ref{eq:stiffness_approx-result}). At high
temperatures $T\sim\Delta$, our theory is not applicable due to the
complete lack of quasiparticles within the proposed model.

However, even in the absence of the latter, we find within the same
formalism for the case of small disorder $\kappa\ll1$ the activation
behavior $\delta\Theta/\Theta\sim\exp\left\{ -2\left\langle \Delta\right\rangle /T\right\} $,
but with twice the standard value of the gap, see \figref{typical-Q_small-disorder}.
The effective gap is doubled since the energy $2\Delta$ corresponds
in our case to a \emph{single} excitation instead of two quasi-particles
in the usual BCS theory.

As a result, one can claim qualitative agreement between theory and
experiment, while a quantitative match is currently beyond reach due
to numerous simplifications employed in the theoretical model. The
quantitative result of our theory (summarized on \figref{mean-order-parameter_numerics-vs-theory})
is only valid in the model~(\ref{eq:pseudo-spin_Hamiltonian}) with
weak statistical fluctuations of the magnitude of interaction matrix
elements $J_{ij}$. As discussed in Ref.~\citep{op-distribution-paper},
strong fluctuations of the matrix elements render the low-value tail
of the distribution of the order parameter even more pronounced and
can thus increase the reported suppression of $\Theta$ as well as
change the overall shape of the temperature dependence. As a result,
the latter will still resemble a power-law, probably with a smaller
exponent $b$, due to a substantial fraction of system sites with
$\Delta_{i}<\left\langle \Delta\right\rangle $. An accurate description
of the matrix element $J_{ij}$ and the resulting effects on the superfluid
density and other physical observables are subjects of future work.
We also emphasize that these unusual low-$T$ properties are predicted
to exist even at relatively large $K\geq\lambda\exp(1/2\lambda)$,
where the replica-symmetry-breaking theory of the SIT~\citep{Feigelman_SIT_2010}
is not yet relevant. An additional element missed in our simplified
theory is the implication of the energy dependence of matrix elements~$J_{ij}$,
which is present due to Mott hybridization mechanism, as discussed
in Ref. \citep[Sec. 2.2.5]{Feigelman_Fractal-SC_2010}; this issue
we also leave for future studies.

Among other things to be considered in the future is the issue of
the relation between the macroscopic superfluid stiffness $\Theta$
and the local current response $Q^{\alpha\beta}\left(\omega;r,r'\right)$
discussed in \subsecref{Macroscopic-superfluid-stiffness}. The approximate
Eq.~(\ref{eq:macro-rho-S_via_typical-current-response}) is currently
supported by certain preliminary simulations as well as similarity
to the problem of the macroscopic response of disordered 2D media
discussed in Ref.~\citep{dykhne1971}, while a qualitative solution
to this problem is to be addressed in the near future. Importantly,
the employed approximation does not give access to the superfluid
stiffness itself, providing only its relative suppression with temperature
$\delta\Theta/\Theta$. This difficulty can be traced down to the
fact that kinetic quantities, such as $\Theta$, include information
about the embedding of the model in real space. For instance, the
semiclassical expression $\Theta\left(T=0\right)=\pi\Delta\,\left[\hbar/(2e)^{2}\right]/R_{\square}$
valid for weak disorder contains this information via the normal-state
resistivity per square $R_{\square},$ with the latter typically determined
experimentally. However, sufficiently disordered a:InO films as well
as the theoretical model used in the present paper demonstrate insulating
behavior in the normal state~\citep{Charpentier2023_thesis}, so
experimentally measured $R_{\square}$ at $T>T_{c}$ might be delivered
by a different mechanism than the one responsible for the formation
of finite superfluid stiffness $\Theta$ for $T<T_{c}$, rendering
$R_{\square}$ irrelevant for the value of $\Theta$. A consistent
treatment of this issue would include a more elaborate account of
the microscopic properties of the underlying single-particle Anderson
localization problem, as the latter unavoidably enter the kinetic
quantities. 

Another important point worth examining more thoroughly is the validity
of description of the local current response proposed in \subsecref{Local-current-response}.
This comprises neglecting \emph{i)~}the quantum fluctuation and \emph{ii)}~the
role of short loops that might be present in the interaction graph
of the system. There does exist certain evidence in favor of the first
approximation~\citep{Feigelman_SIT_2010}, but a more detailed analysis
is required. The second of these approximations is at least partially
controllable~\citep{bollobas2001random}, but because our description
involves dealing with extreme value statistics, small loops might
be important for certain physical quantities. However, estimations
of Ref.~\citep{op-distribution-paper} indicate that finite concentration
of short loops does not change the distribution of the order parameter
at low temperature, leaving the effect on physical quantities for
future studies.

On purely phenomenological grounds, the presence of nearly power-law
behavior~(\ref{eq:stiffness_approx-result}) indicates the existence
of low-energy excitations with energies much below~$\left\langle \Delta\right\rangle $.
Our theoretical approach allows us to reason about the nature of those
excitations, as explained in more detail in \subsecref{Analytical-approach}.
Our conjecture is that the relevant modes are a specific kind of two-level
systems (TLS) corresponding to hopping of a Cooper pair between two
neighboring single-particle localized states close to Fermi surface
with anomalously low order parameter, i.e., $\left|\xi_{i}\right|,\Delta_{i}\apprle T\ll\left\langle \Delta\right\rangle $.
We expect that the spatial size of these modes is of the order of
the typical interaction length between preformed Cooper pairs (up
to few localization lengths). Because a Cooper pair carries charge
$2e$, these modes possess a dipole moment $\sim\left(4-6\right)e\xi_{\text{loc}}$.
One experimental indication of the presence of such TLSs is the non-monotonous
temperature dependence of the microwave resonance quality factor $\mathcal{Q}(T)$,
with a maximum at some nonzero temperature~\citep{Charpentier2023_thesis,Charpentier_in_prep}.
However, to infer any certain conclusions, a detailed study of the
dissipation in the presented model is required, which is a subject
of future studies.

Notably, an experimental observation of TLSs with large dipole moments
(up to $3\,\text{nm}\times e$) was recently reported for thin films
of granular Aluminum (grAl)~\citep{Rotzinger-grAl2023}. These are
of the same scale as we would expect for $\text{InO}_{x}$, according
to the arguments above. While we do not have a detailed theory of
high-resistance grAl yet, preliminary consideration points to a qualitative
similarity between this material and amorphous~$\text{InO}_{x}$.
Indeed, given the ultra-small sizes of Al grains ($3-4\,\text{nm}$),
the superconducting coherence can only appear due to inter-grain tunneling.
In that sense, the effect of partial localization of electrons within
ultra-small grains is similar to the corresponding effect due to Anderson
localization in amorphous $\text{InO}_{x}$.

At first glance, the experimental data on $\text{InO}_{x}$ films
does not rule out more conventional origins of dipole-like TLSs: surface
atomic degrees of freedom in amorphous solids~\citep{Gao2008} or
localized quasiparticles~\citep{deGraaf2020}. However, the latter
mechanism is not present in strongly disordered superconductors at
low temperature due to the pseudogap $\Delta_{p}\apprge T_{c}\gg T$.
Further insights are obtained from the dependence of $\mathcal{Q}$
on the participation ration of the electric field~$p_{E}$ in the
resonator \citep[sec. 7.3]{Charpentier2023_thesis}\citep{Charpentier_in_prep}:
should the atomic surface TLSs be the dominant dissipation mechanism,
the quality factor would increase with the decrease of $p_{E}$, but
this does not seem to be the case for $\text{a:InO}_{x}$ films. This
observation suggests that the dissipation is indeed caused by the
bulk degrees of freedom, such as the ones described in this work.

Finally, the described dissipation mechanism assumes equilibrium thermal
dynamics of the discussed TLSs, which is not at all guaranteed, as
one needs to analyze the relevant \emph{inelastic} relaxation times.
The nonequilibrium situation would result in a dissipation mechanism
similar to the Debye relaxation recently discussed in relation with
superconductors in the mixed state~\citep{Debye1,Debye2}. Experimentally,
such a mechanism would manifest itself via diminished values of $\mathcal{Q}$
in comparison to the equilibrium case as well as via a different frequency
dependence of $\mathcal{Q}$.
\begin{acknowledgments}
The authors would like to thank Denis Basko for numerous fruitful
discussions. A.V.K. is grateful for the support by Laboratoire d’excellence
LANEF in Grenoble (ANR-10-LABX-51-01). B.S. has received funding from
the European Union's Horizon 2020 research and innovation program
under the ERC grant \emph{SUPERGRAPH} No. 866365. N.R. has received
funding from the European Union’s Horizon 2020 research and innovation
program under the ERC grant \emph{SuperProtected} No. 101001310. N.R.
and B.S. acknowledge funding from the ANR agency under the 'France
2030 plan', with reference ANR-22-PETQ-0003. Th.C. and B.S. acknowledge
funding from the ANR project ANR-19-CE30-0014 - CP-Insulators.
\end{acknowledgments}

\bibliographystyle{unsrt}
\bibliography{Bibliography}

\begin{thebibliography}{10}

\bibitem{Sacepe_2011_for-pair-preformation}
B.~Sacépé, T.~Dubouchet, C.~Chapelier, M.~Sanquer, M.~Ovadia, D.~Shahar, M.V.
  Feigel'man, and L.B. Ioffe.
\newblock Localization of preformed cooper pairs in disordered superconductors.
\newblock {\em Nature Physics}, 7(3):239–244, 2011.

\bibitem{Mooij05}
J.~E. Mooij and C.~J. P.~M. Harmans.
\newblock Phase-slip flux qubits.
\newblock {\em New Journal of Physics}, 7:219–219, 2005.

\bibitem{Mooij06}
J.~E. Mooij and Yu.~V. Nazarov.
\newblock Superconducting nanowires as quantum phase-slip junctions.
\newblock {\em Nature Physics}, 2(3):169–172, 2006.

\bibitem{Doucot12}
B.~Douçot and L.~B. Ioffe.
\newblock Physical implementation of protected qubits.
\newblock {\em Reports on Progress in Physics}, 75(7):072001, 2012.

\bibitem{Kitaev13}
P.~Brooks, A.~Kitaev, and J.~Preskill.
\newblock Protected gates for superconducting qubits.
\newblock {\em Phys. Rev. A}, 87(5):052306, 2013.

\bibitem{Groszowski18}
P.~Groszkowski, A.~{Di Paolo}, A.~L. Grimsmo, A.~Blais, D.~I. Schuster, A.~A.
  Houck, and J.~Koch.
\newblock Coherence properties of the 0-$\pi$ qubit.
\newblock {\em New Journal of Physics}, 20(4):043053, 2018.

\bibitem{grunhaupt2019granular}
L.~Grünhaupt, M.~Spiecker, D.~Gusenkova, N.~Maleeva, S.~T. Skacel,
  I.~Takmakov, F.~Valenti, P.~Winkel, H.~Rotzinger, W.~Wernsdorfer, A.~V.
  Ustinov, and I.~M. Pop.
\newblock Granular aluminium as a superconducting material for high-impedance
  quantum circuits.
\newblock {\em Nature Materials}, 18:816, 2019.

\bibitem{Astafiev12}
O.~V. Astafiev, L.~B. Ioffe, S.~Kafanov, Yu.~A. Pashkin, K.~Yu. Arutyunov,
  D.~Shahar, O.~Cohen, and J.~S. Tsai.
\newblock Coherent quantum phase slip.
\newblock {\em Nature}, 484:355, 2012.

\bibitem{Mooij16}
H.~Rotzinger, S.~T. Skacel, M.~Pfirrmann, J.~N. Voss, J.~Münzberg, S.~Probst,
  P.~Bushev, M.~P. Weides, A.~V. Ustinov, and J.~E. Mooij.
\newblock Aluminium-oxide wires for superconducting high kinetic inductance
  circuits.
\newblock {\em Superconductor Science and Technology}, 30:025002, 2016.

\bibitem{DeGraaf18}
S.E. de~Graaf, S.~T. Skacel, T.~Hönigl-Decrinis, R.~Shaikhaidarov,
  H.~Rotzinger, S.~Linzen, M.~Ziegler, U.~Hubner, H.-G. Meyer, V.~Antonov,
  E.~Il'ichev, A.~V. Ustinov, A.~Ya. Tzalenchuk, and O.~V. Astafiev.
\newblock Charge quantum interference device.
\newblock {\em Nature Physics}, 14:590, 2018.

\bibitem{Bylander19}
D.~Niepce, J.~Burnett, and J.~Bylander.
\newblock High kinetic inductance nbn nanowire superconductors.
\newblock {\em Phys. Rev. Applied}, 11:044014, 2019.

\bibitem{Gershenson19}
W.~Zhang, K.~Kalashnikov, W.-S. Lu, P.~Kamenov, T.~DiNapoli, and M.~E.
  Gershenson.
\newblock Microresonators fabricated from high-kinetic-inductance aluminum
  films.
\newblock {\em Phys. Rev. Applied}, 11:011003(R), 2019.

\bibitem{Astafiev22}
R.~S. Shaikhaidarov, K.~H. Kim, J.~W. Dunstan, I.~V. Antonov, S.~Linzen,
  M.~Ziegler, D.~S. Golubev, V.~N. Antonov, E.~V. Il'ichev, and O.~V. Astafiev.
\newblock Quantized current steps due to the a.c. coherent quantum phase-slip
  effect.
\newblock {\em Nature}, 608:45, 2022.

\bibitem{SFK-review-2020}
B.~Sacépé, M.V. Feigel'man, and T.~Klapwijk.
\newblock Quantum breakdown of superconductivity in low-dimensional materials.
\newblock {\em Nature Physics}, 16(7):734, 2020.

\bibitem{Sacepe10}
B.~Sacépé, C.~Chapelier, T.~I. Baturina, V.~M. Vinokur, M.~R. Baklanov, and
  M.~Sanquer.
\newblock Pseudogap in a thin film of a conventional superconductor.
\newblock {\em Nature Commun.}, 1:140, 2010.

\bibitem{NbN}
M.~Mondal, A.~Kamlapure, S.~C. Ganguli, J.~Jesudasan, V.~Bagwe, L.~Benfatto,
  and P.~Raychaudhuri.
\newblock Enhancement of the finite-frequency superfluid response in the
  pseudogap regime of strongly disordered superconducting films.
\newblock {\em Scientific Reports}, 3:1357, 2013.

\bibitem{GrAl}
F.~Levy-Bertrand, T.~Klein, T.~Grenet, O.~Dupré, A.~Benoît, A.~Bideaud,
  O.~Bourrion, M.~Calvo, A.~Catalano, A.~Gomez, J.~Goupy, L.~Grünhaupt,
  U.~v.~Luepke, N.~Maleeva, F.~Valenti, I.~M. Pop, and A.~Monfardini.
\newblock Electrodynamics of granular aluminum from superconductor to
  insulator: Observation of collective superconducting modes.
\newblock {\em Phys. Rev. B}, 99:094506, Mar 2019.

\bibitem{Klapwijk2013}
P.~C. J.~J. Coumou, E.~F.~C. Driessen, J.~Bueno, C.~Chapelier, and T.~M.
  Klapwijk.
\newblock Electrodynamic response and local tunneling spectroscopy of strongly
  disordered superconducting tin films.
\newblock {\em Phys. Rev. B}, 88:180505(R), 2013.

\bibitem{MattisBardeen}
D.~C. Mattis and J.~Bardeen.
\newblock Theory of the anomalous skin effect in normal and superconducting
  metals.
\newblock {\em Phys. Rev.}, 111:412–417, 1958.

\bibitem{Dubouchet_2018_Preformation-of-Pairs}
T.~Dubouchet, B.~Sacépé, J.~Seidemann, D.~Shahar, M.~Sanquer, and
  C.~Chapelier.
\newblock Collective energy gap of preformed cooper pairs in disordered
  superconductors.
\newblock {\em Nature Physics}, 15(3):233–236, 2018.

\bibitem{Feigelman_Microwave_2018}
M.V. Feigel'man and L.B. Ioffe.
\newblock Microwave properties of superconductors close to the
  superconductor-insulator transition.
\newblock {\em Phys. Rev. Lett.}, 120(3):037004, 2018.

\bibitem{op-distribution-paper}
A.~V. Khvalyuk and M.~V. Feigel'man.
\newblock Distribution of the order parameter in strongly disordered
  superconductors: An analytic theory.
\newblock {\em Phys. Rev. B}, 104:224505, 2021.

\bibitem{Grunhaupt2018}
L.~Grünhaupt, N.~Maleeva, S.~T. Skacel, M.~Calvo, F.~Levy-Bertrand, A.~V.
  Ustinov, H.~Rotzinger, A.~Monfardini, G.~Catelani, and I.~M. Pop.
\newblock Loss mechanisms and quasiparticle dynamics in superconducting
  microwave resonators made of thin-film granular aluminum.
\newblock {\em Phys. Rev. Lett.}, 121(11), 2018.

\bibitem{Day2003}
P.~K. Day, H.~G. LeDuc, B.~A. Mazin, A.~Vayonakis, and J.~Zmuidzinas.
\newblock A broadband superconducting detector suitable for use in large
  arrays.
\newblock {\em Nature}, 425(6960):817–821, 2003.

\bibitem{Kulik1973}
I.~O. Kulik.
\newblock Surface-charge oscillations in superconductors.
\newblock {\em Zh. Ekps. Teor. Fiz}, 65:2016–2022, 1973.

\bibitem{Mooij1985}
J.~E. Mooij and Gerd Schön.
\newblock Propagating plasma mode in thin superconducting filaments.
\newblock {\em Phys. Rev. Lett.}, 55:114--117, 1985.

\bibitem{Camarota2001}
B.~Camarota, F.~Parage, F.~Balestro, P.~Delsing, and O.~Buisson.
\newblock Experimental evidence of one-dimensional plasma modes in
  superconducting thin wires.
\newblock {\em Phys. Rev. Lett.}, 86:480--483, 2001.

\bibitem{Charpentier2023_thesis}
T.~Charpentier.
\newblock {\em {Quantum circuits and the superconductor-insulator transition in
  a strongly disordered superconductor}}.
\newblock Phd thesis, {Université Grenoble Alpes}, 2023.

\bibitem{Sacepe2015}
B.~Sacépé, J.~Seidemann, M.~Ovadia, I.~Tamir, D.~Shahar, C.~Chapelier,
  C.~Strunk, and B.~A. Piot.
\newblock {High-field termination of a Cooper-pair insulator}.
\newblock {\em Phys. Rev. B}, 91:220508(R), Jun 2015.

\bibitem{Maki1964}
K.~Maki.
\newblock The behavior of superconducting thin films in the presence of
  magnetic fields and currents.
\newblock {\em Progress of Theoretical Physics}, 31(5):731–741, 1964.

\bibitem{weissl2015kerr}
T.~Weißl, B.~Küng, É. Dumur, A.~K. Feofanov, I.~Matei, C.~Naud, O.~Buisson,
  F.W.J. Hekking, and W.~Guichard.
\newblock Kerr coefficients of plasma resonances in josephson junction chains.
\newblock {\em Phys. Rev. B}, 92(10):104508, 2015.

\bibitem{Krupko2018}
Yu. Krupko, V.~D. Nguyen, T.~Weißl, É. Dumur, J.~Puertas, R.~Dassonneville,
  C.~Naud, F.~W.~J. Hekking, D.~M. Basko, O.~Buisson, N.~Roch, and
  W.~Hasch-Guichard.
\newblock Kerr nonlinearity in a superconducting josephson metamaterial.
\newblock {\em Phys. Rev. B}, 98(9), 2018.

\bibitem{Feigelman_Fractal-SC_2010}
M.~V. Feigel'man, L.B. Ioffe, V.E. Kravtsov, and E.~Cuevas.
\newblock Fractal superconductivity near localization threshold.
\newblock {\em Annals of Physics}, 325(7):1390–1478, 2010.

\bibitem{Feigelman_SIT_2010}
M.V. Feigel'man, L.B. Ioffe, and M.~Mézard.
\newblock Superconductor-insulator transition and energy localization.
\newblock {\em Phys. Rev. B}, 82(18):184534, 2010.

\bibitem{bollobas2001random}
B.~Bollobás.
\newblock {\em Random graphs}.
\newblock Number~73 in Camridge studies in advanced mathematics. Cambridge
  university press, second edition, 2001.

\bibitem{Mirlin_1991}
A.~D. Mirlin and Y.~V. Fyodorov.
\newblock Localization transition in the anderson model on the bethe lattice:
  Spontaneous symmetry breaking and correlation functions.
\newblock {\em Nuclear Physics B}, 366(3):507–532, 1991.

\bibitem{spivak_1988_mesoscopic_rho-s_fluctuations}
B.Z. Spivak and A.~Yu. Zyuzin.
\newblock Mesoscopic fluctuations of the superfluid current density in
  disordered superconductors.
\newblock {\em JETP Letters}, 47(4), 1988.

\bibitem{AGD}
A.A. Abrikosov, L.P. Gorkov, and I.E. Dzyaloshinski.
\newblock {\em Methods of quantum field theory in statistical physics}.
\newblock American Institute of Physics, 1964.

\bibitem{dykhne1971}
A.M. Dykhne.
\newblock Conductivity of a two-dimensional two-phase system.
\newblock {\em Sov. Phys. JETP}, 32(1):63–65, 1971.

\bibitem{MezardParisi_cavity0}
M.~Mézard and G.~Parisi.
\newblock The bethe lattice spin glass revisited.
\newblock {\em The European Physical Journal B}, 20(2):217–233, 2001.

\bibitem{MezardParisi_cavity1}
M.~Mézard and G.~Parisi.
\newblock The cavity method at zero temperature.
\newblock {\em Journal of Statistical Physics}, 111(1/2):1–34, 2003.

\bibitem{yedidia_belief-propagation}
J.~S. Yedidia, W.~T. Freeman, and Y.~Weiss.
\newblock Understanding belief propagation and its generalizations.
\newblock {\em Exploring artificial intelligence in the new millennium},
  8(236-239):0018–9448, 2003.

\bibitem{BiroliCulia}
G.~Biroli and L.~F. Cugliandolo.
\newblock Quantum thouless-anderson-palmer equations for glassy systems.
\newblock {\em Phys. Rev. B}, 64:014206, 2001.

\bibitem{TAP_1977}
D.~J. Thouless, P.~W. Anderson, and R.~G. Palmer.
\newblock Solution of 'solvable model of a spin glass'.
\newblock {\em Philosophical Magazine}, 35(3):593–601, 1977.

\bibitem{feigelman_superfluid}
M.~V. Feigel'man and L.~B. Ioffe.
\newblock Superfluid density of a pseudogapped superconductor near the
  superconductor-insulator transition.
\newblock {\em Phys. Rev. B}, 92:100509(R), 2015.

\bibitem{Ma_Lee_1985_Ref-to-pseudospins}
M.~Ma and P.A. Lee.
\newblock Localized superconductors.
\newblock {\em Phys. Rev. B}, 32(9):5658, 1985.

\bibitem{Charpentier_in_prep}
Thibault~Charpentier et~al.
\newblock {In preparation}.

\bibitem{Rotzinger-grAl2023}
Maximilian Kristen, Jan~Nicolas Voss, Micha Wildermuth, Alexander Bilmes,
  J\"{u}rgen Lisenfeld, Hannes Rotzinger, and Alexey~V. Ustinov.
\newblock Observation of giant two-level systems in a granular superconductor.
\newblock 2023.

\bibitem{Gao2008}
J.~Gao, M.~Daal, A.~Vayonakis, S.~Kumar, J.~Zmuidzinas, B.~Sadoulet, B.~A.
  Mazin, P.~K. Day, and H.~G. Leduc.
\newblock Experimental evidence for a surface distribution of two-level systems
  in superconducting lithographed microwave resonators.
\newblock {\em App. Phys. Lett.}, 92:152505, 2008.

\bibitem{deGraaf2020}
S.~E. de~Graaf, L.~Faoro, L.~B. Ioffe, S.~Mahashabde, J.~J. Burnett,
  T.~Lindström, S.~E. Kubatkin, A.~V. Danilov, and A.~Ya. Tzalenchuk.
\newblock Two-level systems in superconducting quantum devices due to trapped
  quasiparticles.
\newblock {\em Sci. Adv.}, 6:eabc5055, 2020.

\bibitem{Debye1}
M.~Smith, A.~V. Andreev, and B.~Z. Spivak.
\newblock Debye mechanism of giant microwave absorption in superconductors.
\newblock {\em Phys. Rev. B}, 101:134508, 2020.

\bibitem{Debye2}
B.~V. Pashinsky, M.~V. Feigel’man, and A.V. Andreev.
\newblock Microwave response of type-{II} superconductors at weak pinning.
\newblock {\em SciPost Physics}, 14(5):096, 2023.

\end{thebibliography}

\appendix

\appendix

\section{Electrodynamics of a disordered superconductor\protect\label{app:disordered-sc_electrodynamics}}

In this Appendix, we formulate the low-frequency description of the
electromagnetic response of a disordered superconductor, with the
particular aim of deriving Eq.~(\ref{eq:macro-rho-S_problem}). 

Consider the Fourier transform of the Maxwell equations for the electromagnetic
potentials $\left(\varphi,\boldsymbol{A}\right)$ in the Coulomb gauge
$\text{div}\boldsymbol{A}=0$:
\begin{equation}
\begin{cases}
-\Delta\varphi=4\pi\rho/\varepsilon\left(\omega\right),\\
\left(-\Delta+\left(\frac{i\omega}{c}\right)^{2}\right)\boldsymbol{A}+\nabla\left(\frac{i\omega}{c}\varphi\right)=\frac{4\pi}{c}\boldsymbol{j},\\
\boldsymbol{E}=-\nabla\varphi-\frac{i\omega}{c}\boldsymbol{A}.
\end{cases}
\label{eq:maxwell-eqs_exact}
\end{equation}
where $\varepsilon\left(\omega\right)$ is the dielectric permittivity
due to the media surrounding the superconductor and the electrons
deep within the Fermi surface (for simplicity, we neglect its spatial
dependence, it can straightforwardly be restored). All fields here
depend on the frequency $\omega$ and on the coordinate~$\boldsymbol{r}$.
The material response fields $\rho,\boldsymbol{j}$ are to be determined
from
\begin{equation}
\begin{cases}
j^{\alpha}\left(\boldsymbol{r}\right)=\frac{c}{i\omega}\intop d^{3}r'\,Q^{\alpha\beta}\left(\omega;\boldsymbol{r},\boldsymbol{r}'\right)\,E^{\beta}\left(\boldsymbol{r}'\right),\\
\rho\left(\boldsymbol{r}\right)=-\frac{c}{i\omega}\intop d^{3}r'\,R^{\beta}\left(\omega;\boldsymbol{r},\boldsymbol{r}'\right)E^{\beta}\left(\boldsymbol{r}'\right),
\end{cases}
\label{eq:maetrial-response}
\end{equation}
and are bound to satisfy the charge conservation law
\begin{equation}
i\omega\rho+\text{div}\boldsymbol{j}=0.
\label{eq:charge-conservation}
\end{equation}
Here, the kernels $Q$ and $R$ are the Fourier transforms of the
material responses of the medium to the vector potential computed
as direct variational derivatives:
\begin{align}
 & Q^{\alpha\beta}\left(t-t';\boldsymbol{r},\boldsymbol{r}'\right)=-\frac{\delta\left\langle j^{\alpha}\left(t,\boldsymbol{r}\right)\right\rangle }{\delta A^{\beta}\left(t',\boldsymbol{r}'\right)},\nonumber \\
 & R^{\beta}\left(t-t';\boldsymbol{r},\boldsymbol{r}'\right)=\frac{\delta\left\langle \rho\left(t,\boldsymbol{r}\right)\right\rangle }{\delta A^{\beta}\left(t',\boldsymbol{r}'\right)}.
\label{eq:material-response-functions}
\end{align}
As per usual, the system (\ref{eq:maxwell-eqs_exact}-\ref{eq:charge-conservation})
is over-complete because the second equation in Eq.~(\ref{eq:maetrial-response})
describing the charge density response is actually satisfied automatically
by the true configuration of the electromagnetic fields due to the
charge conservation law, Eq.~(\ref{eq:charge-conservation}), so
we can discard the charge response relation in Eq.~(\ref{eq:maetrial-response}).
We can then exclude the charge variable with the help of Eq.~(\ref{eq:maxwell-eqs_exact})
and obtain
\begin{equation}
\begin{cases}
\text{div}\boldsymbol{j}=i\omega\,\frac{\Delta\varphi}{4\pi\varepsilon\left(\omega\right)}\\
\left(-\Delta+\left(\frac{i\omega}{c}\right)^{2}\right)\boldsymbol{A}+\nabla\left(\frac{i\omega}{c}\varphi\right)=\frac{4\pi}{c}\boldsymbol{j},\\
\text{div}\boldsymbol{A}=0,\\
\boldsymbol{E}=-\nabla\varphi-\frac{i\omega}{c}\boldsymbol{A},\\
j^{\alpha}\left(\boldsymbol{r}\right)=\frac{c}{i\omega}\intop d^{3}r'\,Q^{\alpha\beta}\left(\omega;\boldsymbol{r},\boldsymbol{r}'\right)E^{\beta}\left(\boldsymbol{r}'\right).
\end{cases}
\label{eq:maxwell-eqs_exact_no-charge}
\end{equation}

We now want to calculate the superconducting response to a given external
electromagnetic field $E_{\text{ext}}\left(\boldsymbol{r},t\right)$,
which is created by currents other than those in the material in question.
Certainly, this external field automatically satisfies the first four
relations in system~(\ref{eq:maxwell-eqs_exact_no-charge}), but
with the current density $\boldsymbol{j}$ produced by some external
sources (whose currents are obviously unaltered by the response of
the medium). Therefore, to analyze the response of the target medium
we have to substitute the full electric field $\boldsymbol{E}_{\text{full}}=\boldsymbol{E}+\boldsymbol{E}_{\text{ext}}$
in the last expression describing the current response of the medium.
In this way, the Eqs.~(\ref{eq:maxwell-eqs_exact_no-charge}) are
rendered inhomogeneous, so the response problem is well-formulated.

The main difference of the resulting system of equations~(\ref{eq:maxwell-eqs_exact_no-charge})
with the standard semiclassical description is that due to the inhomogeneity
of the superconducting state the current conservation law is not satisfied
``automatically'' by an appropriate choice of gauge. Instead, the
system adjusts the values of microscopic currents and induced electromagnetic
field in order to satisfy the charge conservation, making the resulting
value of the macroscopic response much more complicated than simply
the mean value of the miscroscopic response. In this way, the problem
is similar to the one of the macroscopic conductance of a disordered
media, where the potential distribution is determined from the charge
conservation (or, equivalently, the Kirchhoff's rule for a discrete
system).

We are interested in the low-frequency limit of the system~(\ref{eq:maxwell-eqs_exact_no-charge})%
. Because the system is superconducting, the zero-frequency response
function $Q^{\alpha\beta}\left(\omega=0;\boldsymbol{r},\boldsymbol{r}'\right)$
does not vanish. Eq.~(\ref{eq:maxwell-eqs_exact_no-charge}) then
suggests the following substitution to reproduce the correct low-frequency
behavior of the response:
\begin{equation}
\boldsymbol{j}=\frac{\boldsymbol{u}}{i\omega},\,\,\,\boldsymbol{A}=-\frac{c}{i\omega}\boldsymbol{\mathcal{E}}
\end{equation}
which then leads to the following equivalent system:
\begin{equation}
\begin{cases}
\text{div}\boldsymbol{u}=\left(i\omega\right)^{2}\,\frac{\Delta\varphi}{4\pi\varepsilon\left(\omega\right)},\\
\left(-\Delta+\left(\frac{i\omega}{c}\right)^{2}\right)\left(-c\boldsymbol{\mathcal{E}}\right)+\nabla\left(\frac{\left(i\omega\right)^{2}}{c}\varphi\right)=\frac{4\pi}{c}\boldsymbol{u},\\
\text{div}\boldsymbol{\mathcal{E}}=0,\\
\boldsymbol{E}=-\nabla\varphi+\boldsymbol{\mathcal{E}},\\
u^{\alpha}\left(\boldsymbol{r}\right)=c\intop d^{3}r'\,Q^{\alpha\beta}\left(\omega;\boldsymbol{r},\boldsymbol{r}'\right)\,E_{\text{full}}^{\beta}\left(\boldsymbol{r}'\right).
\end{cases}
\label{eq:supercond-response_exact-eqs}
\end{equation}
One can then take the zero-frequency limit of these equations by simply
putting $\omega=0$, which renders
\begin{equation}
\begin{cases}
\text{div}\boldsymbol{u}=0,\\
\Delta\boldsymbol{\mathcal{E}}=\frac{4\pi}{c^{2}}\boldsymbol{u},\\
\text{div}\boldsymbol{\mathcal{E}}=0,\\
\boldsymbol{E}=-\nabla\varphi+\boldsymbol{\mathcal{E}},\\
u^{\alpha}\left(\boldsymbol{r}\right)=c\intop d^{3}r'\,Q^{\alpha\beta}\left(0;\boldsymbol{r},\boldsymbol{r}'\right)\,E_{\text{full}}^{\beta}\left(\boldsymbol{r}'\right).
\end{cases}
\label{eq:supercond-response_exact-eqs_zero-freq}
\end{equation}

As we can see, even in the limit of vanishing frequency, the medium
generally responds by inducing both electric and magnetic fields,
as $\text{rot}\boldsymbol{E}=\text{rot}\boldsymbol{\mathcal{E}}\neq\boldsymbol{0}$.
However, there's a natural division of scales in this equations, which
allows one to simplify the system significantly as far as the value
of the superfluid stiffness is concerned. Let's decompose all fields
into a sum of fast and slow components, with the distinction made
along the scale $l_{\text{m}}$ at which one can consider the superfluid
stiffness as a self-averaging quantity. The order of $l_{\text{m}}$
can roughly be estimated by the size of the tree-like structure of
the underlying interaction graph (see \subsecref{Model-Hamiltonian}
of the main text), so $l_{\text{m}}\sim C\xi_{\text{loc}}$, where
$C$ is a number of order 2%
. In this case, all equations except the last in system~(\ref{eq:supercond-response_exact-eqs_zero-freq})
independently describe slow and fast components, while the last equation
in system~(\ref{eq:supercond-response_exact-eqs_zero-freq}) mixes
both slow and fast components of the field $\boldsymbol{E}$ due to
small spatial scale of the change in the $Q$ kernel. However, from
the second equation in system~(\ref{eq:supercond-response_exact-eqs_zero-freq})
the fast component of $\boldsymbol{\mathcal{E}}$ can be estimated
as
\begin{equation}
\boldsymbol{\mathcal{E}}^{\left(\text{fast}\right)}\sim\frac{4\pi}{c^{2}}l_{\text{m}}^{2}\boldsymbol{u}\sim\frac{4\pi}{c^{2}}l_{\text{m}}^{2}\rho_{S}\boldsymbol{E}=\frac{l_{\text{m}}^{2}}{\lambda_{L}^{2}}\boldsymbol{E},
\end{equation}
where $\lambda_{L}=\sqrt{c^{2}/4\pi\rho_{S}}$ is the London penetration
depth. Because we expect $\lambda_{L}\gg l_{m}$, this contribution
to the electric field is small, so we can neglect it and obtain
\begin{equation}
\begin{cases}
\text{div}\boldsymbol{u}=0,\\
\boldsymbol{E}=-\nabla\varphi,\\
u^{\alpha}\left(\boldsymbol{r}\right)=c\intop d^{3}r'\,Q^{\alpha\beta}\left(0;\boldsymbol{r},\boldsymbol{r}'\right)E_{\text{full}}^{\beta}\left(\boldsymbol{r}'\right),
\end{cases}
\label{eq:supercond-response_potential-approx-eqs}
\end{equation}
which is equivalent to Eq.~(\ref{eq:macro-rho-S_problem}) presented
in the main text. Note that this only works for the response at small
scales, while at large scales one still has to solve the exact system~(\ref{eq:supercond-response_exact-eqs_zero-freq}).
However, because we expect that the superfluid responses averages
over scales $l_{\text{m}}$, one can replace the last equation with
a simple London-type relation at large scales:
\begin{equation}
\begin{cases}
\text{div}\boldsymbol{u}=0,\\
\Delta\boldsymbol{\mathcal{E}}=\frac{4\pi}{c^{2}}\boldsymbol{u},\\
\text{div}\boldsymbol{\mathcal{E}}=0,\\
\boldsymbol{E}=-\nabla\varphi+\boldsymbol{\mathcal{E}},\\
u^{\alpha}\left(\boldsymbol{r}\right)=\rho_{S}E_{\text{full}}^{\beta}\left(\boldsymbol{r}\right),
\end{cases}
\end{equation}
which can be cast in the form of the standard London equations because
for $\rho_{S}=\text{const}$ the solution for $\varphi$ is trivial,
viz. $\varphi\left(\boldsymbol{r}\right)=0$:
\begin{equation}
\begin{cases}
\text{div}\boldsymbol{j}=0,\\
-\Delta\boldsymbol{A}=\frac{4\pi}{c}\boldsymbol{j},\\
\text{div}\boldsymbol{A}=0,\\
\boldsymbol{j}=-\frac{1}{c}\rho_{S}\boldsymbol{A}_{\text{full}}.
\end{cases}
\end{equation}

Another case where the approximate Eqs.~(\ref{eq:supercond-response_potential-approx-eqs})
are applicable is that of a thin film, for which the magnetic part
of the response is small regardless of the scale, so one neglect it
and put $\boldsymbol{\mathcal{E}}=\boldsymbol{0}$. 

As a result, the system~(\ref{eq:supercond-response_exact-eqs})
accurately describes the full electromagnetic response of a disordered
superconductor, with the system~(\ref{eq:supercond-response_exact-eqs_zero-freq})
being the corresponding low-frequency limit. This includes both the
microscopic effects arising from strongly inhomogeneous superconducting
state and the macroscopic effects such as the Meissner effect. At
the same time, the division of scales or smallness of the absolute
value of the currents in the material (such as in case of a thin film)
allow one to determine the superfluid density of such a superconductor
by using the potential approximation, Eqs.~(\ref{eq:supercond-response_potential-approx-eqs}),
i.e., neglecting the magnetic part of the response.%

\section{Spatial structure of the current response to a potential electric
field\protect\label{app:Current-response_to_potential-field}}

In this Appendix, we analyze the spatial structure of the low-frequency
current response to a potential electric field. In particular, we
derive Eq.~(\ref{eq:macro-response_discrete-problem}) used to determine
the superfluid density in the system via the characteristics of the
problem on a graph. We start from verifying the discrete current conservation
identity. According to Eq.~(\ref{eq:real-space-current_via_edges-currents}),
the charge conservation condition $\text{div}\boldsymbol{j}=0$ corresponds
to
\begin{equation}
0=\frac{1}{2}\sum_{e}\text{div}\boldsymbol{D}_{e}\left(\boldsymbol{r}\right)\,I_{e},
\end{equation}
where $I_{e}$ is the current along directed edge $e$. Due to identity~(\ref{eq:div-D_condition}),
this equation reduces to
\begin{equation}
0=\sum_{i}\left|\psi_{i}\left(\boldsymbol{r}\right)\right|^{2}\,\sum_{j\in\partial i}I_{i\rightarrow j},
\end{equation}
where we have used current anti-symmetry $I_{i\rightarrow j}=-I_{j\rightarrow i}$.
This should be true for any point $\boldsymbol{r}$, which is only
possible if the coefficient in front of each $\left|\psi_{i}\left(\boldsymbol{r}\right)\right|^{2}$
vanishes, rendering the discrete charge conservation, Eq.~(\ref{eq:macro-response_discrete-problem}).

We should now establish the connection between the current response
function in real space and that on the graph. The vector potential
is coupled to the system as $H_{\text{EM}}=-\frac{1}{c}\intop d\boldsymbol{r}'\,\left(\boldsymbol{j}\left(\boldsymbol{r}\right),\boldsymbol{A}\left(\boldsymbol{r}\right)\right)$,
which, together with Eq.~(\ref{eq:real-space-current_via_edges-currents}),
implies that the response of $I_{e}\left(t\right)$ to $\boldsymbol{A}\left(t,\boldsymbol{r}\right)$
reads
\begin{equation}
I_{e}\left(t\right)=-\frac{1}{2}\sum_{e'}\,\intop dt'\,Q_{ee'}\left(t,t'\right)\,A_{e'}\left(t'\right),
\label{eq:edge-current_response_to-edge-potential}
\end{equation}
where the discrete edge potential $A_{e'}$ is given by~(\ref{eq:edge-vector-potential_definition}),
and $Q_{ee'}\left(t,t'\right)$ is the nonlocal current response on
the graph, including the diamagnetic term due to the explicit dependence
of the current operator on the vector potential, Eq~(\ref{eq:current-operator_definition}). 

According to Eq.~(\ref{eq:macro-rho-S_problem}), we are interested
in the current response to a potential vector field $\boldsymbol{A}=c/i\omega\,\nabla\phi$,
for which Eqs.~(\ref{eq:edge-vector-potential_definition})~and~(\ref{eq:div-D_condition})
allow to express the corresponding discrete edge potential as
\begin{equation}
A_{e}=\frac{c}{i\omega}\,\frac{\phi_{\text{end}\left(e\right)}-\phi_{\text{beg}\left(e\right)}}{2},
\end{equation}
where for a directed edge $e=i\rightarrow j$ we denoted $\text{end}\left(e\right)=j$
and $\text{beg}\left(e\right)=i$, and the \emph{discrete }field $\phi_{i}$
is defined according to Eq.~(\ref{eq:discrete-potential-def}) of
the main text. This expression can further be reformulated in terms
of the edge potential induced by a scalar potential applied to a given
site $j$:
\begin{equation}
A_{e}=\sum_{j}A_{e}^{\left(j\right)},\,\,\,A_{e}^{\left(j\right)}=\begin{cases}
\frac{c}{i\omega}\frac{\phi_{j}}{2}, & \text{end}\left(e\right)=j,\\
-\frac{c}{i\omega}\frac{\phi_{j}}{2}, & \text{beg}\left(e\right)=j,\\
0, & \text{otherwise}.
\end{cases}
\label{eq:edge-potential_corresp-to_scalar-potential}
\end{equation}
Our next step will be to compute the response of the current along
a given edge $e$ to the edge potential of the form~(\ref{eq:edge-potential_corresp-to_scalar-potential})
within the AQBP scheme described in \subsecref{Local-current-response}
of the main text. Together with Eqs.~(\ref{eq:edge-current_response_to-edge-potential})~and~(\ref{eq:real-space-current_via_edges-currents}),
this allows us to calculate the density current $\boldsymbol{j}\left(\boldsymbol{r}\right)$
in the response to the potential field at low frequencies and eventually
derive Eq.~(\ref{eq:macro-response_discrete-problem}) the main text.

\begin{figure*}[t]
\begin{centering}
\includegraphics[scale=0.35]{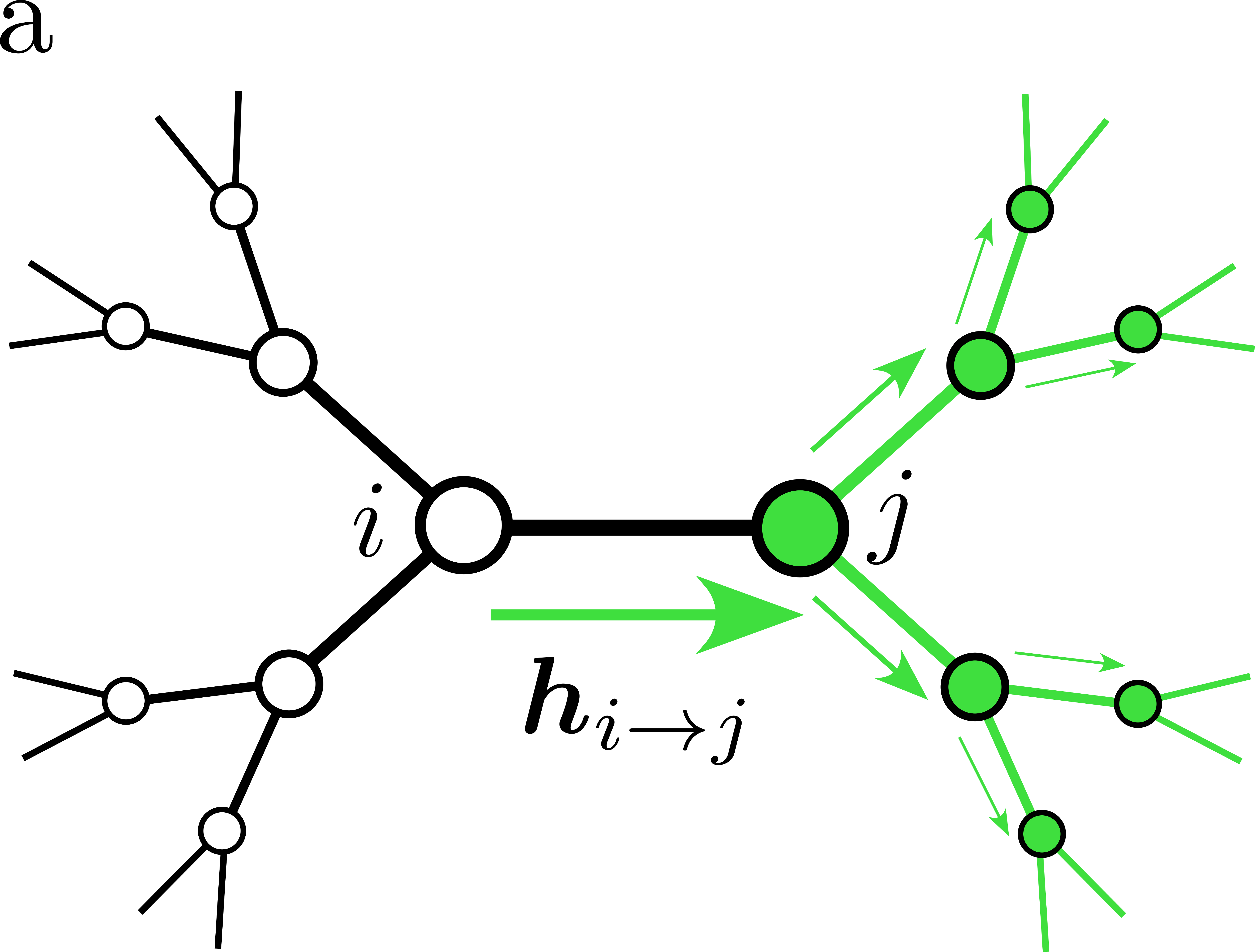}~~~\includegraphics[scale=0.35]{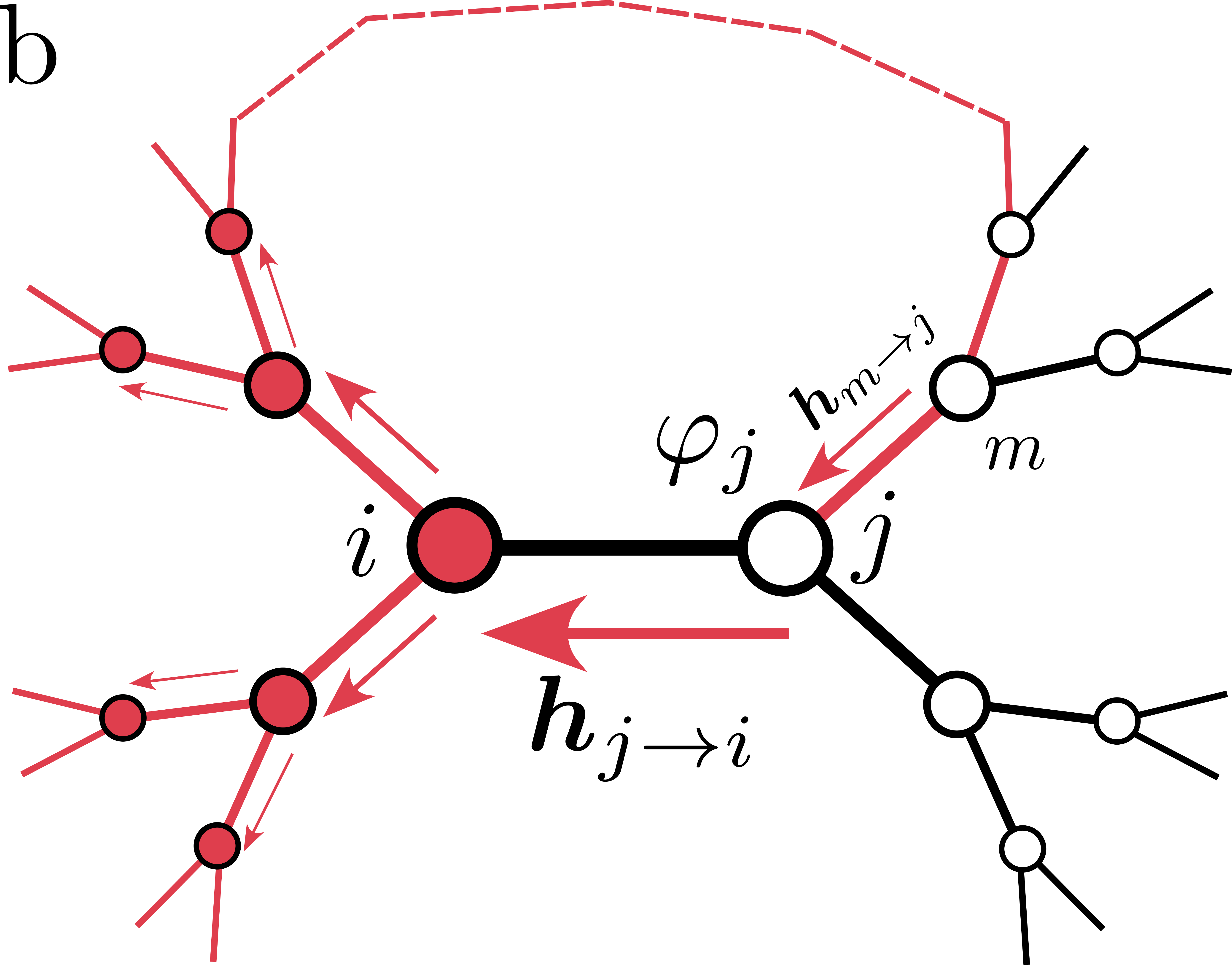}~~~\includegraphics[scale=0.35]{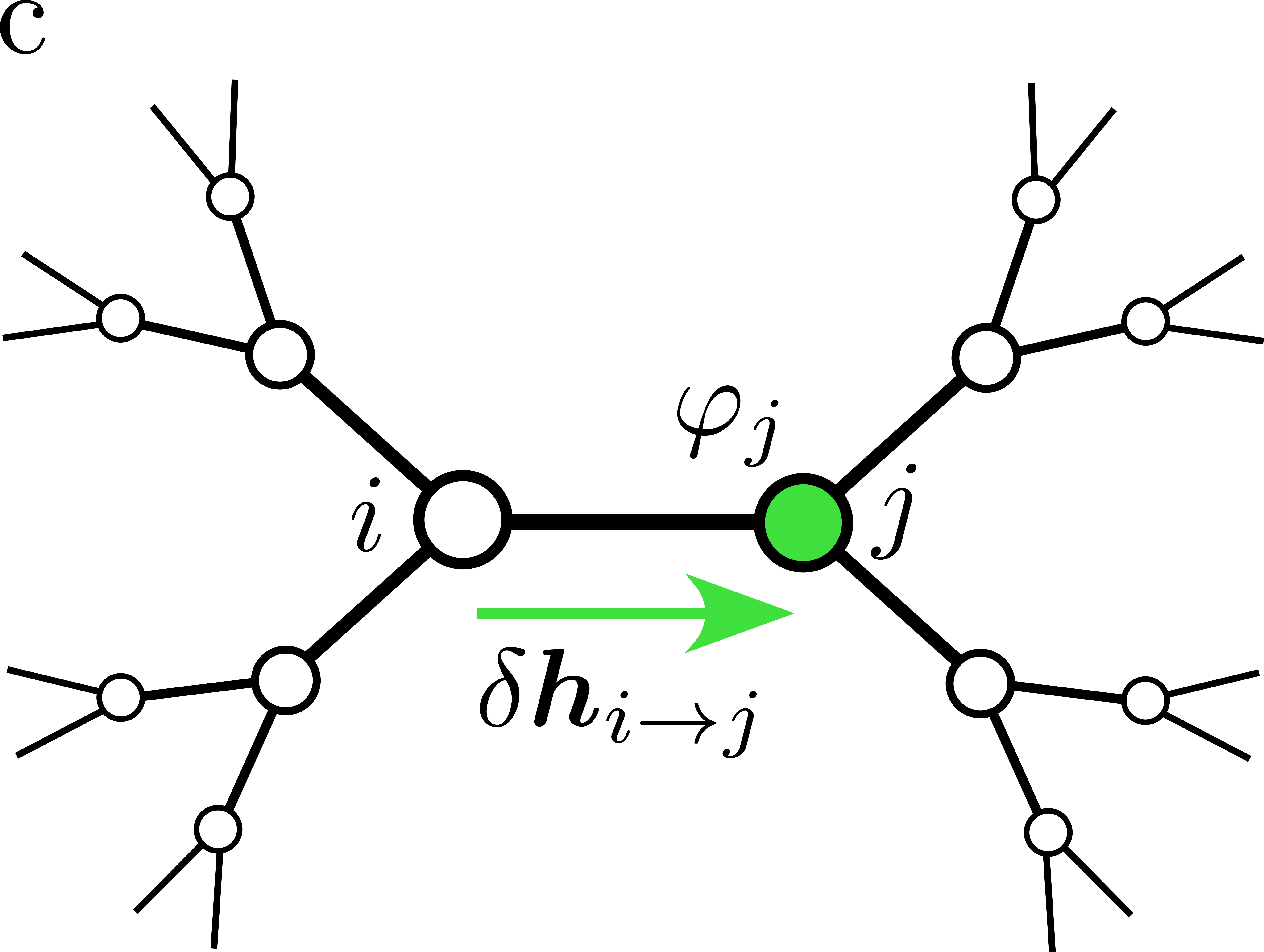}~~~\includegraphics[scale=0.35]{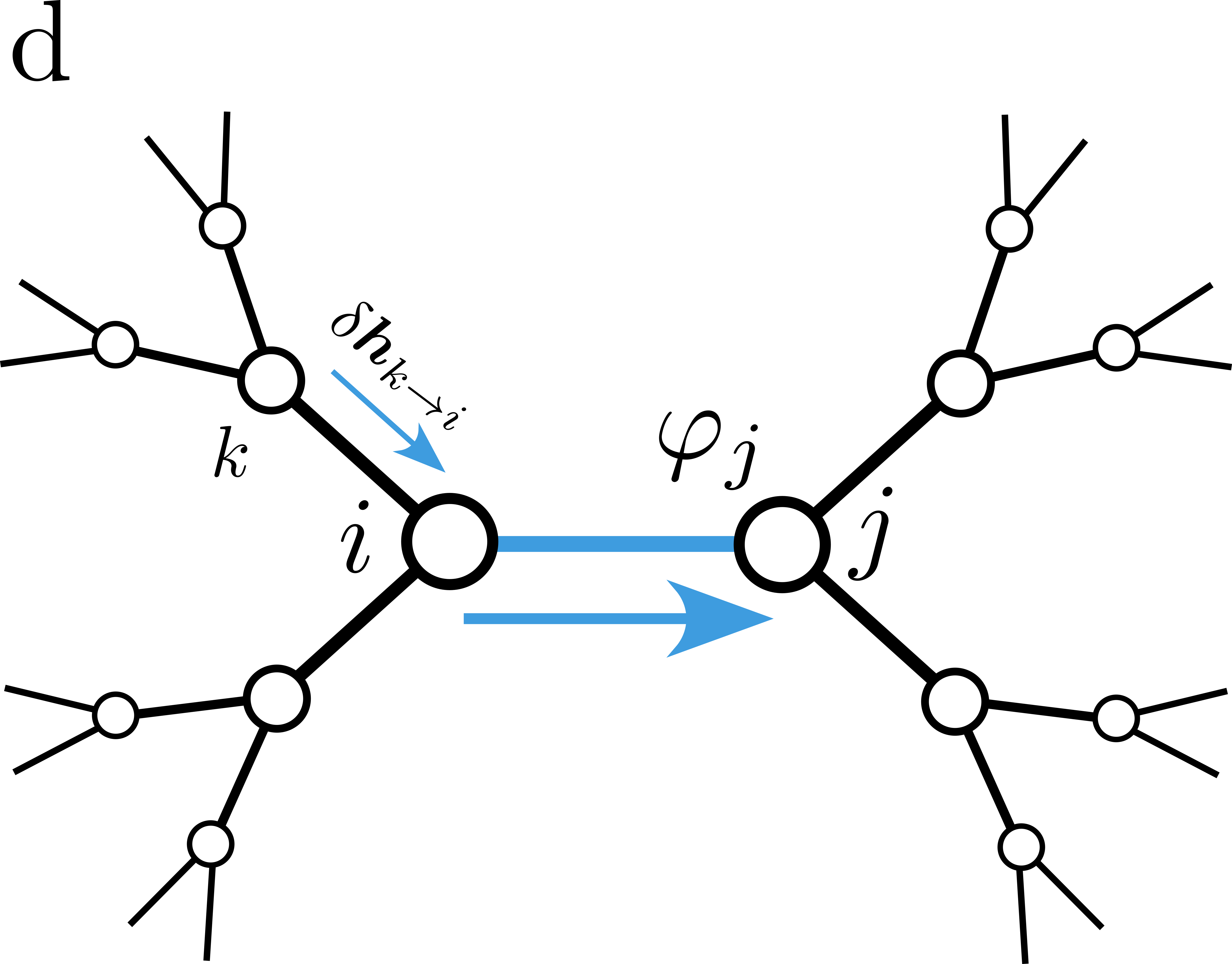}
\par\end{centering}
\caption{Visualization of the structure of dependencies in the self-consistency
equation~(\ref{eq:time-dependent_self-cons-eq}) on a locally tree-like
graph with coordination number $Z=3$. \emph{a)} the value of the
$h$ associated with a given directed edge is influenced by the branch
originating from the end vertex of this directed edge (colored in
green), and only the values of $h$ along the edges oriented in the
same direction contribute to the equation. \emph{b) }Due to the presence
of a large loop (red dashed line), the other branch of the graph also
has some influence over the value of $h$, but according to Eq.~(\ref{eq:time-dependent_self-cons-eq}),
each edge in this loop contributes with an additional factor of $JP\sim1/Z$
to the magnitude of this influence. As a result, the \emph{change}
of $h$ practically vanishes for all edges pointing away from the
source site $j$. \emph{c) }The \emph{change} of $h$ on the edge
ending at the source site $j$ is thus only sensitive to the source
term itself \emph{d) }The \emph{change }of $h$ on the edge at distance
greater than 1 away from the source site $j$ is sensitive only to
the \emph{change} of $h$ on the edge that precedes it on the path
to the source. \protect\label{fig:app_graph-demo}}
\end{figure*}

The first step is to figure out the way to calculate the non-local
response, by which we refer to the case when the edge $e$ in Eq.~(\ref{eq:edge-current_response_to-edge-potential})
does not belong the to the nearest neighborhood of site $j$ from
Eq.~(\ref{eq:edge-potential_corresp-to_scalar-potential}). Within
the AQBP scheme, any change in the operator on edge $e$ by a perturbation
on a different edge $e'\neq\pm e$ is produced by the response of
the order parameter $\delta h_{e}$, so one has
\begin{widetext}
\begin{equation}
Q_{ee'}\left(t,t'\right)=-\intop dt''\,\left[\frac{\delta\left\langle I_{e}\left(t\right)\right\rangle }{\delta\boldsymbol{h}_{e}\left(t''\right)}\cdot\frac{\delta\left\langle \boldsymbol{h}_{e}\left(t''\right)\right\rangle }{\delta A_{e'}\left(t'\right)}+\frac{\delta\left\langle I_{e}\left(t\right)\right\rangle }{\delta\boldsymbol{h}_{-e}\left(t''\right)}\cdot\frac{\delta\left\langle \boldsymbol{h}_{-e}\left(t''\right)\right\rangle }{\delta A_{e'}\left(t'\right)}\right],
\label{eq:current-nonlocal-response}
\end{equation}
\end{widetext}

\noindent where the two terms in the sum correspond to contribution
from the two directions of the target edge. The change of the order
parameter is, in turn, described by the time-dependent generalization
of the self-consistency equation~(\ref{eq:h-equations}) of the main
text:
\begin{widetext}
\begin{equation}
\delta\boldsymbol{h}_{k\rightarrow i}\left(t\right)=\sum_{j\in\partial i\backslash\left\{ k\right\} }J_{ij}\left[\intop dt\hat{P}_{i\rightarrow j}\left(t,t'\right)\cdot\delta\boldsymbol{h}_{i\rightarrow j}\left(t'\right)+\frac{2e}{c}A_{i\rightarrow j}\left(t\right)\left\langle 2S^{x}\right\rangle _{i\rightarrow j}\boldsymbol{e}_{y}\right],
\label{eq:time-dependent_self-cons-eq}
\end{equation}
\end{widetext}

\noindent where $\boldsymbol{e}_{y}$ is the unit vector along $y$
direction, $\left\langle 2S^{x}\right\rangle _{i\rightarrow j}$ is
the static spin average given by Eq.~(\ref{eq:spin-average}) of
the main text, and $P_{i\rightarrow j}^{\alpha\beta}\left(t,t'\right)=\left\langle 2S^{\alpha}\left(t\right)2S^{\beta}\left(t'\right)\right\rangle -\left\langle 2S^{\alpha}\left(t\right)\right\rangle \left\langle 2S^{\beta}\left(t'\right)\right\rangle $
is the polarization operator of a single spin with the Hamiltonian
given by Eq.~(\ref{eq:single-spin_Ham}), whose Fourier transform
reads
\begin{equation}
\hat{P}_{i\rightarrow j}\left(\omega\right)=\frac{\tanh\beta B_{i\rightarrow j}}{B_{i\rightarrow j}\left(B_{i\rightarrow j}^{2}-\omega^{2}/4\right)}\begin{pmatrix}\xi_{j}^{2} & \frac{i\omega\xi_{j}}{2}\\
-\frac{i\omega\xi_{j}}{2} & B_{i\rightarrow j}^{2}
\end{pmatrix},
\label{eq:spin-polarization-op}
\end{equation}
with $B_{i\rightarrow j}=\sqrt{\xi_{j}^{2}+h_{i\rightarrow j}^{2}}$,
and $\omega$ shifted by $+i0$ to restore the retarded structure
of the response. In Eqs.~(\ref{eq:time-dependent_self-cons-eq}-\ref{eq:spin-polarization-op})
we have set the direction of the unperturbed order parameter $h$
along the $x$ axis, analogous to the real order parameter configuration
in conventional superconductors. 

The first important thing to note about Eq.~(\ref{eq:time-dependent_self-cons-eq})
for the edge potential $A_{e}^{\left(j\right)}$ given by Eq.~(\ref{eq:edge-potential_corresp-to_scalar-potential})
is that due to the locally tree-like structure of the graph and directed
nature of this equation, $\delta\boldsymbol{h}_{e}=\boldsymbol{0}$
for all edges that are directed away from the site $j$ where the
potential is applied. For instance, one has $\delta\boldsymbol{h}_{j\rightarrow i}=0$
for any neighboring site $i\in\partial j$, whereas $\delta\boldsymbol{h}_{i\rightarrow j}\neq\boldsymbol{0}$,
as we will see later. This is apparent from the fact that for all
edges $e$ directed away from site $j$, the homogeneous part of Eq.~(\ref{eq:time-dependent_self-cons-eq})
contains only edges $e'$ that originate from $\text{end}\left(e\right)$
and point away from $j$ as well, as explained on \figref{app_graph-demo}.
The same figure also illustrates that the actual value of $\delta\boldsymbol{h}_{j\rightarrow i}$
in the system that contains loops of length at least $l$ can be estimated
as $\delta h_{j\rightarrow i}\sim\left(JP\right)^{l/2}\delta h_{m\rightarrow j}\sim\left(1/Z\right)^{l/2}\delta h_{i\rightarrow j}$
because each additional edge traversed from the source of the perturbation
to the target edge brings in additional power of $J\cdot P\sim\frac{1}{Z}$,
i.e. it is indeed negligible.

Upon application of the edge potential of the form~(\ref{eq:edge-potential_corresp-to_scalar-potential}),
$\delta\boldsymbol{h}_{i\rightarrow j}$ for any $i\in\partial j$
is described only by the source term of Eq.~(\ref{eq:time-dependent_self-cons-eq})
and reads
\begin{equation}
\delta\boldsymbol{h}_{i\rightarrow j}=-\frac{2e}{i\omega}\sum_{s\in\partial j\backslash\left\{ i\right\} }J_{js}\left\langle 2S^{x}\right\rangle _{j\rightarrow s}\,\frac{\phi_{j}}{2}\,\boldsymbol{e}_{y}.
\end{equation}
We then immediately recognize the r.h.s of the static self-consistency
equation~(\ref{eq:h-equations}) for $h_{i\rightarrow j}$, which
leads to
\begin{equation}
\delta\boldsymbol{h}_{i\rightarrow j}=-\frac{2e}{i\omega}h_{i\rightarrow j}\frac{\phi_{j}}{2}\boldsymbol{e}_{y},
\label{eq:local_h-field_response}
\end{equation}
i.e. the time-dependent change $\boldsymbol{h}_{i\rightarrow j}$
of the order parameter for edges pointing at $j$ is transverse to
the direction of the order parameter itself and directly proportional
to the external field $\varphi_{j}$. This result is otherwise expected
from the gauge invariance: one can move the potential from site $j$
to all sites of the system that are contained in branch rooted at
$i$ by a time-depend gauge transform. Because there's no direct response
of $\boldsymbol{h}_{i\rightarrow j}$ to this new field due to the
tree-like structure, as described above, the only thing that remains
is the gauge change of $\boldsymbol{h}_{i\rightarrow j}$, which is
given exactly by Eq.~(\ref{eq:local_h-field_response}). 

Consider now an edge $e=k\rightarrow i$ that is within distance 2
from site $j$, i.e. $\text{end}\left(e\right)\in\partial j$ but
$\text{beg}\left(e\right)\neq j$ (see also \figref{app_graph-demo}).
This time around, both terms in Eq.~(\ref{eq:time-dependent_self-cons-eq})
contribute to the result, but the sum contains only one term corresponding
to edge $i\rightarrow j$ that connects the target edge $e$ to the
source site $j$:
\begin{align}
\delta\boldsymbol{h}_{k\rightarrow i} & =J_{ij}\left[\hat{P}_{i\rightarrow j}\left(\omega\right)\cdot\delta\boldsymbol{h}_{i\rightarrow j}+\frac{2e}{c}A_{i\rightarrow j}\left\langle 2S^{x}\right\rangle _{i\rightarrow j}\boldsymbol{e}_{y}\right]\nonumber \\
 & =\frac{2e}{i\omega}J_{ij}\left[-h_{i\rightarrow j}\hat{P}_{i\rightarrow j}\left(\omega\right)\cdot\boldsymbol{e}_{y}+\left\langle 2S^{x}\right\rangle _{i\rightarrow j}\boldsymbol{e}_{y}\right]\frac{\phi_{j}}{2},
\end{align}
where we have used the Fourier representation of Eq.~(\ref{eq:time-dependent_self-cons-eq}).
Remarkably, in the $\omega\rightarrow0$ limit, we observe from Eq.~(\ref{eq:spin-polarization-op})
that
\begin{equation}
h_{i\rightarrow j}\hat{P}_{i\rightarrow j}\left(0\right)\cdot\boldsymbol{e}_{y}\equiv\left\langle 2S^{x}\right\rangle _{i\rightarrow j}\boldsymbol{e}_{y},
\end{equation}
so at zero frequency we arrive at
\begin{equation}
\delta\boldsymbol{h}_{k\rightarrow i}\left(\omega=0\right)=0.
\end{equation}
Moreover, due to the fact that Eq.~(\ref{eq:time-dependent_self-cons-eq})
contains only the first term for any edge that is not in the nearest
neighborhood of site $j$, we arrive to
\begin{equation}
\delta\boldsymbol{h}_{e}\left(\omega=0\right)=0,
\end{equation}
for any edge $e$ that is not in the nearest neighborhood of site
$j$. This and Eq.~(\ref{eq:current-nonlocal-response}) finally
implies that in response to a potential on site $j$ the current is
induced only along edges that are adjacent to $j$. 

Finally, let's compute the current response directly along the adjacent
edge, i.e $e=i\rightarrow j$ for some $i\in\partial j$. This time
round, the response of $I_{e}$ to the edge potential~(\ref{eq:edge-potential_corresp-to_scalar-potential})
contains both the nonlocal term from the edge potential on adjacent
\emph{edges }as well as the explicit current response to the vector
potential on the same edge:
\begin{widetext}
\begin{equation}
I_{e}=\intop dt'\,\left[-Q_{ee}\left(t,t'\right)\,A_{e}^{\left(j\right)}\left(t'\right)+\intop dt''\,\,\frac{\delta\left\langle I_{e}\left(t\right)\right\rangle }{\delta\boldsymbol{h}_{e}\left(t''\right)}\cdot\sum_{e'\neq-e:\text{beg}\left(e'\right)=j}\frac{\delta\left\langle \boldsymbol{h}_{e}\left(t''\right)\right\rangle }{\delta A_{e'}\left(t'\right)}\,A_{e'}^{\left(j\right)}\left(t'\right)\right].
\label{eq:nonlocal-response_to-potential-field_temp}
\end{equation}
\end{widetext}

We now have to compute the current response to a change of the order
parameter along the same edge. The current itself is described by
the time-dependent version of the two-spin Hamiltonian~(\ref{eq:two-sping_effective_Hamiltonian}),
with one of the two $h$ fields now being the function of time. We
first note that the current response to the change of the longitudinal
component of the $\delta\boldsymbol{h}\propto\boldsymbol{e}_{x}$
field vanishes due to the symmetry of the two-spin Hamiltonian. Indeed,
consider the transformation $S^{x}\mapsto-S^{x}$, $S^{y}\mapsto-S^{y}$
followed by $h^{x}\left(t\right)\mapsto-h^{x}\left(t\right)$, with
both transformations applied to both sites. In the absence of the
transverse field $h^{y}$, this transformation leaves both the two-spin
Hamiltonian and the current intact, so we conclude that
\begin{equation}
\frac{\delta\left\langle I_{e}\left(t\right)\right\rangle }{\delta h_{e}^{x}\left(t''\right)}\equiv-\frac{\delta\left\langle I_{e}\left(t\right)\right\rangle }{\delta h_{e}^{x}\left(t''\right)}=0.
\end{equation}
The response to the transverse component $\delta\boldsymbol{h}\propto\boldsymbol{e}_{y}$
can be calculated by performing a gauge transform on the corresponding
spin, which delivers the following identity:
\begin{equation}
\frac{\delta\left\langle I_{e}\left(t\right)\right\rangle }{\delta\boldsymbol{h}_{e}\left(t''\right)}=\frac{1}{\left(2e\right)/c}\frac{Q_{ee}\left(t,t'\right)}{h_{e}}\boldsymbol{e}_{y}^{T},
\label{eq:current-response_to_order-parameter}
\end{equation}
This and Eq.~(\ref{eq:nonlocal-response_to-potential-field_temp})
the renders the following identity for the value of the response to
the potential field at the adjacent edge:
\begin{equation}
I_{e}=-\frac{1}{i\omega}\,cQ_{ee}\left(\omega\right)\,\phi_{j}.
\label{eq:current-local-response}
\end{equation}

According to Eqs.~(\ref{eq:edge-current_response_to-edge-potential})-(\ref{eq:edge-potential_corresp-to_scalar-potential}),
to compute the current response from arbitrary configuration of the
scalar potential, one should sum the responses over all sites $j$,
while taking into account the anti-symmetry property $I_{j\rightarrow i}=-I_{i\rightarrow j}$.
This finally renders for low frequencies
\begin{equation}
I_{i\rightarrow j}=-\frac{1}{i\omega}\,cQ_{ee}\left(0\right)\,\left(\phi_{j}-\phi_{i}\right),
\label{eq:current-local-linear-response}
\end{equation}
where we neglected the frequency dependence of the $Q$ kernel. As
a result, the current response to a potential field at zero frequency
is present only along the edges adjacent to the site were the potential
is applied, and the magnitude of this response coincides with the
one calculated from the direct effect of the applied potential on
this adjacent edge. Together with Eq.~(\ref{eq:real-space-current_via_edges-currents})
this yields Eq.~(\ref{eq:macro-response_discrete-problem}) of the
main text.

We finally note that the result~(\ref{eq:current-local-linear-response})
is only valid in the linear response regime. Based on its simplicity,
it is tempting to replace $\phi=-\frac{i\omega}{2e}\varphi_{i}$ and
generalize the resulting current-phase relation to the nonlinear response
in a simple fashion akin to the Josephson's relation, viz., $I_{i\rightarrow j}=I_{0}\,\sin\varphi_{j}-\varphi_{i}$,
and thus describe the macroscopic \emph{nonlinear} response in strongly
disordered superconductors. However, this would constitute an unjustified
simplification. Indeed, the AQBP scheme beyond the linear response
essentially involves two steps: \emph{i)}~calculating the nonlinear
current response of the local two-spin problem, which would be the
generalization of the first term in Eq.~(\ref{eq:nonlocal-response_to-potential-field_temp}),
and \emph{ii)}~computing the effect of the external field on the
order parameter field~$h_{i\rightarrow j}$, thus generalizing the
second term in Eq.~(\ref{eq:nonlocal-response_to-potential-field_temp}).
One then has serious grounds to expect that the simple Josephson-like
relation will produce quantitatively incorrect results. First of all,
step~\emph{i)} already prohibits a sinusoidal relation, as the current
response of a two-spin Hamiltonian to external phase difference deviates
from a sinusoidal relation, as can be checked by direct diagonalization.
Secondly, within step\emph{~ii)}, the effect of the external potential
on the order parameter is not additive across different sites, and
it also might additionally becomes nonlocal even in the limit $\omega\rightarrow0$.
Given that the problem features strong disorder (which by itself produces
another step of complication even in the linear response, as discussed
in \subsecref{Macroscopic-superfluid-stiffness}), one would then
have to solve a nonlinear system of nonlocal equations describing
the superconducting phase distribution, so the result is even more
unlikely to be reproduced by a system of local Josephson's relations.
This complicated question thus remains for future study.

\section{Verification of Dykhne's law for a discrete system on a locally tree-like
graph\protect\label{app:Verification-of-Dykhne-law}}

The set of numerical experiments presented below demonstrates the
approximate validity of the relation~(\ref{eq:macro-rho-S_via_typical-current-response})
in the following sense: the macroscopic response $\rho_{S}$ and the
typical local response $Q_{\text{typ}}$ obey this relation as disorder
strength $\kappa$ is varied within a fixed topology of the graph.

\subsubsection{Generating the interaction graph.}

For a given length $L$ and width $w$, $N\gg1$ points were spread
universally over a rectangle of size $L\times w$, with $L\gg w$.
For each point, $Z$ neighbors were selected in the neighborhood of
radius $\xi_{\text{loc}}\ll L,w$ (periodic boundary conditions along
the $w$ directions where used). To ensure the locally tree-like structure,
the condition $Z\ll n\xi_{\text{loc}}^{D}$ was also ensured (here
$D=2$ is the dimensionality of the problem, and $n=N/Lw$ is the
concentration). In case the condition of exactly $Z$ neighbors could
not be satisfied for all $N$ sites, the number of such sites was
controlled to be negligible (to maintain the average number of neighbors
close enough to $Z$). In this way, the instance of a locally-tree
like graph described in \subsecref{Model-Hamiltonian} of the main
text was obtained.

\begin{figure}
\begin{centering}
\includegraphics[scale=0.3]{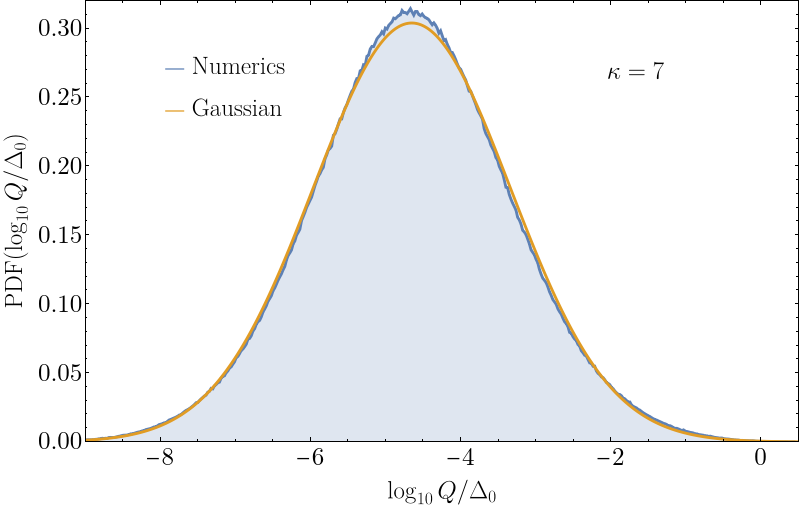}
\par\end{centering}
\caption{The distribution of the decimal logarithm of the current response
functions, as found from the Eq.~(\ref{eq:discrete-current-response_QBP})
of the main text. The parameters of the model are $T=0$, $\kappa=7,$
$\lambda=0.165$, $Z=6$. The orange curve corresponds to the Gaussian
distribution with the parameters inferred from actual distribution:
$\left\langle \log_{10}Q\right\rangle =-4.66$ and $\sqrt{\left\langle \left(\log_{10}Q\right)^{2}\right\rangle -\left\langle \log_{10}Q\right\rangle ^{2}}=1.31$.
\protect\label{fig:app_q-distribution-demo}}
\end{figure}

\subsubsection{Determining the macroscopic response in a given disorder realization.}

The next step was to determine the configuration of the order parameter
according to Eq.~(\ref{eq:h-equations}) and calculate the local
edge responses according to Eq.~(\ref{eq:discrete-current-response_QBP}).
Note that this procedure keeps track of the configuration of the $h$
fields and $Q$ values \emph{in a given disorder realization}, in
contrast to the MPD described in \subsecref{Statistical-properties-of-the-current-response}
of the main text that addresses only the distribution of those quantities.
\figref{app_q-distribution-demo} shows the distribution of the local
responses $Q_{ij}$ in a typical disorder realization.

Eqs.~(\ref{eq:macro-response_discrete-problem}) (discarding the
irrelevant $-c/i\omega$ factor in front of the current) were then
solved numerically with the following boundary conditions: $\varphi_{i}=0$
for a strip of size $d\times w$ with $n^{-1/D}\ll d\ll L$ near one
of the system's boundaries, and $\varphi_{i}=\varphi_{0}$ for a strip
of the same size situated at the other boundary, thus arranging the
geometry of the two-contact measurement. The resulting distribution
of the potential was used to determine the distribution of current
$I_{i\rightarrow j}$. The total current $I_{\text{total}}$ was then
calculated as a sum of the currents along all edges that connected
a site inside the strip of fixed potential with a site outside of
this strip. The resulting value of the superfluid density $\rho_{S}=\varphi_{0}w/I_{\text{total}}L$
was then recorded for a given disorder realization.

\begin{figure}[t]
\begin{centering}
\includegraphics[scale=0.3]{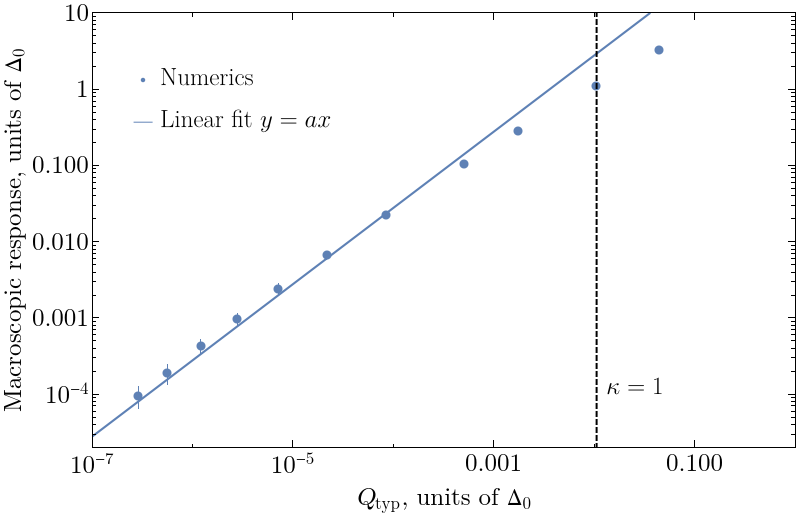}
\par\end{centering}
\caption{Set of points corresponding to pairs $\left(Q_{\text{typ}},\rho_{S}\right)$
for various values of the coupling constant $\lambda$, while maintaining
$Z=6$ and $T=0$. The corresponding values of $\kappa$ are (from
left to right): 17, 15, 13, 11, 9, 7, 5, 3, 2, 1, and 0.5. The parameters
of the numerical experiment of \emph{Steps a.-b}.\emph{ }are $L=20$,
$w=2$, $\xi_{\text{loc}}=0.05$, and $N=960000$. The inhomogeneity
scale mentioned in \emph{Step c.} is estimated as $w_{0}\sim0.8$.
For these values of the concentration $n=N/Lw$, $Z$ and $\xi_{\text{loc}}$,
the approximate depth of the tree like structure is $m\sim6$, i.e.,
the neighborhood of each site of depth $\le6$ is nearly tree-like,
while for larger depths numerous loops start to emerge, indicating
that the graph is embedded in the 2D space. \protect\label{fig:dykhne-law_verification}}
\end{figure}

\subsubsection{Verification of macroscopic limit.}

To ensure that system size $L$ was sufficient, the procedure described
in \emph{Step b.} was repeated for a set of systems with lengths $L'<L$
obtained by truncating the original system, while still controlling
the relation $L'\gg w$. The resulting function $\rho_{S}\left(L'\right)$
was then fitted by the dependence $A+B/\left(L'-2d\right)$, with
the values of the parameters $A,B$ used to ensure good convergence
of the estimation $\rho_{S}\left(L\right)$ to the macroscopic quantity.
In particular, the length scale $w_{0}=B/A$ was interpreted as the
typical scale of the inhomogeneity of the solution, so the condition
$w>2w_{0}$ has been verified as well. 

\subsubsection{Averaging over disorder realizations.}

\emph{Steps a. and b.} described above were repeated for 10 disorder
realizations. The average value of $\rho_{S}$ and its statistical
error represent a single point on \figref{dykhne-law_verification}.
Other points where obtained by changing the value of the coupling
constant $\lambda$ (while preserving the structure of the graph)
and performing \emph{Step b}.

\subsubsection{The effect of correlations.}

The distribution of $Q_{ij}$ possesses short-range correlations that
might influence the resulting macroscopic response. To estimate this
influence, \emph{Step b. }was repeated for a system where the values
of the local responses $Q_{ij}$ had been randomly reshuffled, so
the short-range correlations in the values of $Q_{ij}$ had been destroyed
while preserving the distribution of $Q_{ij}$ itself. The resulting
macroscopic response $\rho_{S}^{\text{shuffled}}$ turned out to be
smaller than the correct one $\rho_{S}$ by a disorder-dependent factor
of the order of unity: for $\kappa=7$ we obtained $\rho_{S}/\rho_{S}^{\text{shuffled}}\sim4.5$,
while $\kappa=0.5$ rendered $\rho_{S}/\rho_{S}^{\text{shuffled}}\sim1.1$.

\subsubsection{Conclusion.}

The numerical experiment shown above is a satisfactory demonstration
of the following statement: the Dykhne's law~(\ref{eq:macro-rho-S_via_typical-current-response})
is approximately correct as a function of dimensionless disorder strength
$\kappa$ in 2D systems at zero temperature for a certain locally
tree-like graph. Because neither the graph's topology nor the qualitative
shape of the $Q$ distribution change at low but finite temperatures,
we thus also expect this dependence to correctly describe the connection
between $\rho_{S}$ and $Q_{\text{typ}}$ as a function of temperature.
It is not clear whether the result holds for the 3D problem, which
is relevant for the actual experimental setup, as the thickness~$d$
of the films used in the experiment (see \secref{Experiment}) is
much larger than, e.g., the localization length corresponding to the
$\xi_{\text{loc}}$ value at \emph{Step a}.

\section{Activation-law representation of the numerical and analytical data\protect\label{app:Other-plots-for-the-data}}

The reported dependence of the superfluid density on temperature can
be interpreted in multiple ways depending on the presentation. \figref{typical-Q_numerical-data}
of the main text plots the dependence $\delta Q_{\text{typ}}/Q_{\text{typ}}$
in log-log scale, echoing the ``power law'' point of view on the
experimental data in \secref{Experiment}. The theoretical description
challenges this picture by showing that the apparent power law actually
depends on temperature and makes sense only in a certain temperature
interval (see \figref{perturbative-estimate-plot}). In this Appendix,
we also present the theoretical data of \secref{Results} in the form
that illustrates the degree of applicability of the activation-law
fit $\delta Q_{\text{typ}}\left(T\right)\propto\exp\left\{ -T_{1}/T\right\} $.

\begin{figure}[h]
\begin{centering}
\includegraphics[scale=0.3]{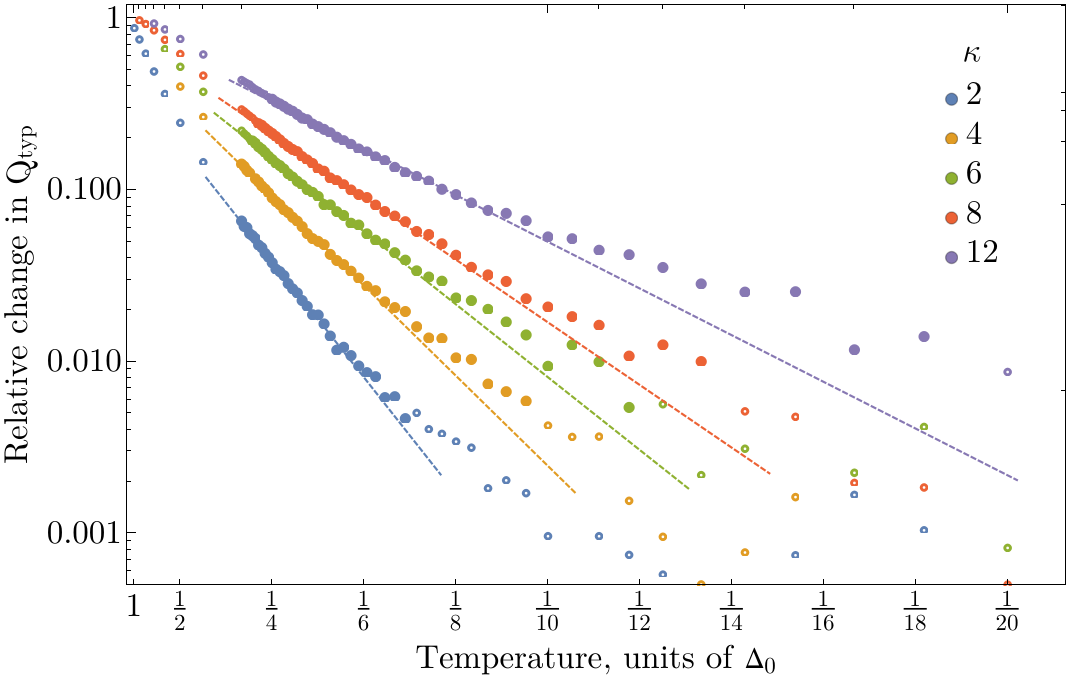}
\par\end{centering}
\caption{The temperature dependence of the relative change of typical local
current response $1-Q_{\text{typ}}\left(T\right)/Q_{\text{typ}}\left(T=0\right)$
in logarithmic scale along the vertical axis and reciprocal ($1/T$)
scale along the horizontal axis. Various colors correspond to various
dimensionless disorder strength~$\kappa$ and $K=15$. The temperature
is measured in units of $\Delta_{0}$. The points with solid filling
are selected according to the criteria $T\in\left[\lambda\left\langle \Delta\right\rangle ,\lambda\left\langle \Delta\right\rangle +0.2\Delta_{0}\right]$
with $\left\langle \Delta\right\rangle $ being the mean order parameter
at zero temperature. These highlighted points are used for exponential
fitting $A\exp\left\{ -T_{1}/T\right\} $, with the latter plotted
with dashed lines. \protect\label{fig:typical-Q_numerical-data_reciprocal-scale}}
\end{figure}

\begin{figure}[h]
\begin{centering}
\includegraphics[scale=0.3]{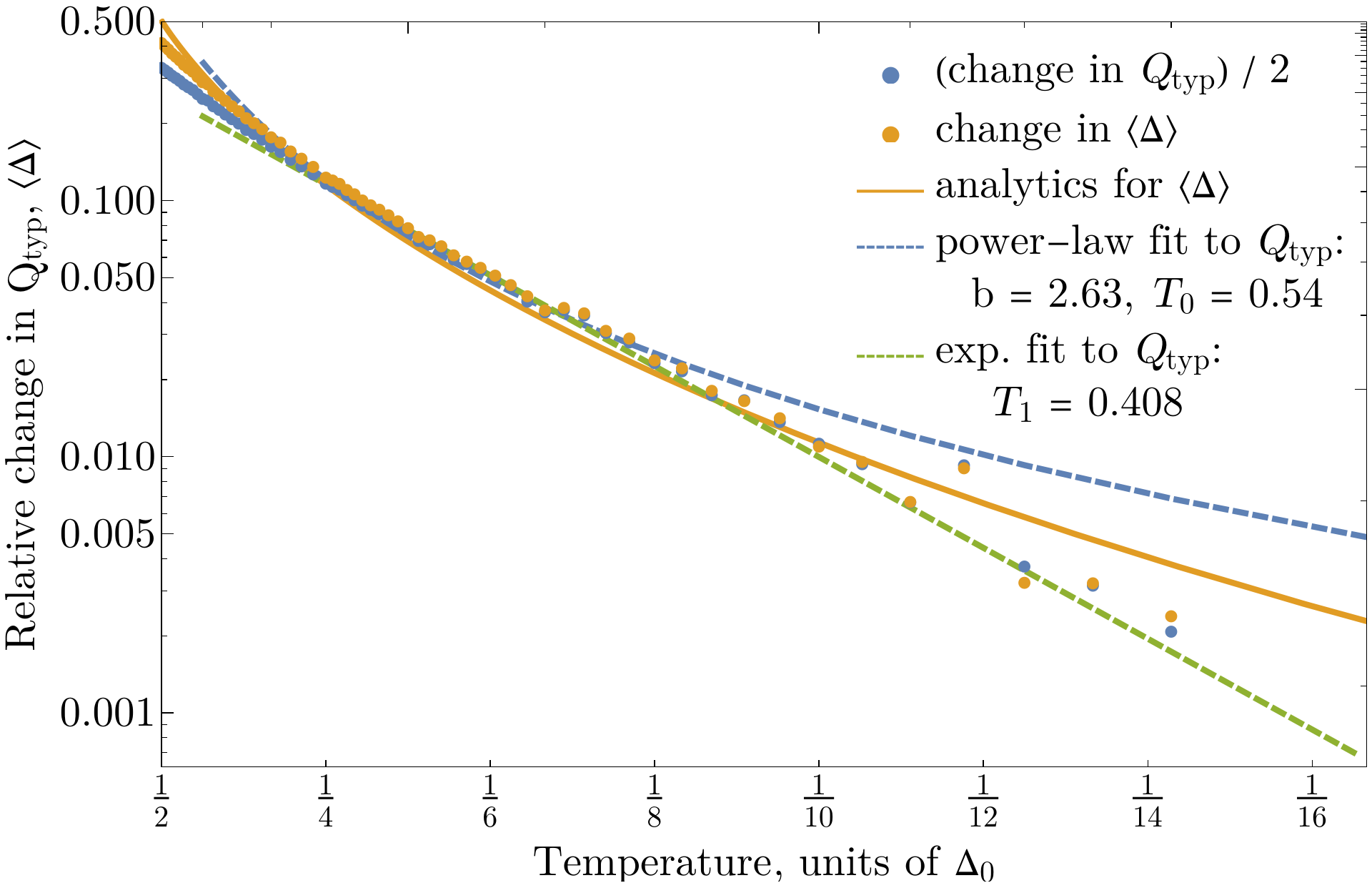}
\par\end{centering}
\caption{Low-temperature dependence of the relative changes in the typical
local current response $1-Q_{\text{typ}}\left(T\right)/Q_{\text{typ}}\left(T=0\right)$
and the mean order parameter $1-\left\langle \Delta\right\rangle \left(T\right)/\left\langle \Delta\right\rangle \left(T=0\right)$
according to the numerical MPD (solid dots) and theoretical description
given by Eqs.~(\protect\hphantom{}\ref{eq:A-correction}\protect\nobreakdash-\ref{eq:mean-h-equation_correction}\protect\hphantom{})
(solid line). The values of $Q_{\text{typ}}$ are divided by two,
according to relation~(\ref{eq:delta-q_to_delta-op}). The dashed
lines correspond to the power-law $\left(T/T_{0}\right)^{b}$ (blue)
and activation-law $A\exp\left\{ -T_{1}/T\right\} $ (green) fits
and of numerical data for $Q_{\text{typ}}$ in the range $T\in\left[\lambda\left\langle \Delta\right\rangle ,\lambda\left\langle \Delta\right\rangle +0.2\Delta_{0}\right]$,
with $\left\langle \Delta\right\rangle =0.49\,\Delta_{0}$. The parameters
of the model are $K=20,\,\kappa=10$. \protect\label{fig:mean-order-parameter_numerics-vs-theory_reciprocal-scale}}
\end{figure}

We start by plotting on \figref{typical-Q_numerical-data_reciprocal-scale}
the numerical data of \figref{typical-Q_numerical-data} in logarithmic
scale along the vertical axis and reciprocal scale $1/T$ along the
horizontal axis. Similarly to the data processing of the main text,
we then select a subset of points to apply the activation-law fit
$\delta Q_{\text{typ}}/Q_{\text{typ}}=A\exp\left\{ -T_{1}/T\right\} $
and plot the resulting curve. On \figref{mean-order-parameter_numerics-vs-theory_reciprocal-scale}
we also plot the data of \figref{mean-order-parameter_numerics-vs-theory}
(both experimental and theoretical) with reciprocal scale of temperatures.
The plot also features the activation-law fit of the numerical data.
The latter appears to describe the data fairly well in a certain range
of temperatures, similarly to what is seen on \figref{typical-Q_numerical-data}
in the main text. 

\begin{figure}[h]
\begin{centering}
\includegraphics[scale=0.3]{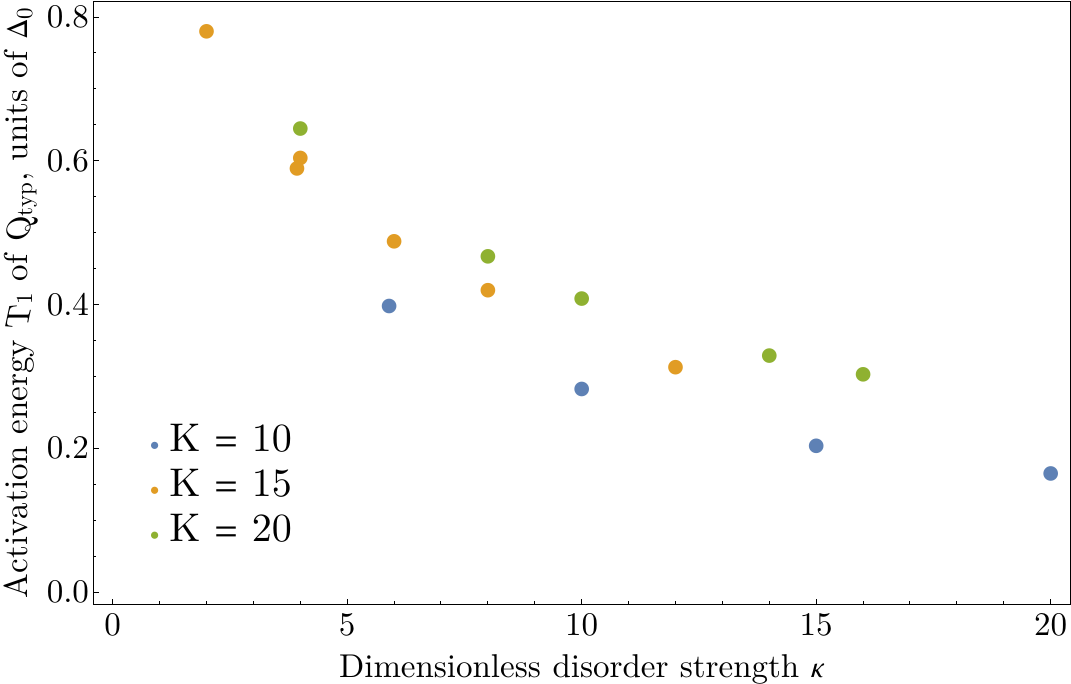}
\par\end{centering}
\caption{Dependence of the $T_{1}$ parameter of the activation-law fit $A\exp\left\{ -T_{1}/T\right\} $
of the portion of numerical data for $\delta Q_{\text{typ}}\left(T\right)/Q_{\text{typ}}\left(0\right)$.
For each set of parameters, the points for fitting are selected according
to the criteria $T\in\left[\lambda\left\langle \Delta\right\rangle ,\lambda\left\langle \Delta\right\rangle +0.2\Delta_{0}\right]$,
with $\left\langle \Delta\right\rangle $ being the mean order parameter
at zero temperature, so the dataset for $K=15$ corresponds to highlighted
points on \figref{typical-Q_numerical-data_reciprocal-scale}. The
value of $\kappa$ was varied by changing the value of $\lambda$
while keeping $K$ constant, and various colors correspond to various
values of $K$. \protect\label{fig:typical-Q_T0-exponent_disorder-dependence}}
\end{figure}

\begin{figure}[h]
\begin{centering}
\includegraphics[scale=0.3]{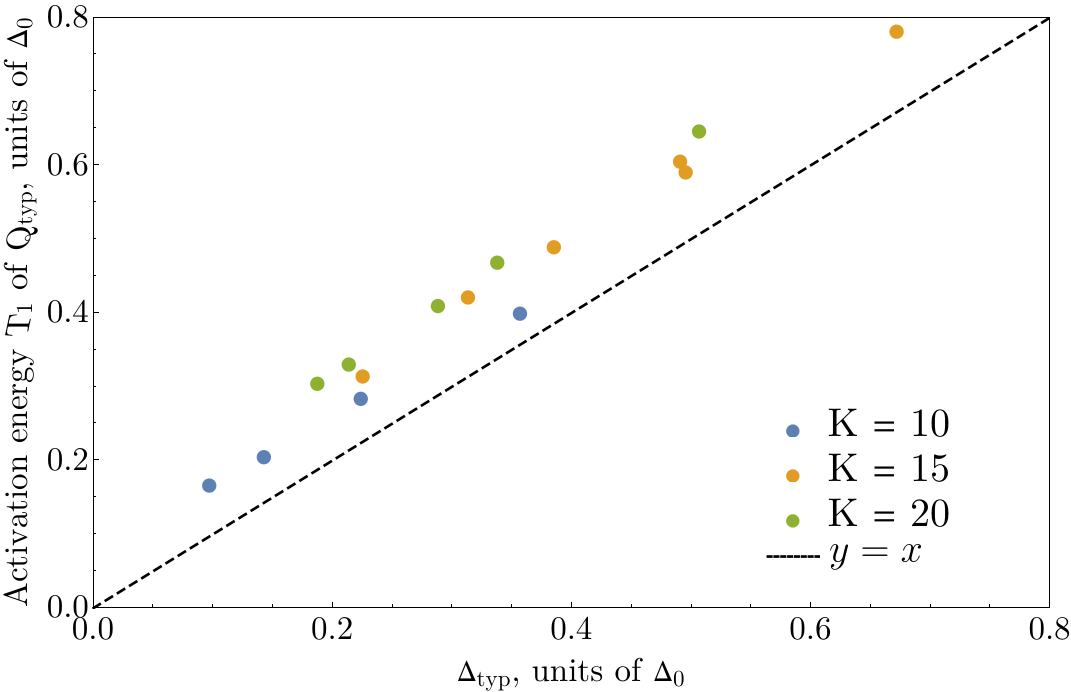}
\par\end{centering}
\caption{Comparison of the $T_{1}$ scale of the activation-law fit $A\exp\left\{ -T_{1}/T\right\} $
from \figref{typical-Q_T0-exponent_disorder-dependence} with the
typical order parameter $\Delta_{\text{typ}}=\exp\left\{ \left\langle \ln\Delta\right\rangle \right\} $
at zero temperature for the same set of parameters $\kappa,K$ and
$T=0$. The data points are the same as those presented on \figref{typical-Q_T0-exponent_disorder-dependence}.
The value of $\Delta_{\text{typ}}$ decreases with the increase of
$\kappa$, thus giving the measure of disorder. The low-disorder case
of \figref{typical-Q_small-disorder} corresponds to $\left(\Delta_{\text{typ}}/\Delta_{0},T_{1}/\Delta_{0}\right)\approx\left(1,2\right)$
and thus does not fall into the apparent linear tendency. \protect\label{fig:activation-T1_vs_typical-delta}}
\end{figure}

The corresponding value of $T_{1}$ for various parameters is shown
on \figref{typical-Q_T0-exponent_disorder-dependence}. As $\kappa$
decreases, the value of $T_{1}$ increases and eventually approaches
$2\Delta_{0}$ at small $\kappa$, as evident from \figref{typical-Q_small-disorder}
in the main text. At large $\kappa$, the value of $T_{1}$ can \emph{roughly}
be described by the typical order parameter $\Delta_{\text{typ}}=\exp\left\{ \left\langle \ln\Delta\right\rangle \right\} $
at zero temperature, as suggested by \figref{activation-T1_vs_typical-delta}.
Note, however, that such a relation is not directly supported by the
analytical approach of \subsecref{Analytical-approach}. Indeed, both
$\delta Q_{\text{typ}}$ and $\Delta_{\text{typ}}$ are complicated
functions of the parameters, and the temperature dependence of $\delta Q_{\text{typ}}$is
not, in general, described by an activation law, so the relation $T_{1}\sim\Delta_{\text{typ}}$
is only empirical.

\section{Restoring the superfluid stiffness $\Theta$ from the limit of small
disorder \protect\label{app:superfluid-stiffness_small-disorder}}

The order of magnitude of the $C$ constant in Eq.~(\ref{eq:macro-rho-S_via_typical-current-response})
can be estimated by examining the limit of weak disorder $\kappa\ll1$
corresponding to large $K$. We first note that $Q$ scales as $1/K$,
as can be seen e.g. from Eq.~(\ref{eq:local-response-estimation}),
so in order for $\Theta$ to remain finite as $K\rightarrow\infty$,
the $C$ value should contain a factor of $K$. Moreover, in the limit
of weak disorder the value of $Q_{\text{typ}}$ in Eq.~(\ref{eq:macro-rho-S_via_typical-current-response})
should be replaced by a simple average~$\left\langle Q\right\rangle $.
Indeed, in the limit of weak disorder one expects $\Theta$ to depend
only on a local self-averaging quantity, and a natural choice for
the latter with the correct dependence on $Q$ and $K$ is the sum
of $Q$s over all neighbors of a given site. 

The value of $\left\langle Q\right\rangle $ can then be evaluated
directly, as it gains its value on configurations with $\xi_{1},\xi_{2}\gg\Delta$,
so Eq.~(\ref{eq:local-response-estimation}) is directly applicable
with $E_{i}=\left|\xi_{i}\right|$. By cutting the logarithmic integrals
over $\xi$ at $\Delta_{0}$ from below and $E_{F}$ from above one
obtains
\begin{equation}
K\left\langle Q\right\rangle =\frac{\Delta_{0}\,\left(2e\right)^{2}}{c}\,\frac{2\nu_{0}\Delta_{0}}{\lambda n},
\end{equation}
where $n$ is the concentration of sites. This results in the following
expression for the superfluid density
\begin{equation}
\rho_{S}=\frac{C}{K}\,Kc\left\langle Q\right\rangle =\frac{C}{K}\,\Delta_{0}\,\left(2e\right)^{2}\,\frac{2\nu_{0}\Delta_{0}}{\lambda n}.
\end{equation}
On the other hand, Ref.~\citep{feigelman_superfluid} provides an
estimation of $\rho_{S}$ in a similar model:
\begin{equation}
\rho_{S}=\frac{2\nu_{0}e^{2}R_{0}^{2}}{\hbar}\left\langle \Delta\right\rangle ^{2},
\end{equation}
where $\left\langle \Delta\right\rangle \approx\Delta_{0}$ is the
mean order parameter, $R_{0}\sim L_{0}\ln\frac{\delta_{\text{loc}}}{\Delta_{0}}\sim L_{0}/\lambda$,
with $L_{0}$ being the interaction radius from \subsecref{Model-Hamiltonian},
and $\delta_{\text{loc}}=\left(\nu_{0}L_{0}^{3}\right)^{-1}$ being
the level spacing in the localization volume. Comparing the two expressions,
one arrives at
\begin{equation}
C\sim nL_{0}^{2}dK.
\end{equation}

\section{The cross-Kerr effect in moderately disordered superconductors for
a strip geometry\protect\label{app:Kerr-effect}}

In this Appendix, we address the cross-Kerr coefficients for the plasmonic
excitations in a moderately disordered superconductor (obeying semiclassical
description). The latter is to be contrasted with the case of strongly
disordered superconductors with a pseudogap discussed in Sections~\ref{sec:Model-and-Theory}~and~\ref{sec:Results}
of the main text. We consider a strip of horizontal sizes $L\times w$
and thickness $d$ with the following hierarchy of scales $L\gg\lambda\gg w\gg d\gg\xi_{0}\gg l$,
with $\lambda$ being the wavelength of the 1D plasmon in question,
$\xi_{0}$ being the zero-temperature superconducting coherence length
in the dirty limit, and $l$ being the mean free path. In particular,
we assume both the charge and current density to be uniform across
the thickness of the film. We also neglect the light retardation,
i.e. the plasmon frequencies satisfies $\omega\ll c/\lambda$. For
simplicity, we restrict our derivation to the case~$T=0$.

The main source of non-harmonicity of the plasmonic modes is the nonlinearity
of the current density response~$\boldsymbol{j}$ to the vector potential
$\boldsymbol{A}$. In the Coulomb gauge $\text{div}A=0$, the latter
can be written as~\citep{Maki1964}
\begin{equation}
j\left(\boldsymbol{A}\right)=-\frac{\pi\sigma\Delta}{\hbar c}\left(1-\frac{4}{3\pi}\xi^{2}\left(\frac{2e}{\hbar c}\boldsymbol{A}\right)^{2}\right)\boldsymbol{A},
\label{eq:app-Kerr_current-density}
\end{equation}
where $\sigma$ is the normal-state conductance, and $\xi=\sqrt{\hbar D/2\Delta}$
is the zero-temperature coherence length in terms of the diffusion
constant $D$. The order parameter $\Delta$ also experiences a correction:
\begin{equation}
\Delta=\Delta_{0}\exp\left\{ -\frac{\pi}{4}\xi^{2}\left(\frac{2e}{\hbar c}\boldsymbol{A}\right)^{2}\right\} ,
\label{eq:app-Kerr_order-parameter}
\end{equation}
with $\Delta_{0}$ being the order parameter at $\boldsymbol{A}=0$
and $T=0$. The density of the kinetic energy is then obtained by
integrating the current density~(\ref{eq:app-Kerr_current-density})
w.r.t the vector potential:
\begin{align}
f_{\text{kin}}\left(\boldsymbol{A}\right) & =-\frac{1}{c}\,\intop_{0}^{A}\left(\boldsymbol{j},d\boldsymbol{A}\right)\nonumber \\
 & =\frac{\rho_{S}\boldsymbol{A}^{2}}{2c^{2}}\left(1-\frac{\alpha\xi^{2}}{2}\left(\frac{2e}{\hbar c}\boldsymbol{A}\right)^{2}+O\left(A^{4}\right)\right),
\label{eq:app-Kerr_kinetic-energy-density}
\end{align}
where in the last line we have used the standard relation $\rho_{S}=\pi\sigma\Delta_{0}$
for the dirty superconductor and expanded the result up to the leading
powers of $\boldsymbol{A}^{2}$, yielding $\alpha=\frac{4}{3\pi}+\frac{\pi}{4}$,
as a result of combining Eqs.~(\ref{eq:app-Kerr_current-density})~and~(\ref{eq:app-Kerr_order-parameter}).
For the case of a thin film, the corresponding free energy reads
\begin{equation}
F_{\text{kin}}=\intop_{\text{volume}}d^{3}\boldsymbol{r}\,f_{\text{kin}}=d\,\intop_{\text{film}}d^{2}\boldsymbol{r}\,f_{\text{kin}},
\end{equation}
where in the second line we assumed that $\boldsymbol{A}$ is uniform
across the film's thickness, and the integrations in the two expressions
go over the film's volume and surface, respectively.

The potential part of the plasmonic Hamiltonian is described by the
electrostatic energy for a given charge density:
\begin{equation}
F_{\text{pot}}=\frac{1}{2}\iintop_{\text{film}}d^{2}\boldsymbol{r}d^{2}\boldsymbol{r}'\,\,\rho\left(\boldsymbol{r}\right)V\left(\boldsymbol{r}-\boldsymbol{r}'\right)\rho\left(\boldsymbol{r}'\right),
\end{equation}
where the integration goes over the surface of the film, $\rho$ is
the 2D charge density, and $V\left(\boldsymbol{r}\right)$ is the
electrostatic potential from a point charge (including the effects
of screening from the ground plate and the substrate if present).
The corresponding electrostatic potential reads~\citep{Krupko2018}:
\begin{align}
V\left(x,y\right) & =\frac{2}{1+\varepsilon}\frac{1}{\sqrt{x^{2}+y^{2}}}\nonumber \\
 & -\frac{2}{1+\varepsilon}\,\frac{2\varepsilon}{1+\varepsilon}\sum_{j=1}^{\infty}\frac{\left(\frac{1-\varepsilon}{1+\varepsilon}\right)^{j-1}}{\sqrt{x^{2}+y^{2}+\left(2hj\right)^{2}}}
\label{eq:app-Kerr_stirp-electristatic-potential}
\end{align}
where $h\gg d$ is the distance from the film to the ground plane,
and $\varepsilon$ is the dielectric permittivity of the substrate.
In what follows, we will neglect the screening, for simplicity, which
amounts to assuming $h\sim L\gg w$.

The Hamiltonian of the plasmonic modes in the absence of retardation
then reads
\begin{align}
H & =\frac{1}{2}\intop d^{2}\boldsymbol{r}d^{2}\boldsymbol{r}'\,\rho\left(\boldsymbol{r}\right)V\left(\boldsymbol{r}-\boldsymbol{r}'\right)\rho\left(\boldsymbol{r}'\right)\nonumber \\
 & +d\intop d^{2}\boldsymbol{r}\,f_{\text{kin}}\left(\frac{\hbar c}{2e}\nabla\varphi\right),
\label{eq:app-Kerr_Hamiltonian-1}
\end{align}
where $d$ is the film thickness, and the superconducting phase $\varphi$
is connected to the vector potential as~$\boldsymbol{A}=\frac{\hbar c}{2e}\nabla\varphi$
(and also assumed to be uniform across the thickness of the film).
The canonical commutation relations for the fields $\varphi,\rho$
take place:
\begin{equation}
\left[\varphi\left(\boldsymbol{r}\right),\rho\left(\boldsymbol{r}'\right)\right]=\left(2e\right)i\,\delta\left(\boldsymbol{r}-\boldsymbol{r}'\right).
\end{equation}
At the boundary of the film, the field $\varphi$ should satisfy the
condition $\left(\boldsymbol{n},\boldsymbol{j}\right)=0$, corresponding
to the absence of the current through the edges of the film:
\begin{equation}
\left(n,\nabla\varphi\right)_{\boldsymbol{r}\in\text{boundary}}=0,
\label{eq:app-Kerr_boundary-conditions}
\end{equation}
where $n$ is the normal vector to the boundary of the film. The Heisenberg's
equations of motion associated with Eq.~(\ref{eq:app-Kerr_Hamiltonian-1})
correspond to the charge conservation law and the Josephson's relation:
\begin{equation}
\dot{\rho}=-\text{div}\boldsymbol{j},\,\,\,\dot{\varphi}\left(\boldsymbol{r}\right)=\frac{2e}{\hbar}\,\intop d^{2}\boldsymbol{r}'\,V\left(\boldsymbol{r}-\boldsymbol{r}'\right)\rho\left(\boldsymbol{r}'\right).
\label{eq:app-Kerr_eqs-of-motion}
\end{equation}
By using the expansion~(\ref{eq:app-Kerr_kinetic-energy-density})
of $f_{\text{kin}}$ in powers of $\boldsymbol{A}$, we can rewrite
the Hamiltonian~(\ref{eq:app-Kerr_Hamiltonian-1}) as
\begin{align}
H & =\frac{1}{2}\,\intop d^{2}\boldsymbol{r}d^{2}\boldsymbol{r}'\,\rho\left(\boldsymbol{r}\right)V\left(\boldsymbol{r}-\boldsymbol{r}'\right)\rho\left(\boldsymbol{r}'\right)\nonumber \\
 & +\frac{1}{2}\Theta\intop d^{2}\boldsymbol{r}\left(\nabla\varphi\right)^{2}\left[1-\frac{\alpha}{2}(\xi\nabla\varphi)^{2}+O\left((\xi\nabla\varphi)^{4}\right)\right],
\label{eq:app-Kerr_Hamiltonian-2}
\end{align}
where $\Theta=\rho_{S}d\,\left(\hbar/2e\right)^{2}$ is the 2D superfluid
stiffness. Strictly speaking, the energy of phase fluctuations in
Eqs.~(\ref{eq:app-Kerr_Hamiltonian-1})~and~(\ref{eq:app-Kerr_Hamiltonian-2})
also contains terms with higher derivatives of $\varphi$, such as,
e.g., $\left(\xi\nabla^{2}\varphi\right)^{2}$, but those are responsible
for the relative corrections of the order $O\left((k\xi)^{2}\right)$
to the plasmonic dispersion law, whereas we are interested in the
leading nonlinearity w.r.t $\varphi$ that produces interaction of
plasmons. As as a result, such terms can be neglected as far as the
small frequencies and wave numbers are concerned.

\subsection{Low-lying normal modes}

To calculate the cross-Kerr coefficients, one first has to extract
the normal modes of the Hamiltonian~(\ref{eq:app-Kerr_Hamiltonian-2})
in the absence of nonlinearity. The latter can be done by finding
the oscillating solutions to the classical equations of motions~(\ref{eq:app-Kerr_eqs-of-motion}):
\begin{equation}
\rho=e^{i\omega t}\rho_{\omega}\left(x,y\right),\,\,\,\varphi=e^{i\omega t}\varphi_{\omega}\left(x,y\right),
\end{equation}
rendering
\begin{equation}
i\omega\rho_{\omega}=\rho_{S}d\,\frac{\hbar}{2e}\,\Delta\varphi,\,\,\,i\omega\varphi_{\omega}=\frac{2e}{\hbar}\,V\rho_{\omega},
\end{equation}
where we introduced the shorthand $V\rho_{\omega}$ for the r.h.s.
of the second relation in Eq.~(\ref{eq:app-Kerr_eqs-of-motion})
for brevity. For a given set of solutions $\left\{ \omega,\rho_{\omega},\varphi_{\omega}\right\} $
with positive angular frequencies $\omega>0$, the harmonic part of
the plasmonic Hamiltonian~(\ref{eq:app-Kerr_Hamiltonian-2}) is rewritten
as
\begin{equation}
H=\sum_{\omega}\hbar\omega\,\left(\alpha_{\omega}^{\dagger}\alpha_{\omega}+\frac{1}{2}\right),
\end{equation}
where $\alpha_{\omega}^{\dagger},\,\alpha_{\omega}$ are the creation
and annihilation operators of the plasmonic modes satisfying the standard
bosonic commutation relations, and the field operators are then expressed
as
\begin{equation}
\rho\left(x,y\right)=\sum_{\omega}\sqrt{\frac{\hbar\omega}{\rho_{\omega}V\rho_{\omega}}}\,\frac{\alpha_{\omega}-\alpha_{\omega}^{\dagger}}{\sqrt{2}i}\,\rho_{\omega}\left(x,y\right),
\end{equation}
\begin{equation}
\varphi\left(x,y\right)=\sum_{\omega}\sqrt{\frac{2e}{i\varphi_{\omega}\rho_{\omega}}}\frac{\alpha_{\omega}+\alpha_{\omega}^{\dagger}}{\sqrt{2}\,}\,i\varphi_{\omega}\left(x,y\right).
\end{equation}
where denoted $\varphi_{\omega}\rho_{\omega}=\intop d^{2}\boldsymbol{r}\,\varphi_{\omega}\left(\boldsymbol{r}\right)\rho_{\omega}\left(\boldsymbol{r}\right)$
and $\rho_{\omega}V\rho_{\omega}=\intop d^{2}\boldsymbol{r}\,d^{2}\boldsymbol{r}'\,\rho_{\omega}\left(\boldsymbol{r}\right)\,V\left(\boldsymbol{r}-\boldsymbol{r}'\right)\rho_{\omega}\left(\boldsymbol{r}\right)$
for brevity.

For the low-lying modes in the strip geometry, one can perform the
plane wave ansatz:
\begin{equation}
\rho_{\omega}=\sqrt{\frac{2}{L}}\cos kx\,\rho_{k}\left(y\right),\,\,\,\varphi_{\omega}=\sqrt{\frac{2}{L}}\cos kx\,\varphi_{k}\left(y\right),
\label{eq:app-Kerr_plain-wave-ansatz}
\end{equation}
with $\partial_{y}\varphi_{k}\left(\pm w/2\right)=0$ as a result
of the boundary conditions~(\ref{eq:app-Kerr_boundary-conditions}).
This leads to the following system of equations for the profile of
the plasmon in the perpendicular direction:
\begin{equation}
i\omega\rho_{k}\left(y\right)=\rho_{S}d\,\frac{\hbar}{2e}\,\left(-k^{2}+\partial_{y}^{2}\right)\varphi_{k},
\label{eq:app-Kerr_strip-charge-conserv}
\end{equation}
\begin{equation}
i\omega\varphi_{k}\left(y\right)=\frac{2e}{\hbar}\,\intop_{-w/2}^{w/2}dy'\,V_{k}\left(y-y'\right)\,\rho_{k}\left(y'\right),
\label{eq:app-Kerr_strip-Josephson-rel}
\end{equation}
\begin{equation}
V_{k}\left(y\right)=\intop_{-\infty}^{\infty}dx\,e^{ikx}\,V\left(x,y\right).
\label{eq:app-Kerr_strip-Coulomb-kernel}
\end{equation}

The system (\ref{eq:app-Kerr_strip-charge-conserv}-\ref{eq:app-Kerr_strip-Coulomb-kernel})
can be solved analytically for the case of the potential~(\ref{eq:app-Kerr_stirp-electristatic-potential})
in the limit $kw\ll1$ and $w\ll h$. We start by discussing the static
limit $k\rightarrow0$, as it formally corresponds to the electrostatic
problem for the same geometry. The charge distribution is then found
from the electroneutrality condition $\boldsymbol{E}=0$:
\begin{equation}
\partial_{y}\,\intop_{-w/2}^{w/2}dy'\,V_{0}\left(y-y'\right)\,\rho_{0}\left(y'\right)=0,
\label{eq:app-Kerr_strip-electro-neutrality}
\end{equation}
where we have used only the $y$-component of the electric field,
as the $x$ component vanishes automatically in the static limit:
$\partial_{x}V\rho\sim k\,V\rho\rightarrow0$. In the limit $k\rightarrow0$,
the Fourier transform~(\ref{eq:app-Kerr_strip-Coulomb-kernel}) of
the electrostatic potential~(\ref{eq:app-Kerr_stirp-electristatic-potential})
reads:

\begin{align}
V_{0}\left(y\right) & =\frac{2}{1+\varepsilon}\left[-\frac{2\varepsilon}{1+\varepsilon}\sum_{j=1}^{\infty}\,\left(\frac{1-\varepsilon}{1+\varepsilon}\right)^{j-1}\,2\ln\frac{w/2}{2hj}\right]\nonumber \\
 & +\frac{2}{1+\varepsilon}\left[2\ln\frac{w/2}{\left|y\right|}+O\left(\left(\frac{y}{h}\right)^{2}\right)\right],
\end{align}
where in the second line we have taken into account that $w\ll h$.

In the absence of screening, i.e., $h\rightarrow\infty$, Eq.~(\ref{eq:app-Kerr_strip-electro-neutrality})
can be solved exactly:
\begin{equation}
\rho_{0}\left(y\right)=\frac{C}{\sqrt{\left(w/2\right)^{2}-y^{2}}},
\label{eq:app-Kerr_charge-profile_k=00003D0}
\end{equation}
where $C$ is a normalization constant. The corrections from finite
$h$ can be retrieved by formal expansion in even Chebyshev polynomials:
\begin{equation}
\rho_{0}\left(y\right)=\frac{1}{\sqrt{\left(w/2\right)^{2}-y^{2}}}\,\sum_{n=0}^{\infty}\rho^{\left(n\right)}T_{2n}\left(2y/w\right),
\end{equation}
with $\rho^{\left(n\right)}=O\left(\left[\frac{w}{2h}\right]^{2n}\right)$.
For instance,
\begin{equation}
\rho^{\left(0\right)}=C,\,\,\,\rho^{\left(1\right)}=-C\left(\frac{w}{2h}\right)^{2}\frac{1}{16}\,\frac{2\varepsilon}{1-\varepsilon}\text{Li}_{2}\left(\frac{1-\varepsilon}{1+\varepsilon}\right),
\end{equation}
where $\text{Li}_{n}\left(z\right)=\sum_{k=1}^{\infty}z^{k}/k^{n}$
is the the polylogarithm function, and $C$ is the normalization constant.
In what follows, we will ignore those corrections.

At finite $k$, the charge distribution $\rho_{k}\left(y\right)$
perturbatively deviates from its static profile $\rho_{0}\left(y\right)$,
so we can find the leading approximation for the phase profile $\varphi_{k}$
by using the Josephson's relation~(\ref{eq:app-Kerr_strip-Josephson-rel})
with the unperturbed value of the charge distribution:
\begin{align}
\varphi_{k} & =\frac{1}{i\omega_{k}}\,\frac{2e}{\hbar}\,\intop_{-w/2}^{w/2}dy'\,V_{k}\left(y-y'\right)\,\rho_{k}\left(y'\right)\nonumber \\
 & \approx\frac{1}{i\omega_{k}}\,\frac{2e}{\hbar}\,\intop_{-w/2}^{w/2}dy'\,V_{k}\left(y-y'\right)\,\rho_{0}\left(y'\right),
\label{eq:app-Kerr_phase-via-Josephson}
\end{align}
By construction, the last expression is constant at $k\rightarrow0$,
so the leading approximation the phase is distributed uniformly. The
exact value of $\varphi_{k}$ can then be found by integrating the
charge conservation~(\ref{eq:app-Kerr_strip-charge-conserv}):
\begin{equation}
\varphi_{k}\approx-\frac{i\omega_{k}}{k^{2}w}\,\intop_{-w/2}^{w/2}dy\,\rho_{0}\left(y\right)=-\frac{i\omega_{k}\,\pi C}{\rho_{S}d\,\frac{\hbar}{2e}\,k^{2}w}.
\label{eq:app-Kerr_via-charge-conserv}
\end{equation}

It is instructive to follow how the charge conservation~(\ref{eq:app-Kerr_strip-charge-conserv})
is respected given the singular profile~(\ref{eq:app-Kerr_charge-profile_k=00003D0})
of the charge density and nearly constant profile of the phase. The
singularity in the charge profile is compensated by a small but singular
corrections to $\varphi_{k}$ at finite $k$:
\begin{equation}
\varphi_{k}=C_{k}+\omega_{k}\left[1-\left(\frac{y}{w/2}\right)^{2}\right]^{3/2}f_{k}\left(y\right),
\end{equation}
where the constant term $C_{k}$ is given by Eq.~(\ref{eq:app-Kerr_via-charge-conserv}),
and $f\left(y\right)$ is a regular function of $y$ that weakly depends
on $k$ at $k\rightarrow0$. Plugging this expression in the r.h.s
Eq.~(\ref{eq:app-Kerr_strip-charge-conserv}) renders a singular
term identical to the one present in the l.h.s. 

Demanding the two Eqs.~(\ref{eq:app-Kerr_phase-via-Josephson})~and~(\ref{eq:app-Kerr_via-charge-conserv})
to be consistent then delivers the plasmonic spectrum:
\begin{equation}
\omega_{k}^{2}=\rho_{S}d\,k^{2}\,\frac{\intop_{-w/2}^{w/2}dy'\,V_{k}\left(y-y'\right)\,\rho_{0}\left(y'\right)}{\intop_{-w/2}^{w/2}dy\,\rho_{0}\left(y\right)},
\end{equation}
where we have retained the $k$-dependence in $V_{k}$ as it contains
the weak logarithmic correction to the leading approximation. Using
the exact profile~(\ref{eq:app-Kerr_stirp-electristatic-potential})
of the electrostatic potential, one obtains asymptotic behavior of
the result in the two limiting cases:
\begin{equation}
\omega_{k}^{2}=\begin{cases}
\frac{4}{1+\varepsilon}\,\rho_{S}dk^{2}w\,\ln\frac{8h}{we^{\delta\left(\varepsilon\right)}}, & k\ll h^{-1},\\
\frac{4}{1+\varepsilon}\,\rho_{S}dk^{2}w\,\ln\frac{8}{kwe^{\gamma}}, & w^{-1}\gg k\gg h^{-1},
\end{cases}
\end{equation}
where $\gamma=0.577...$ is the Euler-Mascheroni constant, and $\delta\left(\varepsilon\right)$
is defined as
\begin{equation}
\delta\left(\varepsilon\right)=\frac{2\varepsilon}{1+\varepsilon}\sum_{j=1}^{\infty}\,\left(\frac{1-\varepsilon}{1+\varepsilon}\right)^{j-1}\,\ln\frac{1}{j}.
\end{equation}
Note, in particular, that the logarithmic correction to the 1D plasmon
dispersion law at $w^{-1}\gg k\gg h^{-1}$ has been predicted theoretically~\citep{Mooij1985}
and observed experimentally \citep{Camarota2001,Charpentier2023_thesis},
with the latter being in perfect agreement, including the numeric
coefficient inside the logarithm. Corrections to $\omega_{k},\,\rho_{k},\,\varphi_{k}$
for $h\sim w$ and/or $kw\sim1$ can be studied numerically, as has
been done to verify the presented results.

\subsection{Cross-Kerr coefficients}

We can now expand the leading anharmonic part of the Hamiltonian in
harmonic modes, according to Eq.~(\ref{eq:app-Kerr_via-charge-conserv}):
\begin{align}
\delta H & =-\frac{\alpha}{4}\,\Theta\xi^{2}\,\intop d^{2}r\,\left|\nabla\varphi\right|^{4}\nonumber \\
 & =-\sum_{\left\{ \omega_{i}\right\} }\eta_{\left\{ \omega_{i}\right\} }\prod_{i}\left(\alpha_{\omega_{i}}+\alpha_{\omega_{i}}^{\dagger}\right)
\label{eq:app-Kerr_Hamiltonian-correction}
\end{align}
where
\begin{align}
\eta_{\left\{ \omega_{i}\right\} } & =\frac{\alpha}{4}\,\Theta\xi^{2}\,\left(\prod_{i=1}^{4}\,\frac{1}{\sqrt{2}}\sqrt{\frac{2e}{i\intop dr\,\varphi_{i}\rho_{i}}}\right)\nonumber \\
 & \,\,\,\intop d^{2}r\,\left(\nabla\varphi_{1},\nabla\varphi_{2}\right)\left(\nabla\varphi_{3},\nabla\varphi_{4}\right).
\end{align}
The calculation of the cross-Kerr effect is then similar to that of
Ref.~\citep[Appendix B]{weissl2015kerr}. We substitute the particular
form, Eqs.~(\ref{eq:app-Kerr_plain-wave-ansatz}),~(\ref{eq:app-Kerr_charge-profile_k=00003D0}),~and~(\ref{eq:app-Kerr_via-charge-conserv}),
of the $\varphi,\rho$ vectors, rendering a simple expression:
\begin{equation}
\eta_{\left\{ \omega_{i}\right\} }=\frac{\alpha}{4}\Theta\,\frac{2\xi^{2}}{Lw}\left(\prod_{i=1}^{4}\sqrt{\frac{\hbar\omega_{i}}{\Theta}}\right)\frac{1}{L}\intop dx\,\prod_{i}\sin k_{i}x.
\end{equation}
In particular, $\eta_{i}$ has complete symmetry w.r.t permutations.
We can then expand the sum in Eq.~(\ref{eq:app-Kerr_Hamiltonian-correction})
according to the number of coinciding frequencies:

\begin{align}
\delta H & =-\sum_{\omega}\eta_{\omega\omega\omega\omega}\left(\alpha_{\omega}+\alpha_{\omega}^{\dagger}\right)^{4}\nonumber \\
 & -3\sum_{\omega\neq\nu}\eta_{\omega\omega,\nu\nu}\left(\alpha_{\omega}+\alpha_{\omega}^{\dagger}\right)^{2}\left(\alpha_{\nu}+\alpha_{\nu}^{\dagger}\right)^{2}\nonumber \\
 & -4\sum_{\omega\neq\nu}\eta_{\omega\omega\omega\nu}\left(\alpha_{\omega}+\alpha_{\omega}^{\dagger}\right)^{3}\left(\alpha_{\nu}+\alpha_{\nu}^{\dagger}\right)\nonumber \\
 & -6\sum_{\omega\neq\nu,\theta\neq\omega,\nu}\eta_{\omega\omega\nu\theta}\left(\alpha_{\omega}+\alpha_{\omega}^{\dagger}\right)^{2}\left(\alpha_{\nu}+\alpha_{\nu}^{\dagger}\right)\left(\alpha_{\theta}+\alpha_{\theta}^{\dagger}\right)\nonumber \\
 & -\sum_{\text{all }\omega_{i}\text{ diff.}}\underset{0,\text{ momentum conservation}}{\underbrace{\eta_{\left\{ \omega_{i}\right\} }}}\prod_{i}\left(\alpha_{\omega_{i}}+\alpha_{\omega_{i}}^{\dagger}\right).
\end{align}
To describe the cross-Kerr effect by this expression, we pick out
only terms that map any two-excitation state to itself. The result
reads

\begin{align}
\delta H & \mapsto-\sum_{\omega}\eta_{\left\{ \omega\right\} }\left[6n_{\omega}^{2}+6n_{\omega}+3\right]\nonumber \\
 & -3\sum_{\omega,\nu\neq\omega}\eta_{\omega\omega,\nu\nu}\left[4n_{\omega}n_{\nu}+2n_{\omega}+2n_{\nu}+1\right]
\end{align}
We can finally rewrite this as
\begin{equation}
\delta H=\sum_{\omega}\delta\omega\,\left(n_{\omega}+\frac{1}{2}\right)-\frac{1}{2}\sum_{\omega,\nu}K_{\omega\nu}n_{\omega}n_{\nu},
\end{equation}
where
\begin{equation}
K_{\omega\nu}=12\left(2-\delta_{\omega\nu}\right)\eta_{\omega\omega\nu\nu},
\end{equation}
\begin{equation}
\delta\omega=-\frac{1}{2}K_{\omega\omega}-\frac{1}{4}\sum_{\nu\neq\omega}K_{\omega\nu}.
\end{equation}
The relevant values of $\eta$ then read
\begin{equation}
\eta_{\omega\omega\nu\nu}=\Theta\,\left(1+\frac{1}{2}\delta_{\omega\nu}\right)\,\frac{1}{4}\,\frac{\alpha}{4}\,\frac{2\xi^{2}}{Lw}\,\frac{\hbar\omega}{\Theta}\,\frac{\hbar\nu}{\Theta},
\end{equation}
rendering the final result for the cross-Kerr coefficients $K_{\omega\nu}$:
\begin{equation}
K_{\omega\nu}=3\alpha\left(1-\frac{1}{4}\delta_{\omega\nu}\right)\,\Theta\,\frac{\xi^{2}}{Lw}\,\frac{\hbar\omega}{\Theta}\,\frac{\hbar\nu}{\Theta},
\label{eq:app-Kerr_cross-Kerr-coeffs}
\end{equation}
where the value of $\alpha=\frac{3}{4\pi}+\frac{\pi}{4}\approx1.02$
originates from the particular form of the nonlinearity of the current-phase
relation~(\ref{eq:app-Kerr_kinetic-energy-density}) in moderately
disordered superconductors.

The result~(\ref{eq:app-Kerr_cross-Kerr-coeffs}) can be qualitatively
compared to that for a chain of Josephson junctions with short-range
capacitive coupling from Ref.~\citep{Krupko2018}. In the latter
case, the nonlinearity in the current-phase relation is described
by $\sin x\approx x-x^{3}/6\Rightarrow\alpha=1/6$, and the factor
$N$ in \citep[Eq. (14)]{Krupko2018} corresponds to $Lw/2\xi^{2}$.

We finally remind the reader that this result is only applicable to
the case of moderately disordered superconductors described by semiclassical
approximation. In contrast, the current-phase nonlinearity in strongly
disordered superconductors with a pseudogap is yet to be described
theoretically.

\end{document}